\documentclass[aps,twocolumn,preprintnumbers,amsmath,amssymb,longbibliography]{revtex4-1}

\usepackage{graphicx}
\usepackage{dcolumn}
\usepackage{bm}
\usepackage{multirow}
\usepackage{color}
\usepackage{amsmath}
\usepackage{gensymb}
\usepackage[colorlinks=true]{hyperref}
\usepackage{textcomp}
\usepackage{placeins}

\usepackage[dvipsnames]{xcolor}
\usepackage{amsmath}
\usepackage{amssymb}
\usepackage{bm}
\usepackage{epsfig}
\usepackage{graphicx}
\usepackage{color}
\usepackage{hyperref}

\newcommand{\specialcell}[2][c]{%
  \begin{tabular}[#1]{@{}c@{}}#2\end{tabular}}

\newcommand{\addAnna}[1]{\textcolor{red}{#1}}

\def\be{\begin{equation}}
\def\ee{\end{equation}}
\def\bea{\begin{eqnarray}}
\def\eea{\end{eqnarray}}

\begin{document}

	\title{Electron, hole and exciton effective $g$-factors \\ in semiconductor nanocrystals}
	\author{M. A. Semina\footnote{semina@mail.ioffe.ru}, A.A. Golovatenko and A. V. Rodina}
	\affiliation{Ioffe Institute, 194021, St.-Petersburg, Russia. }
	
	\begin{abstract}
We review the existing  and present the new results of  ${\bm k} \cdot {\bm p}$ calculations  of the electron, hole, and exciton effective $g$-factors in semiconductor nanocrystals of different shape and  symmetry.  We propose a simple yet accurate method  for calculation of electron $g$-factor size dependence in bare nanocrystals within the eight-band Kane model. Using the spherical approximation for Luttinger Hamiltonian we find the dependence of hole $g$-factor on light to heavy hole effective mass ratio in semiconductor nanostructures with spherical, axial, and cubically symmetric shape. We show that the non-equidistant Zeeman splitting of the four-fold degenerate hole state may take place in cube and spheroidal nanocrystals. We present a comparison of the calculated hole $g$-factors in nanostructures based on II-VI and III-V semiconductors for different sets of the Luttinger parameters and analyze the main effects contributing to the $g$-factor renormalization with the respect to the bulk value.  We discuss different approaches to the definition of the hole and exciton $g$-factors which should be taken into account during the analysis of the experimental data and  compare  our results of $g$-factor calculations with the experimental data for semiconductor spherical nanocrystals and thin nanoplatelets available in the literature. 
%
	\end{abstract}
	
\date{\today}
	
	\maketitle
	
\section{Introduction}
For several decades the synthesis of semiconductor nanocrystals by colloidal chemistry has passed into the state of a mature technology. It provides precise control over the size, shape, and composition of nanocrystals \cite{Doneg2011, Ithurria2008, Kovalenko2015}. Today colloidal semiconductor nanocrystals (NCs) are widely demanded by the market. The area of NCs application includes solar cells, displays, photodetectors, and molecular sensors \cite{GrahamRowe2009, Lohse2012, Freeman2012, Kamat2012, Carey2015, Lhuillier2016, Wang2016}.  All these applications are mostly based on emission or absorption of light by spatially confined electron-hole pairs. Spin properties of confined charge carriers in colloidal nanocrystals are of great interest from both experiment and theory due to the prospect of their use in spintronics and quantum computing devices \cite{Loss1997, Imamoglu1999, NadjPerge2010, Warburton2013, Cao2016}.
The control over the spin state of carriers requires knowledge about their $g$-factors, spin coherence times, the fine structure of their states, and states of their complexes.

The $g$-factor determines the response of the electrons, holes, or their complexes to the external magnetic field including the Zeeman energy splitting between spin sublevels. 
The comprehensive experimental and theoretical studies are being performed in this direction. Using different magneto-optical techniques $g$-factors of electrons, holes and  excitons were measured \cite{Kuno1998,Gupta2002,Htoon2009,Biadala2010, Fernee2012nc, Liu2013}. Theoretical understanding of  experimental data on electron $g$-factor in semiconductor nanostructures of different size, shape, and dimensionality were developed within multiband ${\bm k} \cdot {\bm p}$ theory \cite{Kiselev1992, Kiselev1998,Rodina2003, Yugova2007} and tight-binding calculation \cite{Schrier2003,Tadjine2017}. The surface-induced contribution to the electron $g$-factor in core/shell nanocrystals was studied as well \cite{Rodina2003}. 

Quite often researchers extract from experiments  $g$-factor of excitons \cite{Htoon2009,Biadala2010}, i.e. electron-hole pairs coupled by strong direct and exchange Coulomb interaction. Moreover, colloidal nanocrystals can be in a charged state due to intentional chemical doping, or as a result of photocharging, when  electron or hole is trapping to the surface from photoexcited exciton. The subsequent act of photoexcitation of charged nanocrystal results in the formation of a trion, charged three-particle exciton complex. Both in the case of exciton and trion the knowledge of hole $g$-factor becomes crucial. However, much less attention was devoted to the properties of the hole and exciton $g$-factors in semiconductor nanostructures as compared with the electron $g$-factor \cite{Chen2004, Csontos2009}.

\begin{figure}[!ht]
\includegraphics[width=0.99\columnwidth]{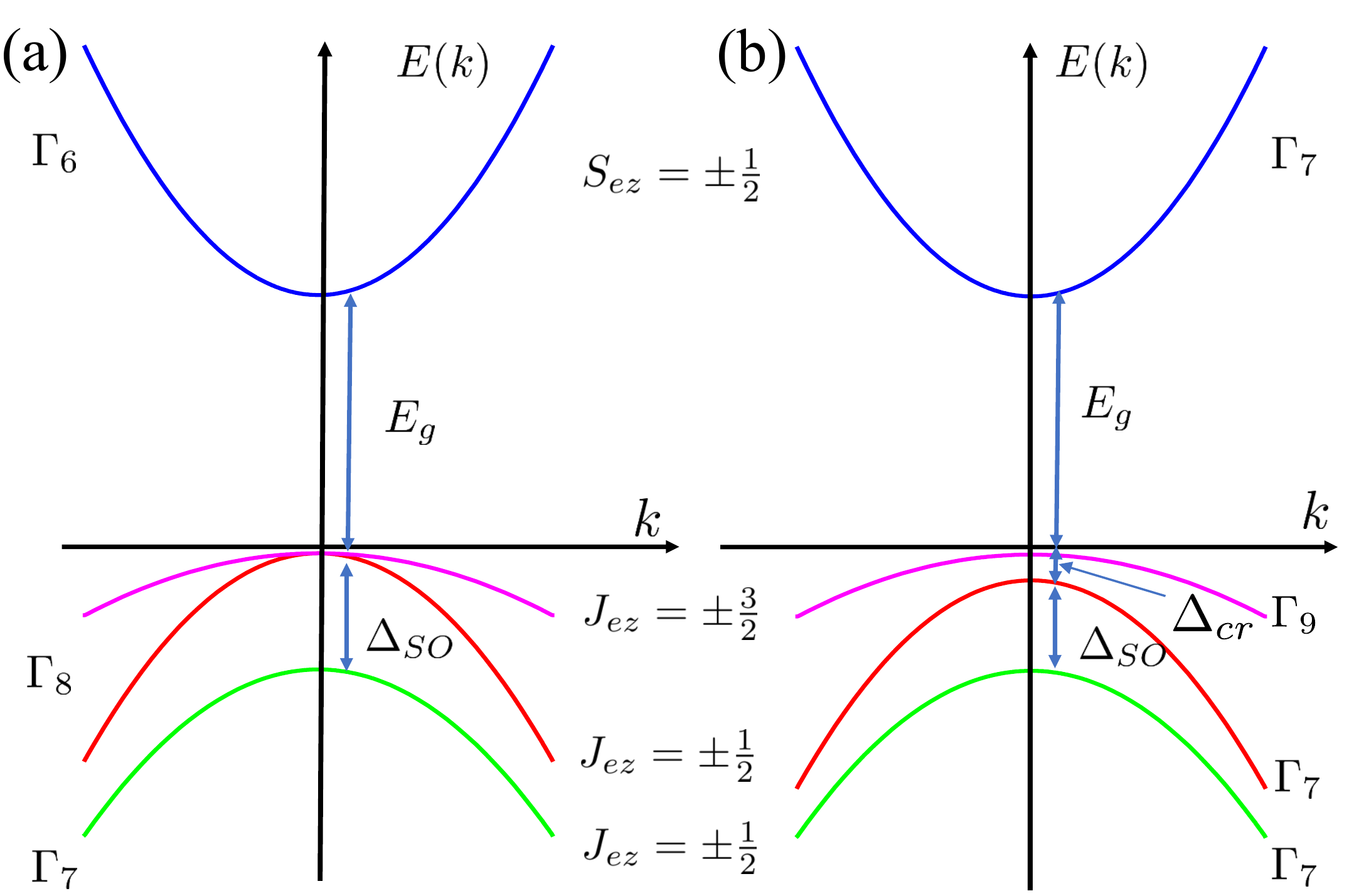}
   \caption{The band structure of the $\text{A}_{\rm II}\text{B}_{\rm VI}$ or $\text{A}_{\rm III}\text{B}_{\rm V}$ typical semiconductor with (a) zinc-blende; (b) wurtzite crystal structure. Note, that for graphic reasons $E_g$, $\Delta_{SO}$ and $\Delta_{cr}$  are shown not to scale.}
   \label{fig:1}
\end{figure}

In present paper we consider nanocrystals based on  direct gap semiconductors with bandgap in $\Gamma$ point of the  Brillouin zone with $\Gamma_6$ conduction band (electron spin $S=1/2, S_z=\pm 1/2$), $\Gamma_8$ topmost valence band  (electron spin $J=3/2, J_{ez}=\pm 1/2,\pm 3/2$) and  $\Gamma_7$ spin-split valence band (electron spin $J=1/2, J_{ez}=\pm 1/2$). The corresponding energy bands are shown schematically in Fig. \ref{fig:1} (a) in the electron representation. It could be, for example, zinc blende modifications of $\text{A}_{\text{II}}\text{B}_{\text{VI}}$ or $\text{A}_{\text{III}}\text{B}_{\text{V}}$ compounds, such as zb-CdSe and GaAs. Spin projections $J_{ez}=\pm 3/2$ and $J_{ez}=\pm 1/2$ describe so-called heavy and light hole subbands, respectively.  In wurtzite modification of CdSe (wz-CdSe), which is widely used for fabrication of colloidal nanocrystals, the effective crystal field splits the $\Gamma_8$ topmost valence subband into two $\Gamma_9$ (heavy holes) and $\Gamma_7$ (light holes) subbands, see scheme in Fig. \ref{fig:1} (b).  The heavy and light hole splitting $\Delta_{cr}$ is typically small enough (25 meV in bulk wz-CdSe) to be treated as a perturbation. 
For these semiconductors dependencies of hole $g$-factor on type and dimensionality of confining potential are calculated with respect to the ratio of light hole and heavy hole masses. We discuss the relationship between various formalisms in defining the hole g-factor, common in various communities. The relation between electron, hole, and exciton $g$-factors in different types of nanostructures is also considered. We make the calculations and compare the results with the experimental data for  colloidal nanocrystals assuming high potential barriers for the electrons and holes at the NC surface.

The rest of the paper is organized as follows: In Sect. \ref{electron} we revisit the electron $g$-factor calculation in the framework of the eight-band ${\bm k} \cdot {\bm p}$ model and compare results with the results of the tight-binding calculations as well as with experimental data. In Sect. \ref{hole} we discuss the definition of hole $g$-factor (\ref{hole_def}), present semi-analytical effective mass methods and results of calculation of hole $g$-factors in spherical and cube NCs, quantum wires and planar two-dimensional nanoplatelets (NPLs)  (\ref{hole_calc}). We compare the results of the hole $g$-factor calculations in different semiconductor nanostructures with published experimental data and discuss the dependence of hole $g$-factor on its size in Sect. (\ref{size}). In Sect. \ref{exciton} we discuss the definition and size dependence of $g$-factor of different exciton states in low-dimensional structures.  We summarize our results in Sect. \ref{sum}. The additional  details of calculations are presented in Appendices \ref{AA}, \ref{AB} and \ref{AC}.

\section{Electron $g$-factor in the eight-band \texorpdfstring{${\bm k}\cdot{\bm p}$}   c -Model}\label{electron}

\begin{figure}[!ht]
	\includegraphics[width=1\columnwidth]{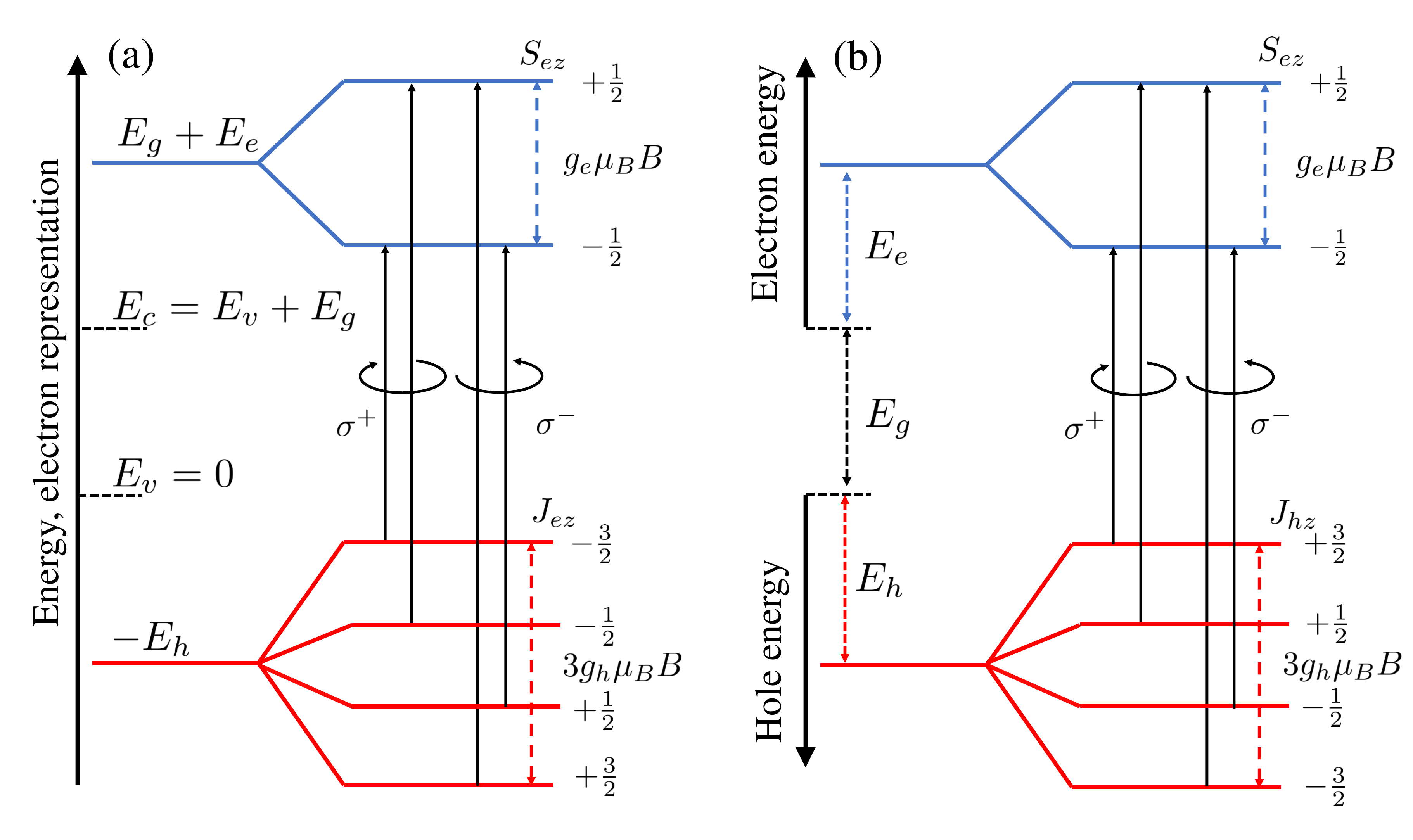}
	\caption{The scheme of  Zeeman energy levels splittings of electrons and $\Gamma_8$ holes in bulk zinc-blende semiconductor (a) in electron and (b) in mixed electron-hole representations. $E_h$ and $E_e$ are the  hole and electron energy levels in zero magnetic field calculated from the top of the valence band, $E_v=0$, and the bottom of the conduction band, $E_c=E_g$, respectively. The order of electron and hole levels is shown for positive $g$-factors both for electrons and holes: $g_e>0$, $g_h>0$.  The circular polarized optical transitions are denoted by $\sigma^+$ and $\sigma^-$. }
	\label{fig:2}
\end{figure}

In the presence of an external magnetic field, electron states with different spin projections on the magnetic field direction split. In the low field regime this  splitting is linear on magnetic field strength and it is defined by electron effective  $g$-factor or Lande factor $g_e$.  Corresponding  Zeeman part of the electron Hamiltonian in the low magnetic field limit can be written as:
\begin{equation}\label{g_e}
\widehat{H}_{Z}^{(e)}=\mu_Bg_e\left(\bm {S}_e \bm B\right), ~g_e=\frac{E_{1/2}-E_{-1/2}}{\mu_B B} \, .
\end{equation}
Here $\mu_B=\frac{e\hbar}{2m_0c}$ is  Bohr magneton,  $m_0$ is free electron mass,  $e=|e|$ is the absolute value of electron charge, and  $E_{\pm 1/2}$ are energies of states with  spin projection $S_{ez}=\pm 1/2$ on the direction of magnetic field.  The scheme of the electron energy levels splitting  is shown in Fig. \ref{fig:2}. 

In bulk semiconductors the spin-orbit interaction is usually parametrically strong and $g_e$  differs significantly from the value of free electron $g$-factor $g_0=2$~\cite{Roth1959}. For example, $g_e = -0.44$ in GaAs~\cite{Weisbuch1977}, $g_e = 0.42$ in zb-CdSe \cite{Karimov2000}, $g_e = 0.68$ in wz-CdSe \cite{Piper1967}, $g_e = -1.66$ in CdTe \cite{Oestreich1996} and $g_e=1.2$ in InP \cite{Oestreich1996}. Spatial localization of electrons in nanostructures leads to additional renormalization of  $g$-factors and even to their spatial anisotropy \cite{Kiselev1992, Kalevich1992, Sirenko1997,Kiselev1998, Yugova2007}. 
The main mechanisms of conduction band electrons $g$-factor renormalization are well-known, and the theoretical calculations quantitatively well describe the experimental data \cite{Sirenko1997,Kiselev1998,Rodina2003,Yugova2007,Tadjine2017}.

The value of electron $g$-factor at the bottom of the conduction band in typical semiconductors with zinc blende crystal structure can be calculated within the second-order ${\bm k} \cdot {\bm p}$ theory. It was first made to account the contributions from the $\Gamma_8$ and $\Gamma_7$ valence subbands  in Ref.~[\onlinecite{Roth1959}]:
\begin{equation}\label{LR_bulk}
g_e\approx g_0-\frac{2E_p}{3}\left(\frac{1}{E_g}-\frac{1}{E_g+\Delta_{SO}}\right) \, .
\end{equation}
Here $E_p=2|\langle X|\hat p_x|S \rangle |^2/m_0$ is the Kane energy expressed via the interband momentum matrix element,  $E_g$  is the bandgap, $\Delta_{SO}$ is the spin-orbit splitting of the valence band. More accurate calculations require consideration of the remote conduction band contributions \cite{Hermann1977}. In most cases, however, it can be accounted for by a fitting  parameter ${g_\text{rb}}$ added to $g_0$ in Eq. \eqref{LR_bulk} in order to obtain the experimental value of the bulk electron $g$-factor $g_e$  when the other parameters $E_p$, $E_g$ and $\Delta_{SO}$ are known.

The eight-band Kane or ${\bm k} \cdot {\bm p}$ model allows one to account for the effect of the electron localization in low-dimensional structures on the $g$-factor value  \cite{ivch_kis_will96,Kiselev98,Rodina2003}. It results in the dependence of the electron $g$-factor on the electron  size quantization energy $E_e$ counted from the bottom of the conduction band as
\begin{equation}\label{geNC}
	g_e(E_e)  = g_0 + \int  (\tilde g_e(E_e) - g_0) | \Psi_e^c({\bm r})|^2 \,d^3 {\bm r} + g_{\rm sur} \, ,
	\end{equation}
	\begin{equation}
\tilde g_e (E_e) \approx g_0+g_{\text{rb}} -\frac{2E_p}{3}\left(\frac{1}{E_g+E_e}-\frac{1}{E_g+\Delta_{SO}+E_e}\right). \label{LR}
\end{equation}
Here $\Psi_e^c({\bm r})$ describes the conduction band contribution to the eight-band electron envelope function  $\Psi_e({\bm r})$  and $ g_{\rm sur}$ describes the surface/interface contribution  proportional to the squared value  $|\Psi_e^c|^2_{\rm sur}$ taken at the surface of the nanostructure or at the interface between two semiconductors in the heterostructure. Importantly, the normalization condition $ \int |\Psi_e({\bm r})|^2 \,d^3 {\bm r}=1$ \cite{Kiselev1998,Rodina2003,Merkulov2010book} implies $ \int |\Psi_e^c({\bm r})|^2 \,d^3 {\bm r}={\cal A}(E_e)<1$ even in bare core nanocrystals. Generally, the renormalization constant ${\cal A}(E_e)$ depends on the electron quantization energy $E_e$ and can be found from the dependence  of electron effective mass on its energy $m_e(E_e)$ reflecting the nonparabolicity of  the conduction band energy dispersion   (see for details the Appendix \ref{AA}). As it was shown in Ref. [\onlinecite{Rodina2008}], for $E_e\ll E_g$ the renormalization of the conduction band contribution can be expressed as  ${\cal A}(E_e) \approx m_e/m_e(E_e)$ where $m_e\equiv m_e(E_e=0)$ is the bulk electron effective mass. Combining together the renormalization of the conduction band contribution in Eqs. \eqref{geNC} and \eqref{LR} we arrive to:
\begin{equation}\label{eqApp2}
	\quad g_e(E_e)=g_0(1-{\cal A}(E_e))+  \tilde g_e(E_e) {\cal A}(E_e) +g_{\rm sur} 
\end{equation} 
for the bare core nanostructures. The distinctive feature of Eq. \eqref{eqApp2} is that it allows one to calculate the electron $g$-factor in nanostructures  knowing the bulk semiconductor parameters $E_p,E_g,\Delta_{SO},m_e,g_e$ and the surface contribution $g_{\rm sur}$.

\begin{figure}[!ht]	\includegraphics[width=0.9\linewidth]{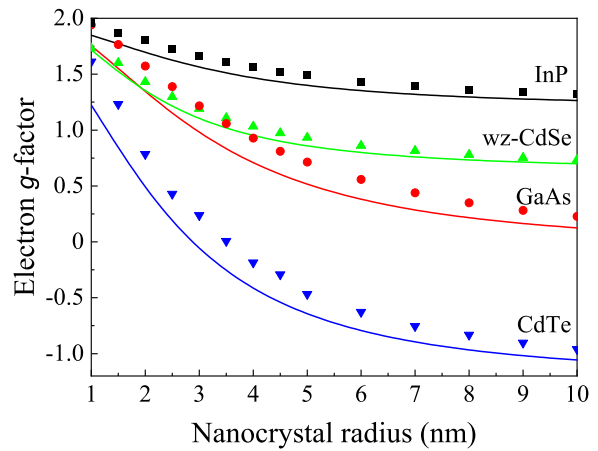}
	\caption{Size dependencies of the electron $g$-factor	in different semiconductors calculated within tight-binding approach, Ref.~[\onlinecite{Tadjine2017}], (symbols) and using the eight-band ${\bm k} \cdot {\bm p}$, Eq. \eqref{eqApp2}, with the same sets of parameters listed in  Table~A1. 
	}\label{gefit_tight}
\end{figure}

\begin{figure}[!ht]	\includegraphics[width=0.9\linewidth]{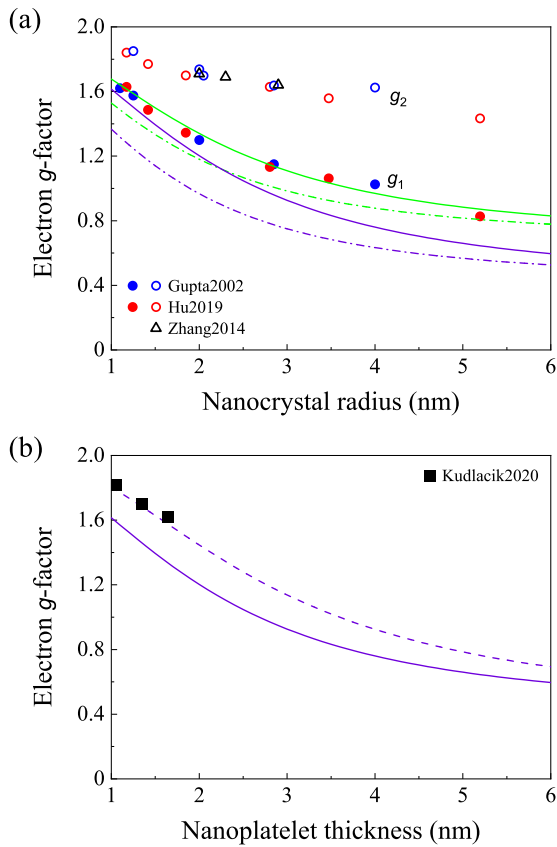}
	\caption{Size dependencies of the electron $g$-factor: (a) in spherical wurtzite (green lines) and zinc blende (purple lines) CdSe nanocrystals.   Solid (dashed) lines  are calculated for the ground $1S_e$ state of the infinite box-like potential according to Eq.~\eqref{eqApp2} (Eq.~\eqref{LR} ) with parameters from  Refs.~[\onlinecite{Ekimov1993,Piper1967}]  and Ref.~[\onlinecite{Karimov2000}] for wz-CdSe and zb-CdSe, respectively. Symbols correspond to experimental $g$-factors  observed via the pump-probe Faraday rotation  in  wz-CdSe NCs,
	Refs.~[\onlinecite{Gupta2002}] (blue circles), [\onlinecite{Hu2019}] (red circles) and  zb-CdSe NCs [\onlinecite{Zhang2014}] (black triangles); (b) in zb-CdSe nanoplatelets calculated for the $1S_e$ state in the infinite box-like (dashed line) and parabolic (solid line) confining potential. Black squares are localized resident electron $g$-factors in zb-CdSe NPLs obtained via the SFRS in Ref.~[\onlinecite{Kudlacik2019}] 
    }\label{gefit}
	\end{figure}

As it follows from Eqs. \eqref{LR} and \eqref{AEe}, additional localization  in nanostructures changes the electron $g$-factor noticeably only if the size-quantization energy $E_e$ is comparable with the bandgap $E_g$. Such a situation is realized in small nanocrystals and,  especially, in nanoplatelets where the localization length of charge carriers,  at least in one dimension, is about 1 or 2 nm  (see Fig. \ref{fig:Eeme}). 
As a result, the contribution of the valence band to the electron  $g$-factor decreases with $\tilde g_e(E_e)$ tending to  $g_0+g_{rb}$ and ${\cal A}(E_e)$ decreasing, so that the $g_e(E_e)$ values become closer to $g_0$.  Importantly,  the eight-band  ${\bm k} \cdot {\bm p}$ calculations of the electron $g$-factor, according to Eq. \eqref{eqApp2} provide a very good agreement with results of tight-binding calculations \cite{Tadjine2017} for spherical NCs made from wz-CdSe, CdTe, GaAs, and InP, see \addAnna{Fig. \ref{gefit_tight}}. We used the same set of bulk parameters for the ${\bm k} \cdot {\bm p}$ calculations in \addAnna{Fig. \ref{gefit_tight}} obtained by the tight-binding calculations in \cite{Tadjine2017}, see Table A1.

In Fig. \ref{gefit} the dependencies of the electron $g$-factor on the radius of wz-CdSe and zb-CdSe nanocrystals calculated according to Eq.~\eqref{eqApp2}  are shown by the green and violet solid lines, respectively. Calculations are made with the set of parameters from Refs. [\onlinecite{Ekimov1993,Piper1967}] and $g_{rb}=0.02$ for wz-CdSe and from Ref.\onlinecite{Karimov2000} and $g_{rb}= -0.125$ for  zb-CdSe, correspondingly, and with $g_{\rm sur}=0$. Dashed lines show the calculations according to Eq.~\eqref{LR}. One can see that the difference between solid, $g_e(E_e)$, and dashed, $\tilde g_e(E_e)$, lines increases with decreasing of NC size related to the ${\cal A}(E_e)$ effect while the difference  between $g$-factors calculated for zb- and wz-CdSe decreases. We show also the values of the $g$-factors measured by the pump-probe Faraday rotation experiment in Refs.  [\onlinecite{Gupta2002,Hu2019}]  for wz-CdSe ($g_1$ and $g_2$ values) and in Ref. \cite{Zhang2014} for zb-CdSe. Calculated $g_e(E_e)$ dependence for wz-CdSe is in a good agreement with the experimental data ($g_1$ values) from Refs. [\onlinecite{Gupta2002,Hu2019}] shown by filled red and blue  circles and attributed the ground state electron $g$-factor. Note, that a good agreement of the calculated size dependence of electron $g$-factor with experiment can be achieved with more than one set of ${\bm k} \cdot {\bm p}$ model parameters. In Refs. \cite{Rodina2003,Gupta2002} another set of  parameters was used together with the general boundary condition assuming non-vanishing electron wave function at NC surface resulting in $g_{\rm sur} \ne 0$.

It is noteworthy, that the  second set of experimental data for wz-CdSe, $g_2$ values, given by open symbols in Fig. \ref{gefit}(a), as well as the data for zb-CdSe,  can not be described by the calculated dependence for electron $g$-factor at lowest quantum size level, $1S_{e}$. There were different assignments of these values to the exciton $g$-factor \cite{Gupta2002} and surface-localized electron $g$-factor \cite{Hu2019} for wz-CdSe and to the electron $g$-factor for zb-CdSe   \cite{Zhang2014}. Wherein, $g_1$ values corresponding to the electron $g$-factor size dependence in zb-CdSe were not observed in spherical NCs. Interestingly, the values larger than $g_1$ and very close to $g_2$ in small NCs can be  obtained from Eq. \eqref{eqApp2} for electrons localized at the excited  $2S_{e}$   size-quantization level. While the calculations of the surface-localized electron $g$-factor are beyond the scope of the present paper, we will discuss the possibility to assign the $g_2$ values to the exciton $g$-factor in Sect. \ref{exciton}.

Recently, $g$-factor of the localized resident electron was measured in zb-CdSe nanoplateletes with thickness of 3, 4 and 5 monolayers (ml) by spin-flip Raman scattering (SFRS) \cite{Kudlacik2019}. The attribution of the obtained $g$-factors to the resident electron (not exciton) is justified by the polarization selection rules \cite{Kudlacik2019,Rodina2020}. In Fig. \ref{gefit}\addAnna{(b)} the measured results are given by black squares. The  solid and dashed lines are results of the calculations according Eq. \eqref{eqApp2} for electrons in the NPLs  with parabolic and box-like confining potentials, respectively   (see Appendix \ref{AA}). One can see that the parabolic potential better describes the experimental data while the measured $g$-factors are about 10 \% larger than calculated for the box-like potential. However, the discrepancy can be related to the effect of $g$-factor anisotropy in the NPL neglected in the presented calculations as well as to the effect of the additional in-plane localization of the resident electron. Therefore, we will examine further  the effect of both confining potentials on the hole and exciton $g$-factors and compare the calculated values with the experimental data.

Besides the renormalization of the   $g_e$-factor controlling the Zeeman splitting of spin sublevels,   electron localization in the nanostructure   also leads to the renormalization of the orbital effective $g$-factor (orbital magnetic momentum)  \cite{Rodina2003,vanBree2014}.  The orbital magnetic field effect comes from the electron kinetic energy term $k^2/2m_e$,   where the electron wave vector  $\bf k$  is replaced by  $\mathbf k + \frac{e}{c}\bm A$, where $\bm A$ is the vector potential of the applied magnetic field.  By this term, one takes into account the velocity generated by the envelope wave function. As it was shown in \cite{vanBree2014}, there is also the second contribution to the orbital magnetic momentum  generated by velocity coming from the Bloch functions. We restrict our consideration hereafter by the account only by the envelope velocity contribution.  The size dependence of the orbital effective $g$-factor is  then controlled by the $m_e/m_e(E_e)$ ratio \cite{Rodina2003}. However, the envelope orbital magnetic field effect does not influence the electron states in bulk semiconductors, and the lowest electron states with zero orbital momentum in most nanostructures.  As we show below, the situation is significantly different for the holes  where the orbital contribution leads to the substantial variation of hole $g$-factors between different nanostructures.

\section{Hole  $g$-factor}\label{hole}
\subsection{Definition}\label{hole_def}

In the present paper, we consider the hole from the four-fold degenerate $\Gamma_8$ valence subband for semiconductors with relatively large spin-orbit splitting (for CdSe, ZnSe $\Delta_{SO}\sim 400$ meV). In these materials, the spin-split valence subband can be safely neglected from consideration.  We also neglect the cubic symmetry of the crystal environment (valence band warping) as it affects the presented results only quantitatively but not strongly.  With these approximations the valence band holes in the external magnetic field can be described by the following Hamiltonian, which we rewrite in the hole representation as the sum of three contributions: \begin{equation}\label{Hamilt}
\widehat{H}=\widehat{H}_L+\widehat{H}^{(h)}_{\text{Z}}+\widehat{H}_{\text{B}}.
\end{equation}
Here
\begin{equation}\label{lutt}
\widehat{H}_L=\frac{\hbar^2}{2 m_0}\left[\left(\gamma_1+\frac{5}{2}\gamma\right)k^2-\gamma\{k_\alpha k_\beta\}\{J_\alpha J_\beta\}\right],
\end{equation}
is the Luttinger Hamiltonian \cite{Luttinger1956,Gelmont1971} in hole representation in zero magnetic field describing hole kinetic energy in spherical approximation.  
 Operator ${\bm J}$ is the hole internal angular momentum operator,  for $\Gamma_8$ valence subband $J=3/2$,  $\gamma_1$ and $\gamma=(2\gamma_2+3\gamma_3)/5$ are Luttinger parameters related to the bulk light-hole, $m_{lh} = m_0/(\gamma_1 + 2 \gamma)$, and heavy- hole, $m_{hh} = m_0/(\gamma_1 - \gamma)$,  effective masses.

Following the classical approach introduced by Luttinger \cite{Luttinger1956}, the Zeeman part of free hole Hamiltonian, $\widehat{H}_{Z}^{(h)}$,  has the form:

\begin{equation}\label{Hzh1}
\widehat{H}_{Z}^{(h)}=-2\mu_B\varkappa\left(\bm J_h\bm B\right)=+2\mu_B\varkappa\left(\bm J_e\bm B\right),
\end{equation}
where $\varkappa$ is the magnetic Luttinger parameter \cite{Luttinger1956}, which  value can be estimated by perturbation theory as \cite{Roth1959}:
\begin{equation}\label{kappa}
\varkappa\approx-2/3+5\gamma/3-\gamma_1/3.
\end{equation}
Here $\bm J_h$ is the hole spin operator in hole representation (in what follows $\bm J\equiv \bm J_h$) and $\bm J_e$ is the electron spin operator in valence band in the electron representation. Both in the electron and hole representations, the sign of $g_h$ and the relative sign between the Zeeman term and kinetic energy remains the same as the change of representation results in sign inversion of both energy and spins. 


The second, orbital, contribution from magnetic field to the hole Hamiltonian, $\widehat{H}_{\text{B}}$, comes from  the hole wave vector  $\bf k$ being replaced  by  $\mathbf k -\frac{e}{c}\bm A$.   The explicit form of $\widehat{H}_{\text{B}}$ can be found in Ref. \cite{Semina2015} (in the units of the heavy hole effective Rydberg energy) and in  Appendix \ref{AB} (in the units of $\mu_{\rm B}B$).  In semiconductors with the cubic lattice symmetry, one can also separate the cubically symmetric contribution to the Zeeman part of hole Hamiltonian, originating from the valence band warping, $\propto q(B_x J_x^3+B_y J_y^3 + B_z J_z^3)$.  In most semiconductors the cubic Luttinger parameter $q$ is small, and the cubic contribution to the hole Zeeman splitting is much smaller then  the isotropic contributions  $\widehat{H}^{(h)}_{\text{Z}}$ and $\widehat{H}_{\text{B}}$ \cite{Marie1999}.  Cubic contribution plays an important role only if symmetric contributions lead to the absence of hole Zeeman splitting. For example, such a situation is realized for heavy holes in in-plane symmetric quantum well-like structures with the magnetic field applied in the structure plane.

 In what follows, we consider the low field regime, and only linear on magnetic field terms are taken into account. In that case, for a bulk hole, the orbital contribution $\widehat{H}_{\text{B}}$ disappears. However, for the localized hole, the states with different orbital momenta are mixed, and $\widehat{H}_{\text{B}}$ becomes important. Also, in high magnetic fields, the contributions from quadratic on magnetic field terms appearing from  $\widehat{H}_{\text{B}}$ become important even for bulk holes. It results in a diamagnetic shift and strong mixing of hole states with $J_{hz}=\pm 3/2$ and $J_{hz}\pm 1/2$ and a consequent possible magnetic field dependence of the hole {effective} $g$-factor.

Here we use the following definition of the hole effective $g$-factor \cite{Gelmont1973,Efros1996}:
\begin{multline}\label{g_h}
 g_h=\frac{E_{-J_{hz}}-E_{+J_{hz}}}{2J_{hz}\mu_B B}=\frac{E_{-3/2}-E_{+3/2}}{3\mu_B B}=
 \\=\frac{E_{-1/2}-E_{+1/2}}{\mu_B B},
\end{multline}
where $J_{hz}$ is the hole  spin projection on the magnetic field direction (which is chosen along the $z$-axis). The positive $g$-factor corresponds to the hole ground state with positive spin projection $J_{hz}$. It is convenient to write the effective Zeeman Hamiltonian \eqref{Hzh1}  as
\begin{equation}\label{Hzh2}
\widehat{H}_{Z}^{(h)}=-\mu_B g_h J_{hz}B.
\end{equation}
Hamiltonian \eqref{Hzh2} describes the splitting of the otherwise four-fold degenerate hole ground state in bulk crystals as well in spherically-symmetric structures having the same $g$-factor for light and heavy holes. For example, in bulk semiconductors,  the  Zeeman effect for both light ($J_{hz}=\pm 1/2$) and heavy ($J_{hz}=\pm 3/2$) holes is characterized by the same $g$-factor $g_h \equiv g_h^{\rm bulk}=2\varkappa$.  Note that the actual Zeeman splitting of heavy holes ($J_{hz}=\pm 3/2$) is three times larger than that of light holes ($J_{hz}=\pm 3/2$) for the same $g$-factor: $\Delta E_{3/2}=3\Delta E_{1/2}$. The scheme of hole energy levels splitting both in the electron and hole representations for non-interacting electron and holes is shown in Fig. \ref{fig:2}.

The definition Eq. \eqref{g_h} and  Hamiltonian Eq. \eqref{Hzh2}  are widely used in the physics of colloidal nanocrystals with  $J_{hz}$  being changed by the total angular momentum projection $M$ \cite{Efros1996,EfrosCh3,Liu2013,Shornikova2020nn,Shornikova2020nl,Shornikova2020acs}. We will use hereafter the  definition of Eqs. \eqref{g_h},\eqref{Hzh2}. 

It has to be noted, that another definition of hole $g$-factor  with the opposite sign  as compared with \eqref{g_h}  and \eqref{Hzh2}, is also widely used in literature, see, for example, Refs. \cite{Kiselev1996,Durnev2012,vanKesteren1990,Rodina2001f}. Moreover, especially for structures with strong light and heavy holes splitting, the heavy hole Zeeman splitting is often defined as $\Delta E_{3/2} = E_{+3/2}-E_{+3/2}= \mu_B g_{hh} B$ with  $g_{hh}=-3g_h=-6\varkappa$ \cite{Kiselev1996,Durnev2012} describing the whole Zeeman splitting of heavy holes. While all definitions of hole $g$-factor follow from the same Zeeman contribution to hole Hamiltonian Eq. \eqref{Hzh1} and describe the same splitting of hole states in the magnetic field, one should carefully consider the chosen definition when comparing of the calculated $g$-factors with experimentally evaluated data.

\subsection{Hole effective {\it g}-factor  in low-dimensional structures}\label{hole_calc}

The symmetry considerations have proven themselves very helpful in the analysis of hole states in the complex valence band.\cite{Broido1985, Broido1985err,SercelPRL90, SercelPRB90,Rego1997} For localized holes, their internal angular momentum and its projections are not good quantum numbers anymore. For example, in spherically symmetric systems, the states could be classified by the hole total angular momentum.\cite{Gelmont1971, Baldereschi1973,SercelPRL90, SercelPRB90} In axially symmetric structures, the good quantum number is the total angular momentum projection on the symmetry axis.\cite{SercelPRB90,Rego1997} In structures with an inversion center, one can use a state "parity''  as a quantum number to classify hole states.\cite{Broido1985,Rego1997} The states with odd and even parity are an analog of electron spin-up and spin-down states and are degenerate in zero magnetic field.

For a localized hole Hamiltonian takes a form
\begin{equation}\label{Hamilt_dot}
\widehat{H}=\widehat{H}_L+\widehat{H}^{(h)}_{\text{Z}}+\widehat{H}_{\text{B}}+V_{\rm ext}({\bf r}),
\end{equation}
where the potential $V_{\rm ext}({\bf r})$ is a quantum structure potential acting on the hole. For a hole in a free  exciton or an acceptor, $V_{\rm ext}({\bf r})$ is the Coulomb potential. The profile, as well as the symmetry of $V_{\rm ext}({\bf r})$, strongly affect the structure of the hole wave function and, therefore, its $g$-factor. 

 In the general case, the problem of hole $g$-factor calculation can not be solved analytically or even semi-analytically. Although, in some special cases, the symmetry of the system allows one to simplify the calculation and obtain the expression for hole $g$-factor in closed semi-analytical form. In this section, we present methods for the hole $g$-factor calculation in such special cases. The main effect of the $V_{\rm ext}({\bf r})$ on the hole $g$-factor is the additional mixing of the states from the heavy and light-hole subbands depending on the mass ratio $\beta=m_{lh}/m_{hh}=(\gamma_1-2\gamma)/(\gamma_1+2\gamma)$. The additional spin-orbit interaction induced by the external potential makes the hole internal angular momentum ${\bm J}_h$ not a good quantum number. We assume in the following that the hole states in each nanostructure are characterized by the total angular momentum projection $M$ on its symmetry axis,  which is parallel to the $z$-direction, and consider the magnetic field along $z$ as well. In this case, one can use the definitions similar to Eqs. \eqref{g_h} and \eqref{Hzh2} for the hole effective $g$-factor and Zeeman term, respectively, with $J_{hz}$ replaced by $M$. However, except for the spherically symmetric confining potential, the Zeeman splitting of the light and heavy subbands  might be controlled by the different effective $g$-factors depending on $|M|$, so that it is instructive to define
 	\begin{equation}\label{g_hM}
 g_{h,|M|} =\frac{E_{-M}-E_{+M}}{2M\mu_B B}, ~ 
 	\widehat{H}_{Z}^{(h)}=-\mu_B g_{h,|M|} M B.
 	\end{equation}

The first correction to the hole effective $g$-factor as compared with bulk comes from the renormalization of the Zeeman term $\propto \varkappa$ of Eq. \eqref{Hzh1}. This renormalization  $\langle \Psi_M | \hat  J_{hz} | \Psi_M \rangle/M$  is {a} function of  $\beta$, { which particular form is controlled by} the shape and type of nanostructure potential as it causes the mixing of holes states with different spin projections on the magnetic field direction.  The second one is related to the orbital contribution $\propto \gamma_1$ and $\gamma$ and is also controlled by some function of $\beta$. For each semiconductor there {are} several parametrizations, ones we use are taken from \cite{Adachi2004, Karazhanov2005, Horodysk2010}. For our estimations, we considered materials with zinc blende crystal structure with the inclusion of wurtzite modification of CdSe.

\subsubsection{Hole {\it g}-factor in spherically symmetric potentials}

The spherically symmetric structures represent the special case of {the highest} possible symmetry. In spherically symmetric external potential ${V}_{\text{ext}}({\bf r}) \equiv V^{\rm sph}(r)$, hole  states can be classified by its total angular momentum.\cite{Gelmont1971, Baldereschi1973,SercelPRL90, SercelPRB90} Its wave function can be written following \cite{Gelmont1971} as
\begin{multline}
\Psi_{M} = \sqrt{2j+1}\sum_{l} (-1)^{l-3/2+M} (i)^l R_{jl}(r)\times \\ \times\sum_{m+\mu = M}
\left(
\begin{array}{ccc}
l & 3/2&j \\ m&\mu&-M
\end{array}
\right)
Y_{l,m} u_\mu \, ,
\label{Gelmont_functions}
\end{multline}
Here $\bm j=\bm J_h+\bm l$ is the hole total angular momentum with $M$ being its $z$-axis projection (in considered case $j=3/2$, $J_h=3/2$), $l$ is the hole orbital momentum, $Y_{lm}$ are spherical harmonics \cite{Edmonds}, $\left(_{m~n~p}^{i~~k~~l}\right)$ are $3j$ Wigner symbols, $u_\mu$ are the Bloch functions of  $\Gamma_8$ valence band with spin $z$-axis projection $\mu$ \cite{ivchenko05a} (for details see Appendix \ref{AB}, Eq. \eqref{Bloch}). 
The hole ground state is four-fold degenerate $SD$-like  state and consists of functions with $l=0$ and $l=2$. For simplicity, we denote the respective radial functions as $R_0$ and $R_2$. Taking into account the spherical symmetry of the structure one can simplify the Schr\"{o}dinger equation and reduce it to the system of two equations for  $R_0$ and $R_2$ \cite{Gelmont1971}.

In external  magnetic field the four-fold degenerate ground state splits into four equidistant levels with $M=\pm 3/2,\pm 1/2$. As in a bulk semiconductor, such a splitting is characterized by the single  $g$-factor $g_{h,3/2}=g_{h,1/2}=g_h \equiv g_{\rm h}^{\text{sph}}$ according to  \eqref{g_hM}. The expression for $g_h \equiv g_{\rm h}^{\text{sph}}$  was first obtained in  Ref. [\onlinecite{Gelmont1973}] for the hole bound to the acceptor Coulomb potential. We rewrite it as
	\begin{eqnarray}\label{Gelmontgen}
&&	g_{\rm h}^{\text{sph}}   = 2\varkappa S(\beta) + \frac{4}{5} \gamma_1 I(\beta)  \, ,  \\
&&	S(\beta) = (1- \frac{4}{5} I_2^{\text{g}} ) \, , \quad 
	I(\beta) = \frac{1-\beta}{1+\beta} I_1^{\text{g}}+ \frac{2\beta}{1+\beta} I_2^{\text{g}} \, .\nonumber
	\end{eqnarray}
Function $S(\beta)$ describes the renormalization of the spin Zeeman contribution $\hat H_{Z}^{\rm h}$,  function $I(\beta)$ describes the orbital contribution to $g$-factor correction stemming from $\hat H_B$.   
Integrals $I_1^{\text{g}}$ and $I_2^{\text{g}}$ were introduced in Ref. [\onlinecite{Gelmont1973}]:
\begin{equation}\label{Gelmont1}
I_1^{\text{g}}=\int\limits_{0}^{\infty}r^3 R_2(r) \frac{d R_0(r)}{ dr} dr,~ I_2^{\text{g}}=\int\limits_{0}^{\infty}r^2 R_2^2(r) dr \, .
\end{equation}
It is worth noting, that if the hole wave function does not vanish on the surface of the nanocrystal or in the case of the hole  confined in the core-shell spherical nanocrystal constructed from the semiconductors with {strongly} different values of {Luttinger parameters}, one has to use another expression for the $I_1^{\text{g}}$ in each material:
\begin{multline}\label{Gelmont+}
I_1^{\text{g}}=\frac{1}{2} \int \Bigg[ r^3 \left( R_2(r) \frac{d R_0(r)}{ dr}  - R_0(r) \frac{d R_2(r)}{ dr} \right) - \\- 3 r^2  R_0(r) R_2(r) \Bigg] dr \, .
\end{multline}
 The exact values of $I_1^{\text{g}}$ and $I_2^{\text{g}}$, and, therefore, $S(\beta)$ and $I(\beta)$, depend only on  profile of the confining potential and on light to heavy hole effective mass ratio $\beta$  but not on the characteristic  size of the localized hole wave function. The expression \eqref{Gelmontgen} is  obtained for low magnetic fields, when only linear on magnetic field terms are taken into account by perturbation theory. Considering quadratic on magnetic field terms requires  numerical solution of Schr{\"o}dinger equation. 

Now we consider $g$-factor of a hole in spherical NC in more detail. For the box-like infinite potential ${V}_{\text{box}}^{\rm sph}({r})$ radial wave functions $R_0$ and $R_2$ can be found analytically, see Refs. [\onlinecite{Efros92,PhysRevB.47.10005}]. For smooth parabolic or Gaussian potentials ${V}_{\text{p,Gauss}}^{\rm sph}({r})$, although the solution can not be found analytically, the variational method can be used. It allows one to obtain the hole wave function for any light to heavy hole effective mass ratio in a simple analytical form, which demonstrates a good agreement with numerical calculations not only for the energy of the ground state but also for the hole wave function itself.\cite{Semina2016} Thereby, the integrals \eqref{Gelmont1} contributing to the hole $g$-factor for parabolic and Gaussian potentials can be easily calculated as well. We have demonstrated that integrals $I_1^g$ and $I_2^g$ almost the same dependence on $\beta$ for both mentioned types of smooth potentials \cite{Semina2016}. 
The wave functions of the hole localized on an acceptor with the Coulomb potential of a charged center also can be found by variational method \cite{Rodina1993} or numerically \cite{Baldereschi1973}. For resting exciton (with the vanishing momentum of the center of masses), the solution is the same as for the acceptor with the renormalized Luttinger parameter $\gamma_1'=\gamma_1+m_0/m_e$, with $m_e$ being electron effective mass. Due to the renormalization, which makes the effective $\beta$ closer to unity as the electron usually is much lighter than the heavy hole, the effects of the complex band structure are substantially weaker for a hole in an exciton than for a hole in an acceptor.

\begin{figure}
\includegraphics[width=0.9\linewidth]{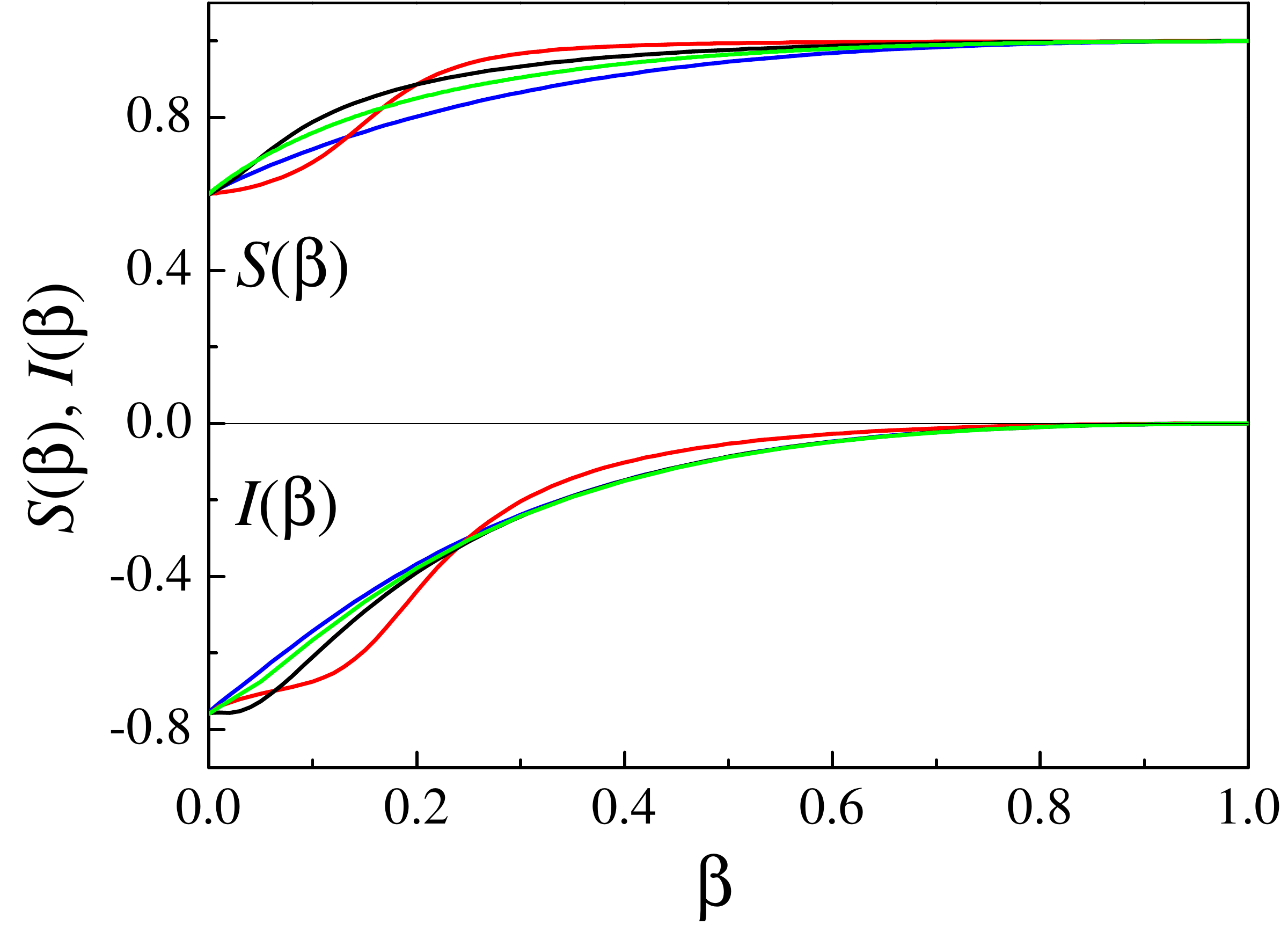}
\caption{  Quantities $S(\beta)$ and $I(\beta)$ for the $g$-factor calculation in spherical potentials: parabolic, weak and strong Gaussian (black), infinite spherical box (red), Coulomb (green), zero radius (blue).}\label{gfact_sph}
\end{figure}

The dependencies of functions $S(\beta)$ and $I(\beta)$ for box-like, parabolic, Gaussian, zero-radius, and Coulomb potentials are shown in Fig.\ref{gfact_sph}. Note, that the sign of $S(\beta)$ remains always the same and its absolute value decreases up to 3/5 of the bulk value for $\beta=0$. In opposite, function   $I(\beta)$ is negative  
and may result in the change of $g_h$ sign as compared with bulk hole $g$-factor $g_h^{\rm bulk}$.
For NCs with parabolic and Gaussian potential profiles the values of $I_1^g$ and $I_2^g$ are almost the same for given material parameters.\cite{Semina2016} Here we denote the strong Gaussian potential as potential, which characteristic size is comparable with hole wave function localization length. The characteristic size of weak Gaussian potential is much smaller that hole wave function localization length.\cite{Semina2016}  Although, for  box-like potential with abrupt infinite barrier  $I_1^g$ and $I_2^g$  can be quite different. The $g$-factors of hole in exciton and acceptor are different as well due to renormalization of Luttinger parameter $\gamma_1$ in exciton.\cite{Baldereschi1973}

\subsubsection{Hole {\it g}-factor in nanocrystals of  spheroidal shape} \label{anis}

Now we consider axially symmetric  NCs with the shape of spheroid (ellipsoid of revolution). We assume that the symmetry axis, which is the short axis in oblate (or long axis in prolate) NC, is directed along $z$. The deviation from the spherical shape is assumed to be small and can be treated as a  perturbation of the spherical NC. Such axially-symmetric perturbation splits the hole ground state into two Kramers doublets with spin projections $M \pm 3/2$ and $M \pm 1/2$ on the crystal axis similar to the effect of the internal crystal field in wurtzite semiconductors.\cite{Efros1993,EfrosCh3,Rodina2010,Semina2016} The total anisotropic splitting of the hole states is given by $\Delta =  v_{cr}(\beta)\Delta_{cr}+\Delta_{sh}$ with the crystal field-induced splitting $v_{cr}(\beta)\Delta_{cr}$ being independent of $a$ and the shape anisotropy induced contribution increasing as  $\Delta_{sh} \propto v_{sh}(\beta) \mu/a^2$ in small NCs. The proportionality coefficient $v_{sh}(\beta)$ is different for the abrupt box-like and smooth  parabolic potentials and may have opposite signs for  $\beta < 0.14$.\cite{Semina2016}  For the vast majority of Luttinger parameters sets in oblate NCs the ground state is a state with the total angular momentum projection $M=\pm 3/2$ on NC axis (it will be referred as heavy holes as it consists mostly of Bloch states with $J_{hz}=\pm 3/2$), and  $M \pm 1/2$ (light holes) in prolate NCs.\cite{Efros1993,Semina2015, Semina2016} Note, that changing $\Delta_{sh}$ one can compensate crystal field related splitting and even reverse the sign of the total splitting. For example, for wz-CdSe with $\beta=0.28$, the ''quasi-spherical" situation with $\Delta=0$ and degenerate hole state can be realized for both types of the confining potentials with the prolate shape.    Below we study the effect of such nanocrystal anisotropy on the heavy and light hole $g$-factors, $g_{h,3/2}$ and  $g_{h,1/2}$, correspondingly.

\begin{figure}[h]
\includegraphics[width=0.95\columnwidth]{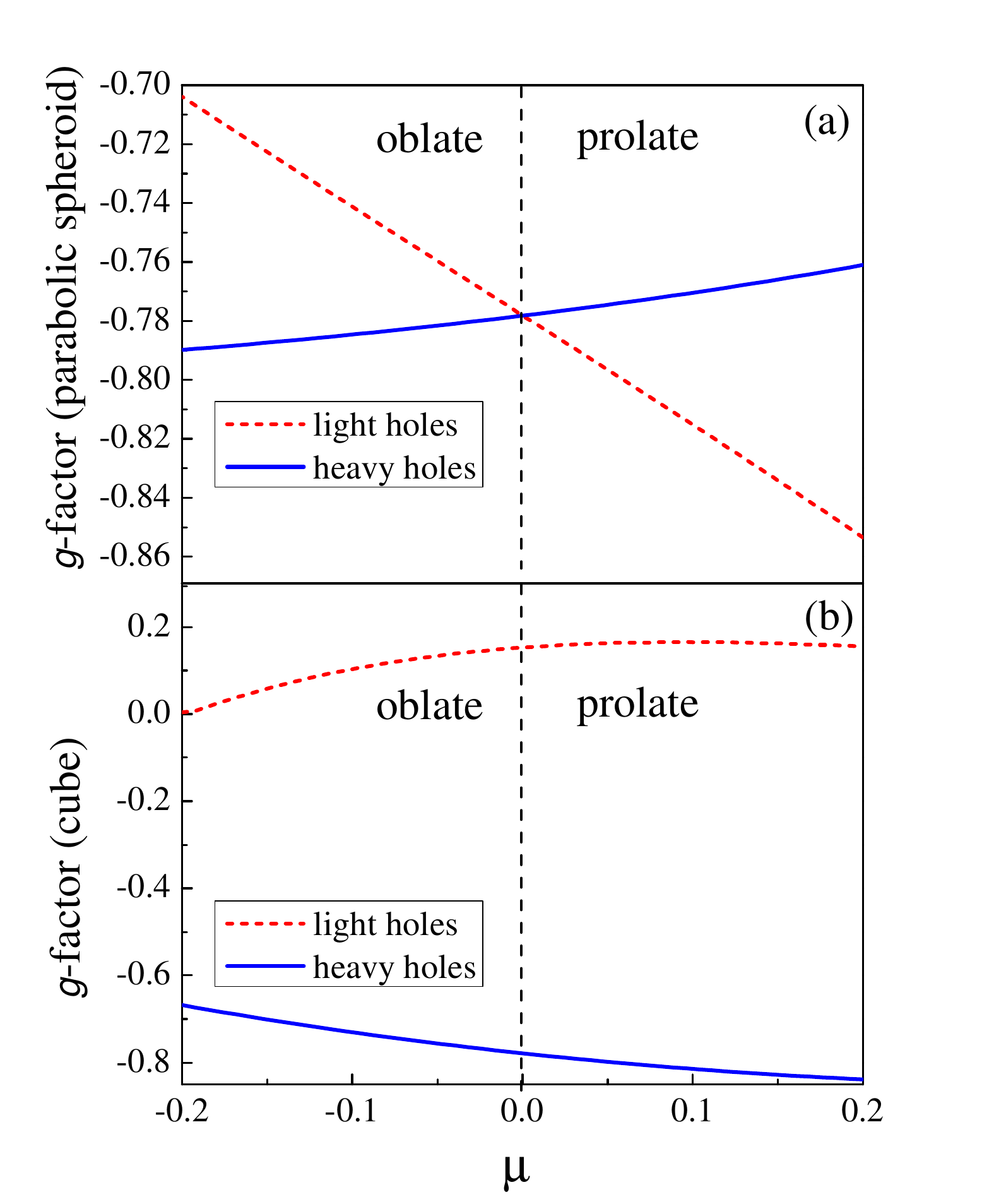}
   \caption{The dependencies of  the light, $g_{h,1/2}$, and heavy hole, $g_{h,3/2}$, $g$-factors on the anisotropy parameter $\mu$ in (a) spheroidal NCs with parabolic potential and (b) cuboid NCs with box-like infinite potential. The calculations are done for zb-CdSe parameters given in Table \ref{Table_g1} with $\beta = 0.22$ and $g_h^{\rm sph}=-0.78$.}
   \label{shape_anis}
\end{figure}

As an example we consider the parabolic confining potential. Following Ref. \cite{Semina2016}, we introduce  the shape anisotropy of quantum dot as anisotropy of its potential:
\begin{eqnarray} \label{an}
{V}_{\text{ext}}({\bf r}) \equiv V_{\rm p}^{\rm an} (\rho,z)&=&\frac{\kappa _{\rho}}{2}\rho^2+\frac{\kappa _{z}}{2}z^2 \\& =& \frac{\kappa}{2}r^2+\Delta V_{\rm p}^{\rm an}(\rho,z,\mu) \, , \nonumber  \\  \Delta V_{\rm p}^{\rm an}(\rho,z,\mu)& =& - \kappa  \mu \left(z^2- \frac{1}{3}r^2 \right) \, . \nonumber
\end{eqnarray}
Here  $\kappa =(2\kappa _\rho+\kappa _z)/3$ is  the average stiffness of parabolic potential and  $\mu$ is the anisotropy parameter:
 \begin{eqnarray}\label{k}
\mu=  \frac{(\kappa_\rho- \kappa_z )}{2\kappa },~\kappa _{\rho} \approx \kappa \left(1+\frac{2}{3}\mu\right),~ \kappa _{z} \approx \kappa \left(1-\frac{4}{3}\mu\right). \nonumber
\end{eqnarray}
The dependence of the light-heavy hole energy splitting $\Delta E_{\rm an} = E_{1/2}-E_{3/2}$ on $\mu$ was studied in Ref.~[\onlinecite{Semina2016}]. At small enough  $\mu$ this dependence is linear and can be calculated within the first-order perturbation theory. The slope $\Delta E_{\rm an}/(\mu E)$ depends only on effective mass ratio $\beta$.   The dependencies of  light and heavy holes $g$-factors on the anisotropy parameter $\mu$ calculated for zinc blende CdSe are shown in  Fig. \ref{shape_anis}(a). In Fig. \ref{shape_anis}(b) the similar dependencies for cubic NCs are shown, for details see section \ref{cubic}.

\begin{figure}[h]
\includegraphics[width=0.85\columnwidth]{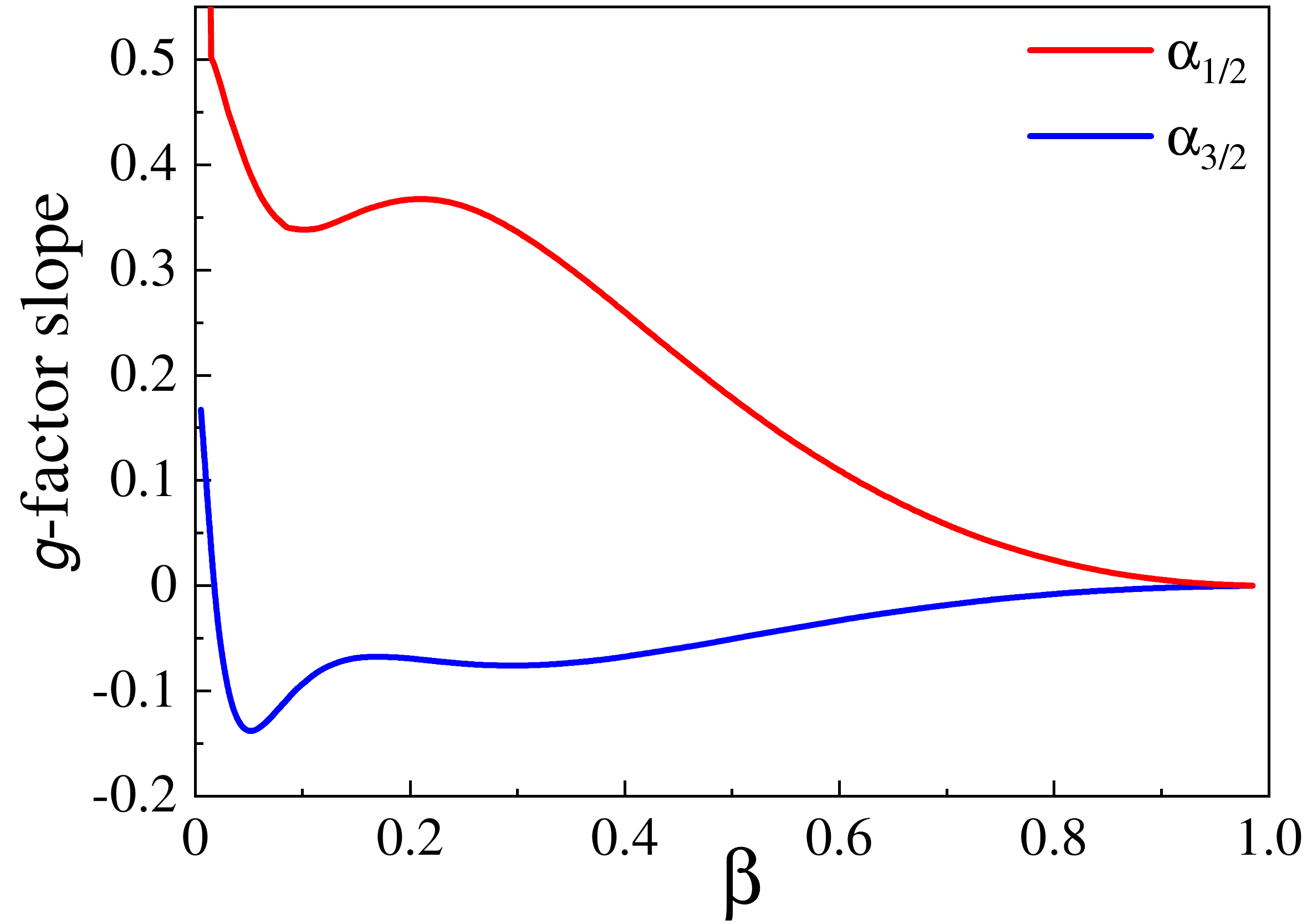}
   \caption{The dependencies of the anisotropy slopes for the light, $\alpha_{1/2}$, and heavy, $\alpha_{3/2}$, holes $g$-factors on the effective mass ratio $\beta$ in spheroidal nanocrystals with parabolic potential.}
   \label{slope}
\end{figure}

One can see, that at small enough $\mu$, the dependencies shown in Fig. \ref{shape_anis}(a) are also linear and can be approximated as
\begin{equation} \label{ghl}
g_{h,3/2}=g^{\rm sph}_{\rm h}-\alpha_{3/2}\mu, \quad g_{h,1/2}=g^{\rm sph}_{\rm h}-\alpha_{1/2}\mu,
\end{equation}
where $\alpha_{3/2}$ and $\alpha_{1/2}$ are some constants. From Fig. \ref{shape_anis}(a) one can see, that  slopes $\alpha_{3/2}$ and $\alpha_{1/2}$ have not only opposite signs but also different absolute values. In other words, in anisotropic nanocrystals the light and heavy holes $g$-factors would be different.  The numerically calculated dependencies of $\alpha_{3/2}$ and $\alpha_{1/2}$ on the light to heavy hole mass ratio $\beta$ are shown in Fig. \ref{slope}. 

As our analysis showed, the coefficients $\alpha_{3/2}$ and $\alpha_{1/2}$ can be calculated by perturbation theory as first-order corrections coming from $H_B$ by using wave functions, which already take into account the shape anisotropy in zero magnetic field.

Alternatively, one can use a coordinates substitution  $x\rightarrow x \sqrt{\kappa/\kappa_\rho}$, $y\rightarrow y \sqrt{\kappa/\kappa_\rho}$, $z\rightarrow z \sqrt{\kappa/\kappa_z}$,  which makes a  quantum dot potential spherical so that $\Delta V_{\rm p}^{\rm an}=0$ but result in additional terms in the kinetic energy Luttinger Hamiltonian $H_L \rightarrow H_L + H_{L}^{\rm an}$ \cite{Rodina1993} and $H_B\rightarrow H_B +\hat H_B^{\rm an}$. The $H_{L}^{\rm an}$ can be found in \cite{Semina2016} and Appendix \ref{AB}, the  $\hat H_B^{\rm an}$ is given by:
\begin{eqnarray}\label{HBa}
\hat H_B^{\rm an}=-2\gamma\mu\mu_B B(J_yJ_z+J_zJ_y)xk_z \, .
\end{eqnarray}
To calculate the corrections to  hole $g$-factors due to the anisotropy in this method, one has to calculate the corrections to the energy splittings $\Delta E_M$ from  Eq. \eqref{HBa} in the first-order of perturbation theory and mutual corrections from the $H_B$ and $H_L^{\rm an}$  in the second-order perturbation theory. The derivation of the analytical expressions for $\alpha_{hh}$ and $\alpha_{lh}$ is, therefore, the problem for a separate paper. 

The  method of the coordinates substitution  $x\rightarrow x b/a$, $y\rightarrow y b/a$, $z\rightarrow z c/a$ can be also used for the spheroidal nanocrystals with abrupt box potential at the surface $x^2/b^2+y^2/b^2+z^2/c^2=1$. In this case, the small anisotropy parameter is given by $\mu=c/b-1$ with $c \approx a(1-2\mu/3)$, $b \approx a(1+\mu/3)$ \cite{Efros1993}. Similarly, this method can be applied to treat the anisotropy in the  NCs of square cuboid shape with abrupt box-like potential and dimensions $2b$ and $2c$. The coordinate substitution  transforms them into the NCs of cubic shape with the cube edge $2a$.  Such nanocrystals with the cubic shape are considered in the next subsection \ref{cubic} and Appendix \ref{AC}.

It is worth noting, that in contrast to the potential shape anisotropy,  the  effective crystal field perturbation $\propto \Delta_{cr}$ in wurtzite semiconductors does not affect the hole $g$-factors (while magnetic field is weak enough). Thus, the crystal field and magnetic field perturbations can be taken into account separately, and both light and heavy hole Zeeman splittings in spherical wurtzite NCs can be described by $g_h^{\rm sph}$. In contrast, on needs two $g$-factors,  $g_{h,3/2} \ne g_{h,1/2}$, for "quasi-spherical" wurtzite NCs.

\subsubsection{Hole {\it g}-factor in nanocrystals of  cubic shape}\label{cubic}

Here we consider nanocrystals with the infinite potential of cubic shape or cube NCs (see details in Appendix \ref{AC}). Note, that we still neglect the effects of cubic symmetry of crystal lattice. In this case, the system under study loses its spherical and even axial symmetry. The total angular momentum and its projection on magnetic field direction (unless the magnetic is directed along one of the cubic axes)  are not good quantum numbers. Importantly, the hole ground state remains four-fold degenerate in zero magnetic field because cubic anisotropy does not split states with total angular momentum less than $5/2$ (as it would be in spherically symmetric NC) \cite{Baldereschi1974}. 

We have shown before \cite{Golovatenko2018}, that in spherical NCs with box-like or parabolic potential the energies of the two lowest hole states with the total angular momentum $j=3/2$ but different orbital momenta ($SD$-like with $l=0,2$ and $PF$-like with $l=1,3$) become close to each other at small values of $\beta$ as well. However, the $SD$-like state remains the lowest energy state for all $\beta$ values in spherically symmetric potentials.  The striking feature of cube nanocrystals is that there is a range of $\beta$ values, where the hole ground state is a $PF$-like ($E_1$) state instead of $SD$-like ($E_0$) one. The dependencies of  the energies of two lowest hole states, $E_0$ and $E_1$, on $\beta$ in zero magnetic field are shown in Fig. \ref{hhcontr} (a)  in Appendix \ref{AC}. 

In the range of $\beta$ where the intersection occurs, the energy of both states are close to each other. This is reflected in the significantly different contributions from different Bloch states to the $E_0$ state comparing to the range of  $\beta$ values, where the $E_0$ and $E_1$ energies are far from each other {(see Fig. \ref{hhcontr}(b) in  Appendix \ref{AC})}. This fact exhibits itself in the values of hole $g$-factors as functions of $\beta$ even for the magnetic field directed along the symmetry axis $z$ of the cubic NC. 
\begin{figure}[h]
\includegraphics[width=0.95\columnwidth]{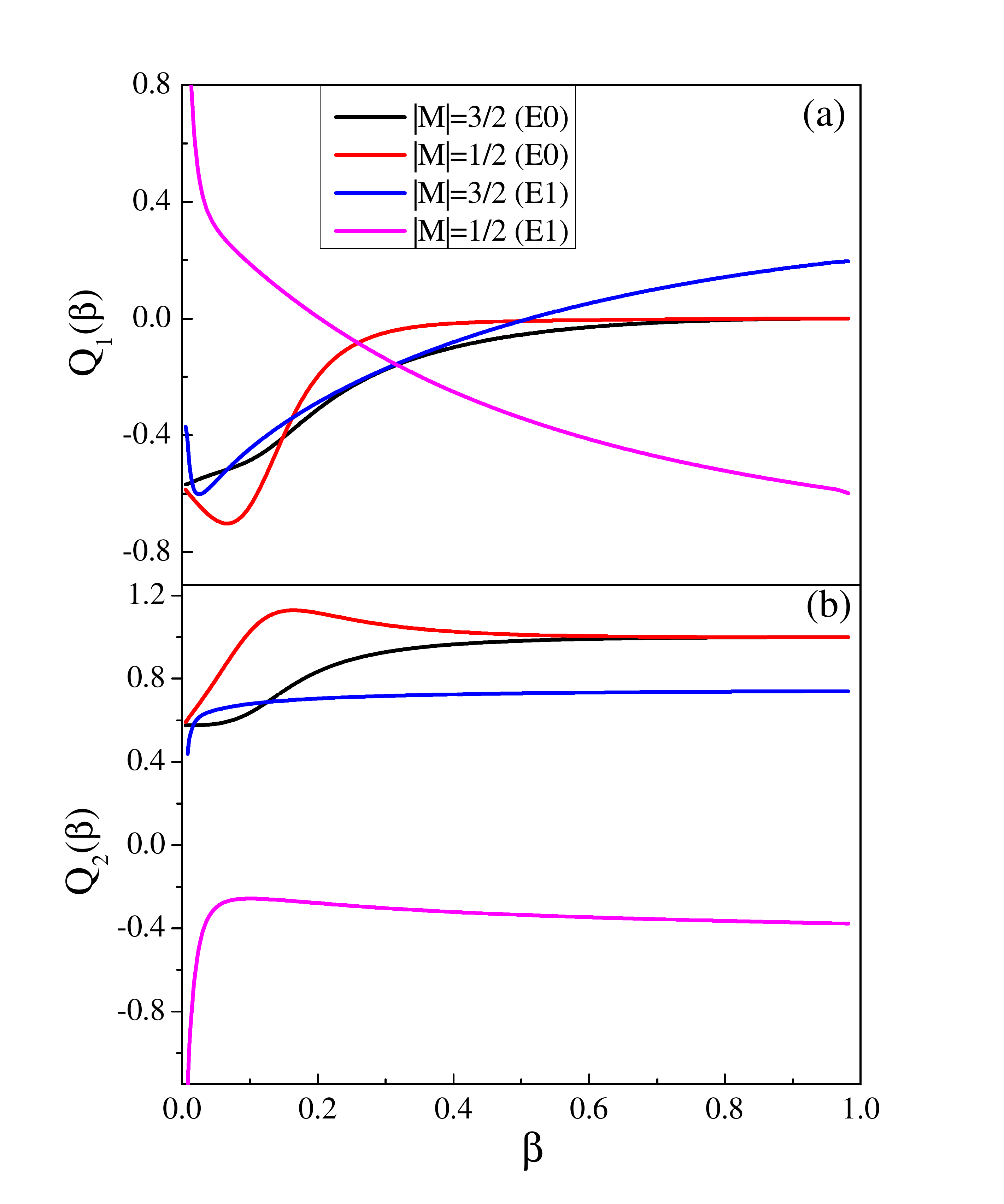}
   \caption{ The dependencies  (a) $Q_{1,|M|}(\beta)$ and (b) $Q_{2,|M|}(\beta)$ for the light ($|M|= 1/2$) and  heavy ($|M|= 3/2$)  hole two first quantum states $E_0$ and $E_1$ in  nanocrystal  of cubic shape calculated numerically. }
   \label{cubic_gfactor}
\end{figure}

The  hole effective $g$-factors for the  light ($|M|= 1/2$) and  heavy ($|M|= 3/2$)  hole states can be written as
\begin{equation} \label{gcube}
g_{h,|M|}=2\varkappa Q_{2,|M|}(\beta)+\gamma_1Q_{1,|M|}((\beta), 
\end{equation} 
where functions $Q_{1,|M|}(\beta)$ and  $Q_{2,|M|}(\beta)$ depend only on {light  to heavy}  hole effective mass ratio $\beta$ and describe the orbital contribution to hole $g$-factor and the renormalization of the spin contribution, respectively. In contrast to the spherical case, these functions are now different for the  heavy and light hole $g$-factors.

The dependencies of the functions  $Q_{1,|M|}(\beta)$ and ($Q_{2,|M|}(\beta)$  for two lowest quantum states of cube NC calculated numerically are shown in Fig.~\ref{cubic_gfactor}. It is interesting, that even in cube nanocrystal, where the hole ground and first excited states in zero magnetic field are four-fold degenerate, the $g$-factors of heavy and light holes are different and at some band parameters can have even opposite signs. Such a situation is demonstrated in Fig. \ref{shape_anis} (b), there the dependencies of light and heavy hole ground state $g$-factors are shown as functions of cuboid CdSe NC anisotropy parameter $\mu$.  This effect is the manifestation of the breaking of spherical symmetry and can be qualitatively understood by considering a cubic contribution $Q(B_x J_x^3+B_y J_y^3 + B_z J_z^3)$ to the spherically-symmetric hole Hamiltonian with a large $Q$ (unlike neglected term with a small $q$ in bulk semiconductors and spherical nanocrystals, which originates from cubic symmetry of crystal lattice itself).  If one would include such a contribution in Hamiltonian \eqref{Hamilt_dot} with a spherical nanocrystal potential, the resulting $g$-factors for light and heavy holes will be different.

\subsubsection{Hole $g$-factor in cylindrical quantum wires}

If the symmetry of nanocrystal potential is lower than spherical or cubic, heavy and light holes states are split even in zero magnetic field.   For cylindrical quantum wires, in which the structure size along the symmetry axis is much larger than in its cross-section, the hole ground state is characterized by the total angular momentum projection $M=\pm 1/2$ and constructed mostly from the light hole states \cite{Semina2015} with $J_{hz}=\pm 1/2$. However,  this state has a small admixture of the valence band states with $J_{hz}=\mp 3/2$ and orbital momentum projection $\pm 2$ even in the 1D limit of the nanowire with infinite length and zero potential along the wire axis.  The corresponding wave functions are:\cite{Semina2015}
\begin{multline}\label{psi_1D}
\Psi_{M=\pm \frac{1}{2}}^{\text{wire}}=\psi_1(\rho)|J_{hz}=\pm \frac{1}{2}\rangle+\\+\psi_2(\rho)\exp(\mp 2\mathrm{i}\varphi)|J_{hz}=\mp \frac{3}{2}\rangle,
\end{multline}
where $\rho$ and $\varphi$ are  polar coordinates in the quantum wire cross section, $\psi_1(\rho)$ and $\psi_2(\rho)$  are in-plane radial wave functions, depending on  $\beta$ and quantum wire potential. In-plane wave functions $\psi_1(\rho)$ and $\psi_2(\rho)$ can not be found analytically in  general case, but can be calculated, for example, by the variational method.\cite{Semina2011}

\begin{figure}
\includegraphics[width=0.9\linewidth]{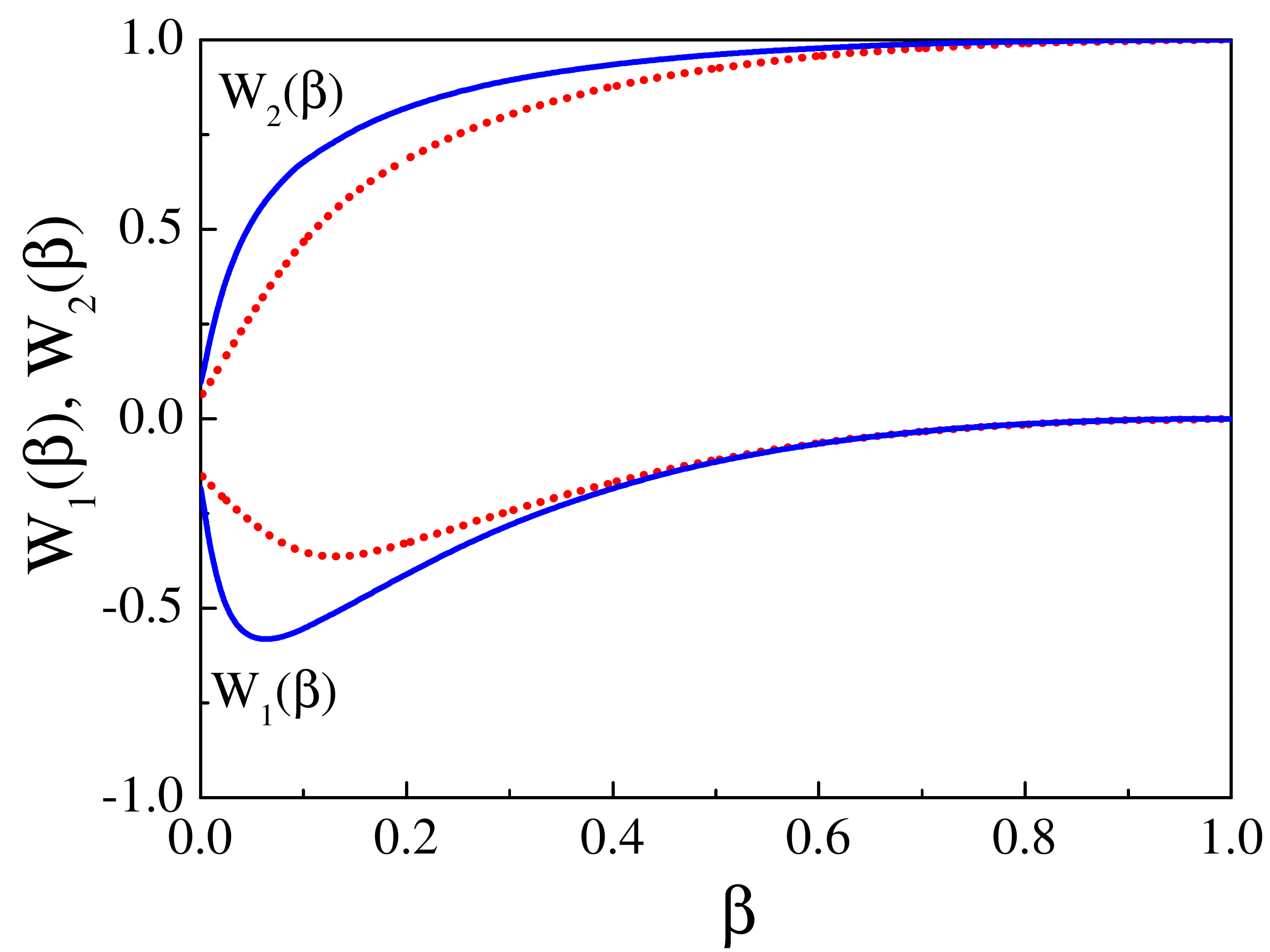}
\caption{Functions $W_1(\beta)$ and $W_2(\beta)$ for $g$-factor in wire-like potentials: parabolic and strong Gaussian (solid line) and weak Gaussian (dotted line).}\label{gfact_wire}
\end{figure}

Magnetic field lifts {Kramers}  degeneracy, and the hole ground state $g$-factor $g_{h,1/2} \equiv g_{h,1/2}^{\rm wire}$ was derived in Ref. [\onlinecite{Semina2015}]. Similarly to \eqref{Gelmontgen} we rewrite it  :
\begin{eqnarray}\label{g_1D}
g_{h,1/2}^{\text{wire}}= 2 \varkappa W_2(\beta) + \gamma_1 W_1 (\beta) \, . 
\end{eqnarray} 
Here 
\begin{eqnarray}\label{g_1D1}
 &&W_1(\beta) = \frac{2(3+\beta)}{1+\beta}F(\beta) -  \frac{1-\beta}{1+\beta}\sqrt{3}H(\beta),\\
 &&W_2(\beta)= 1- 4 F(\beta), \nonumber \\
&& F(\beta)=   \int\limits_0^{\infty}\psi_2^2(\rho)\rho d\rho, ~  H(\beta)= 2 \int\limits_0^{\infty} \psi_2(\rho)\frac{d \psi_1(\rho)}{d\rho}\rho^2 d\rho. \nonumber
\end{eqnarray} 
Function $W_1(\beta)$ describes the orbital contribution to $g$-factor correction stemming from $\hat H_B$ while the function $W_2(\beta)$ describes the renormalization of the spin Zeeman contribution $\hat H_{Z}^{\rm h}$. Note, for the cylindrical core/shell heterostructure or nonvanishing hole wave function at the cylindrical surface, more general expression for each material is:  
\begin{multline}\label{H_1D1}
H(\beta)=\int\limits_0^{\infty}\left(\psi_2(\rho)\frac{d \psi_1(\rho)}{d\rho}-\psi_1(\rho)\frac{d \psi_2(\rho)}{d\rho}\right)\rho^2d\rho-\\-2\int\limits_0^{\infty} \psi_1(\rho)\psi_2(\rho)\rho d\rho . 
\end{multline}

Functions $W_{1,2}(\beta)$  for parabolic and strong Gaussian potentials (solid line) and weak Gaussian potential (dotted line) are shown in Fig. \ref{gfact_wire}.  The accuracy of Eq. \eqref{g_1D} was checked with the numerical calculation. For infinite box-like potential both  $F(\beta)$ and $H(\beta)$ are close to zero due to negligible mixture of hole states and $\psi_2$ in \eqref{psi_1D} being small.

\subsubsection{Hole $g$-factor in a nanoplatelet}
\begin{figure}
\includegraphics[width=\linewidth]{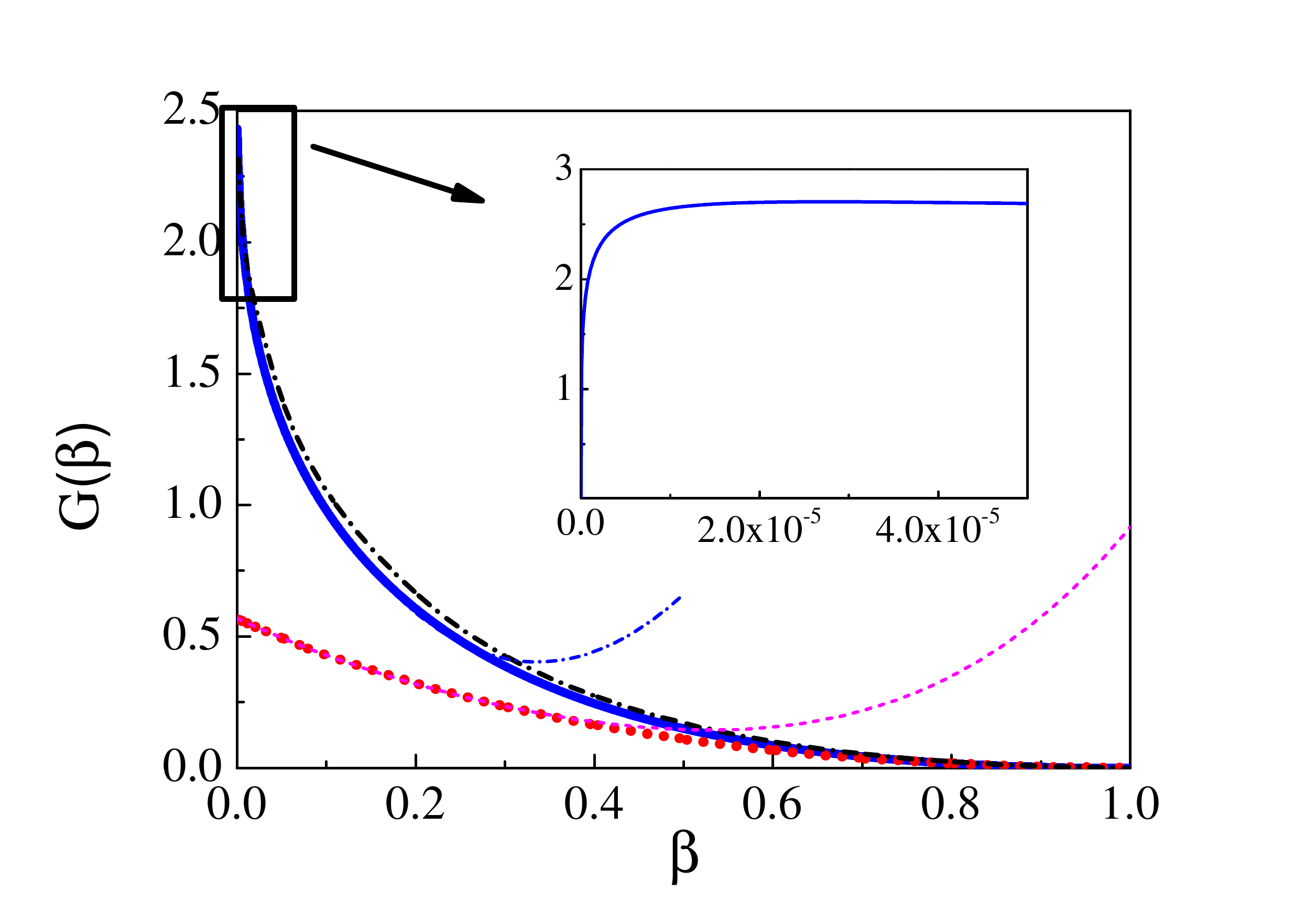}
\caption{$G(\beta)$ for calculation of $g$-factor in well-like potentials for parabolic and strong Gaussian (solid line), weak Gaussian (dash-dotted line), and box-like infinite (dotted line) potentials. Thin dashed line shows asymptotics for box-like potential for $\beta\rightarrow 0$, thin dash-dotted line shows asymptotics for parabolic potential (for details see Appendix \ref{AB}). The inset shows $G(\beta)$ for parabolic potential in vicinity of $\beta=0$.}\label{gfact_well}
\end{figure}
In a quantum well-like nanocrystal, nanoplatelet, where the thickness is much smaller than in-plane dimensions, the heavy and light hole states are strongly split in zero magnetic field (about 150 meV in 4 ml NPL \cite{Shornikova2018}).  In the in-plane isotropic NPL,  the hole ground state is characterized by the total angular momentum projection $M=\pm 3/2$ on the symmetry axis directed perpendicular to the NC plane ($z$-axis).  In the limiting case of an isotropic  NC structure with infinite in-plane size, the hole ground state is composed only of the heavy hole valence band states with $J_{hz}=\pm 3/2$ with a vanishing admixture of light holes.  Therefore, in the limiting case of a two-dimensional (2D) structure with the vanishing width to in-plane size ratio there is no renormalization of the spin Zeeman effect. However, as it is was shown in Ref. \cite{Wimbauer1994},  the orbital corrections to the heavy hole $g$-factor $g_{h,3/2} \equiv g_{h,3/2}^{\text{2D}} $ coming from the $H_B$  in the first order perturbation theory for magnetic field directed along the symmetry axis are present due to the quantization of $k_z$ and   magnetic field induced heavy and light hole mixing (see details in Appendix \ref{AB}):
\begin{equation}\label{g2d}
 g_{h,3/2}^{\text{2D}} =2\varkappa- 4\frac{\hbar^2}{m_0}\sum\limits_{n=1}^\infty \frac{|\langle lh_{2n}|\gamma \hat{k}_z|hh_1\rangle|^2}{E_{lh_{2n}}-E_{hh_1}} \, .
\end{equation}
Here $|hh_1\rangle$ is the envelope wave function of heavy hole ground state of the quantization along $z$-axis, $|lh_{2n}\rangle$ are wave functions of even excited states of a light hole, $E_{hh_1}$ and  $E_{lh_{2n}}$  are the corresponding energies.
After summation Eq. \eqref{g2d} for 2D hole $g$ factor can be written as
\begin{equation}\label{g2d1}
 g_{h,3/2}^{\text{2D}} =2\varkappa-\frac{\gamma_1}{3}G(\beta),
\end{equation}
where $G(\beta)$ depends only on $\beta$ and the type of the localization potential along the $z$ direction. In a similar way, one may obtain the 2D-limiting case expression for the light-hole $g$-factor, $g_{h,1/2}^{2D}$, at the lowest light-hole  size-quantization level, $E_{lh_{1}}$,  (see Eq. \eqref{ghl}). In this case, the corrections to the bulk value $2\varkappa$ come from the admixture of the even excited states of the heavy-hole,  $E_{hh_{2n}}$. Importantly, a proximity of the lowest light-hole state energy with one of the energies  $E_{hh_{2n}}$  of the heavy-hole may result in  a giant enhancement of the light-hole Zeeman splitting   \cite{Durnev2012}. Even in the absence of such a giant enhancement, for example in the case of thin CdSe NPLs,  $g_{h,1/2}^{2D} \ne  g_{h,3/2}^{2D}$.

The dependencies of parameter $G(\beta)$ on $\beta$ for parabolic and strong Gaussian, weak Gaussian, and box-like infinite potentials are shown in Fig.~\ref{gfact_well}.  In the figure, one can see the much stronger dependence for smooth potentials due to smaller distances between hole energy levels as compared with box-like potential. Also here there is almost no difference between parabolic and strong Gaussian potentials and a very slight difference with weak Gaussian potentials.

Here we discussed the case of the magnetic field directed perpendicular to the structure plane. The hole $g$-factor in NPLs and other planar nanostructures is strongly dependent on magnetic field direction \cite{vanKesteren1990,Sirenko1997} as well as on crystallographic orientation of the structure \cite{Kubisa2011}. In the analysis of the experimental data, however, it is possible to take into account the anisotropy of the $g_{h,3/2}$ via the coefficient $\cos \Theta$, where $\Theta$ is an angle between the magnetic field direction and the anisotropic axis \cite{Liu2013,Shornikova2017nl}. In this approach,  the Zeeman splitting of the heavy-hole is zero for the magnetic field directed in the NPL plane unless the effects of the qubic terms or additional light-hole to heavy-hole mixing induced by the in-plane anisotropy are taken into account \cite{Semina2015}. 
\newline

\subsection{Hole $g$-factor in different semiconductor nanostructures: comparison with experimental data.}\label{size} 


 \begin{figure*}[ht]
	\begin{center}
		\includegraphics[width=\textwidth]{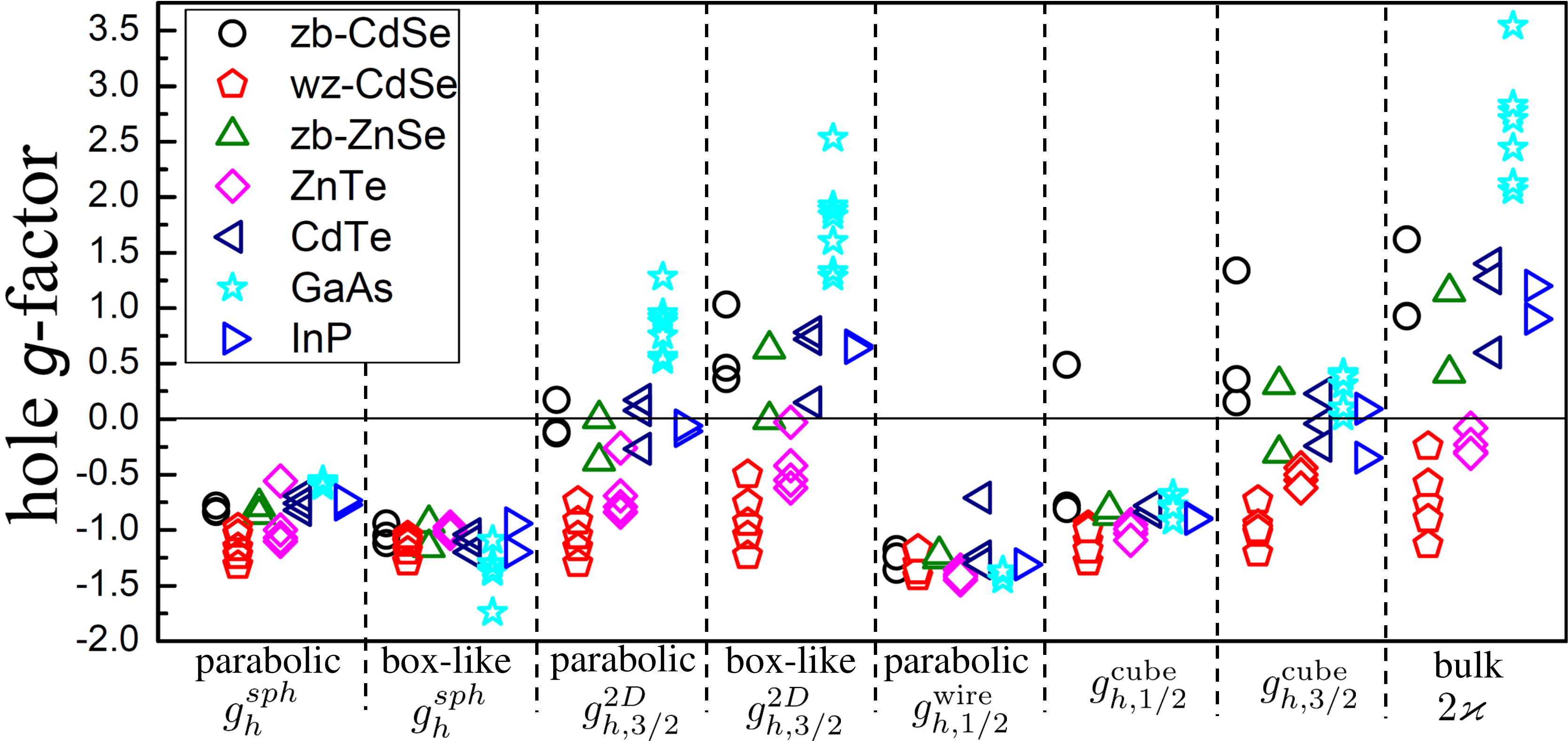}
		\caption{\label{all}Hole $g$-factors for different II-VI  and III-V semiconductor model nanostructures. Confinement potentials notation: sph.par. - spherical parabolic potential, sph.box - spherical infinite box potential, QW par. - NPL with parabolic potential along the $z$-axis, QW box - NPL with infinite box potential, wire par. - nanowire with parabolic potential in the cross-section plane, cubic $g_{h,3/2}$ - cubic potential for heavy hole, cubic $g_{h,1/2}$ - cubic potential for light hole. 
		}
	\end{center}
\end{figure*}

\begin{table*}[ht]
\begin{center}
\caption{Hole effective $g$-factors in nanostructures.}
\label{Table_g1}
   \begin{tabular}{| l | l | l | l |  l | l | l |  l | l |  l | l |l | l |l |}
    \hline
    Material & $\gamma_1$ & $\gamma$ & $\varkappa$&\specialcell{$g_h^{\rm bulk}$\\ $=2\varkappa$ } &$\beta$&\specialcell{$g_h^{\text{sph}},$\\ par. }&\specialcell{$g_h^{\text{sph}},$ \\ box}&
   \specialcell{ $g^{2D}_{h,3/2},$ \\ par. }&\specialcell{ $g^{2D}_{h,3/2},$ \\ box }&\specialcell{ $g^{\text{wire}}_{h,1/2},$ \\ par. }&$g^{\text{cube}}_{h,3/2}$&$g^{\text{cube}}_{h,1/2}$&Refs.$^*$\\ \hline
   zb-CdSe &5.51 & 1.78 &0.46&0.93 &0.22&-0.78&-0.94&-0.11&0.36&-1.36&-0.78&0.15&\cite{Adachi2004}  \\ \hline
    zb-CdSe & 3.27 & 1.33 & 0.46 &0.93 &0.1&-0.83&-1.12&-0.13&0.46&-1.17&-0.81&0.36&\cite{Karazhanov2005}\\ \hline
    zb-CdSe& 3.8 & 1.65 & 0.81&1.62 &0.07&-0.83&-1.05&0.17&1.03&-1.24&0.49&1.34&\cite{Karazhanov2005} \\ \hline
     wz-CdSe& 2.04 & 0.58 &-0.38&-0.76 &0.28&-1.16&-1.12&-1.05&-0.93&-1.3&-1.1&-0.95&\cite{Norris1996} \\ \hline
      wz-CdSe& 2.52 & 0.83 & -0.12&-0.25 &0.2&-0.98&-1.05&-0.74&-0.5&-1.21&-0.96&-0.73&\cite{Fu1998n2} \\ \hline
  wz-CdSe& 1.7 & 0.4 & -0.57&-1.13 &0.36&-1.32&-1.29&-1.3&-1.23&-1.42&-1.29&-1.21&\cite{Kapustina2000} \\ \hline
 wz-CdSe& 2.1 & 0.55 &-0.45&-0.9 &0.31&-1.23&-1.19&-1.16& -1.05&-1.37&-1.18&-1.03&\cite{Ekimov1993} \\ \hline
  wz-CdSe& 1.67 &0.56 &-0.29&-0.58 &0.2&-1.04&-1.09&-0.92&-0.76&-1.17&-1&-1&\cite{Berezovsky2005} \\ \hline
    zb-ZnSe& 3.94 & 1.31 & 0.21&0.41 &0.2&-0.86&-0.94&-0.38&-0.004&-1.26&-0.87&-0.31&\cite{Adachi2004} \\ \hline
      zb-ZnSe&3.77 & 1.5 & 0.57&1.14 &0.11&-0.79&-1.16&0.003&0.63&-1.22&-0.81&0.31&\cite{Lawaetz1971} \\ \hline
  ZnTe& 4 & 1.11 & -0.15&-0.29 &0.28&-0.56&-1.03&-0.26&-0.03&-1.45&-1.03&-0.55&\cite{Stradling1968} \\ \hline
    ZnTe& 3.8 & 1.14 & -0.04&-0.08 &0.25&-1&-0.98&-0.69&-0.42&-1.36&-0.95&-0.44&\cite{Said1990} \\ \hline
      ZnTe& 3.9 & 1.11 & -0.11&-0.22 &0.27&-1.07&-0.97&-0.79&-0.55&-1.4&-0.99&-0.5&\cite{Oka1981}\\ \hline
        ZnTe& 3.8 & 1.07 & -0.15&-0.31 &0.28&-1.1&-1.01&-0.84&-0.62&-1.42&-1.09&-0.62&\cite{Wagner1992} \\ \hline
          CdTe& 4.14 & 1.4 & 0.3&0.6 &0.19&-0.82&-1.04&-0.27&0.15&-1.24&-0.84&0.23&\cite{Friedrich1994} \\ \hline
           CdTe& 5.3 & 1.88 & 0.7&1.4 &0.17&-0.69&-1.1&0.17&0.78&-1.3&-0.84&-0.04&\cite{Said1990}\\ \hline
 CdTe& 4.11 & 1.6 & 0.63&1.26 &0.12&-0.76&-1.2&0.08&0.72&-0.71&-0.78&-0.24&\cite{Neumann1988} \\ \hline
 GaAs& 6.98 &2.63 & 1.39&2.77 &0.14&-0.57&-1.3&0.91&1.88&-1.38&-0.69&0.02&\cite{Skolnick1976} \\ \hline
  GaAs& 6.85 &2.58 & 1.35&2.7 &0.14&-0.58&-1.3&0.87&1.83&-1.37&-0.69&0.02&\cite{Landoldt22a} \\ \hline
  GaAs& 7.17 &2.89 & 1.77&3.54 &0.11&-0.60&-1.38&1.28&2.53&-1.45&-0.68&0.31&\cite{Neumann1988} \\ \hline
  GaAs& 6.79 &2.38 & 1.03&2.06 &0.18&-0.61&-1.36&0.53&1.28&-1.37&-0.78&0.41&\cite{Molenkamp1988} \\ \hline
  GaAs& 6.8 &2.4 & 1.06&2.12 &0.17&-0.60&-1.09&0.56&1.33&-1.37&-0.79&0.38&\cite{Shanabrook1989} \\ \hline
  GaAs& 7.2 &2.69 & 1.42&2.83 &0.15&-0.55&-1.3&0.95&1.92&-1.39&-0.92&0.1&\cite{Said1990} \\ \hline
  GaAs&7.1 &2.55 & 1.22&2.44 &0.16&-0.57&-1.74&0.75&1.6&-1.37&-0.8&0.39&\cite{Bingelli1991} \\ \hline
  InP&5.05 &1.68 & 0.45&0.9 &0.2&-0.77&-0.94&-0.11&0.66&-1.31&-0.89&0.09&\cite{PhysRevB.50.10598} \\ \hline
  InP&4.6&1.68 & 0.6&1.2 &0.15&-0.73&-1.2&-0.06&0.64&-1.25&-0.9&-0.35&\cite{PhysRevB.50.10598} \\ \hline
 \end{tabular}
\end{center}
$^*$ References are given for the $\gamma_{1}$, $\gamma_{2}$  and  $\gamma_{3}$  Luttinger parameters. We use relations $\gamma=(2\gamma_2 +3 \gamma_3)/5$  \cite{Baldereschi1970} and $\varkappa\approx-2/3+5\gamma/3-\gamma_1/3$  \cite{Roth1959}.
\end{table*}

In Table I  and Fig. \ref{all} the hole $g$-factors for discussed above cases in II-VI and III-V semiconductors with different parametrizations for Luttinger parameters are shown. 
One can see in the Fig. \ref{all} that in spherically symmetric potentials, wire potential, and cube potential for a heavy hole, the hole $g$-factor is almost the same for all the studied semiconductors and Luttinger parameters. Contrary, for NPLs with parabolic or box-like potential, and cube NCs for the light hole, hole $g$-factors show distinct dependence on the semiconductor material and Luttinger parameters. Worth to note, that for wz-CdSe localization of a hole results in a minor renormalization of the bulk hole $g$-factor. In contrast, in zb-CdSe, InP, and GaAs nanostructures the hole $g$-factor experiences strong renormalization up to sign inversion.  This strong renormalization is caused by the orbital contribution $\propto \gamma_1$.

In the literature  the following experimental hole $g$-factors were reported:  $g_h=-1.04$ and $g_h=-0.76$ in bare core wz-CdSe NCs with diameters 2.5 and 1.9 nm, respectively \cite{Kuno1998}; $g_h=-0.6$ in CdSe/ZnS NCs \cite{Biadala2010}; $g_h=-0.54$ in CdSe/CdS NCs  with thick shell \cite{Liu2013}. In zb-CdSe based nanoplatelets,   $g_h=-0.4$  was reported in CdSe/CdS NPLs with thick shell (changing  to $g_h=-0.7$ with increase of magnetic field) \cite{Shornikova2017nl},  and $g_h \approx -0.1$ (in the range of -0.03 and -0.2) in bare-core CdSe NPLs \cite{Shornikova2020nn,Shornikova2020nl}. These values were deduced from the circular polarized emission of the negative trions. Interestingly, the hole $g$-factors in CdSe/CdS NPLs, $g_h=-0.4$ and  $-0.7$, are close to the values calculated for a hole in spherically or cubically symmetric potential, while  $g_h=-0.1$ is close to the  value calculated for a hole in 2D parabolic (not box-like) potential (see Table \ref{tab:table1}). A smooth confining potential instead of an abrupt one may be induced in thin NPLs due to the dielectric confinement effect. Indeed, the repulsion of the carries from the image-charge results in the additional repulsing potential $\sim 1/d$, where $d$ is the distance to the surface.

The expressions above for hole $g$-factors in spherical, cube, quasi-1D and ideal 2D structures, Eqs. \eqref{Gelmontgen}, \eqref{gcube}, \eqref{g_1D} and \eqref{g2d}, are written neglecting the quadratic on magnetic field terms. Taking it into account results in the hole $g$-factors nontrivial dependence on the magnetic field. Indeed, the magnetic field substantially modifies light and heavy hole mixing as Zeeman splitting becomes comparable with distances between hole size-quantization energy levels \cite{Rego1997,Kotlyar2001}. 
This effect could be enhanced if the structure is anisotropic in the plane perpendicular to the magnetic field \cite{Kapoor2010}.  The nonlinear dependence of hole Zeeman splitting on the magnetic field was also observed in quantum wells (e.g. \cite{Kotlyar2001,Traynor1995,Potemski2012}).

In the calculations presented in the Table \ref{Table_g1} and in Fig. \ref{all}  the renormalization of Luttinger parameters $\gamma_1$, $\gamma$ and $\varkappa$ caused by the quantum confinement of holes which similar the to effect of the non-parabolic energy dispersion for electrons \cite{Pfeffer1996} was neglected. Such a renormalization can be taken into account within the 8-band Kane model \cite{Kiselev1996,Efros1998} and results in the size-dependence of the hole $g$-factor similar to the effect for electron \cite{Kiselev1996,Sirenko1997,Kiselev1998}.  This effect might be important for small NCs and thin NPLs when the hole energy $E_h$ becomes comparable with $E_g$. For quantum wells, the size dependence of the hole $g$-factor was widely studied experimentally \cite{Sirenko1997,Kotlyar2001,Snelling1992}.  	For quantum wires and spherical NCs with small radius $a$, hole $g$-factors may depend on $a$ because of the admixture of different bands by spin-orbit interaction \cite{Kotlyar2001,Bayer1995}.

Importantly, if the hole quantization energy becomes comparable with the spin-orbit energy $\Delta_{\rm SO}$, the admixture of $\Gamma_7$ spin-orbit split valence band states (as expected in InP where $\Delta_{SO}=100$ meV) must be taken into account as well and may result in additional corrections to the hole $g$-factor and, consequently, to its size dependence.  Recently, $g_h=-1.9$ was reported in InP/ZnSe spherical  NCs \cite{Brodu2019}, which is much larger than the theoretically calculated values for two parameterizations (see Table \ref{Table_g1}). Besides the corrections coming from the admixture of $\Gamma_7$ hole states, the anisotropic corrections to heavy and light hole $g$-factors caused by NC  shape could be also important. Indeed, it was argued in Ref. \cite{Brodu2019} that for zinc-blende InP with $\beta=0.15$ the splitting $\Delta$ between the heavy and light hole energy levels in the spheroidal NCs is very small as $v_{sh}(\beta=0.14)=0$. Therefore,  one may deal with the degenerate hole state (and as consequence, with the isotropic exciton, see the next section) even in prolate and oblate InP NCs. We have shown above, however, that even in this case the shape anisotropy results in the corrections to the $g_{h,3/2} \ne g_{h,1/2}$ which should be taken into account in the analysis of the experimental data. Such a situation can be also expected for the spheroidal NCs based on GaAs where $\beta=0.14$. As we have shown above, a similar situation can be realized in the case of the hole state in cube NCs based on zinc-blende with degenerate hole state ($\Delta=0$) in zero magnetic field and equidistant levels splitting in the external magnetic field with  $g_{h,3/2} = g_{h,1/2}$. A similar situation is also expected for the wz-CdSe "quasi-spherical" NCs, where the anisotropic splitting caused by their prolate shape compensate exactly the crystal field splitting. 

 \begin{figure}[h!]
\includegraphics[width=0.9\linewidth]{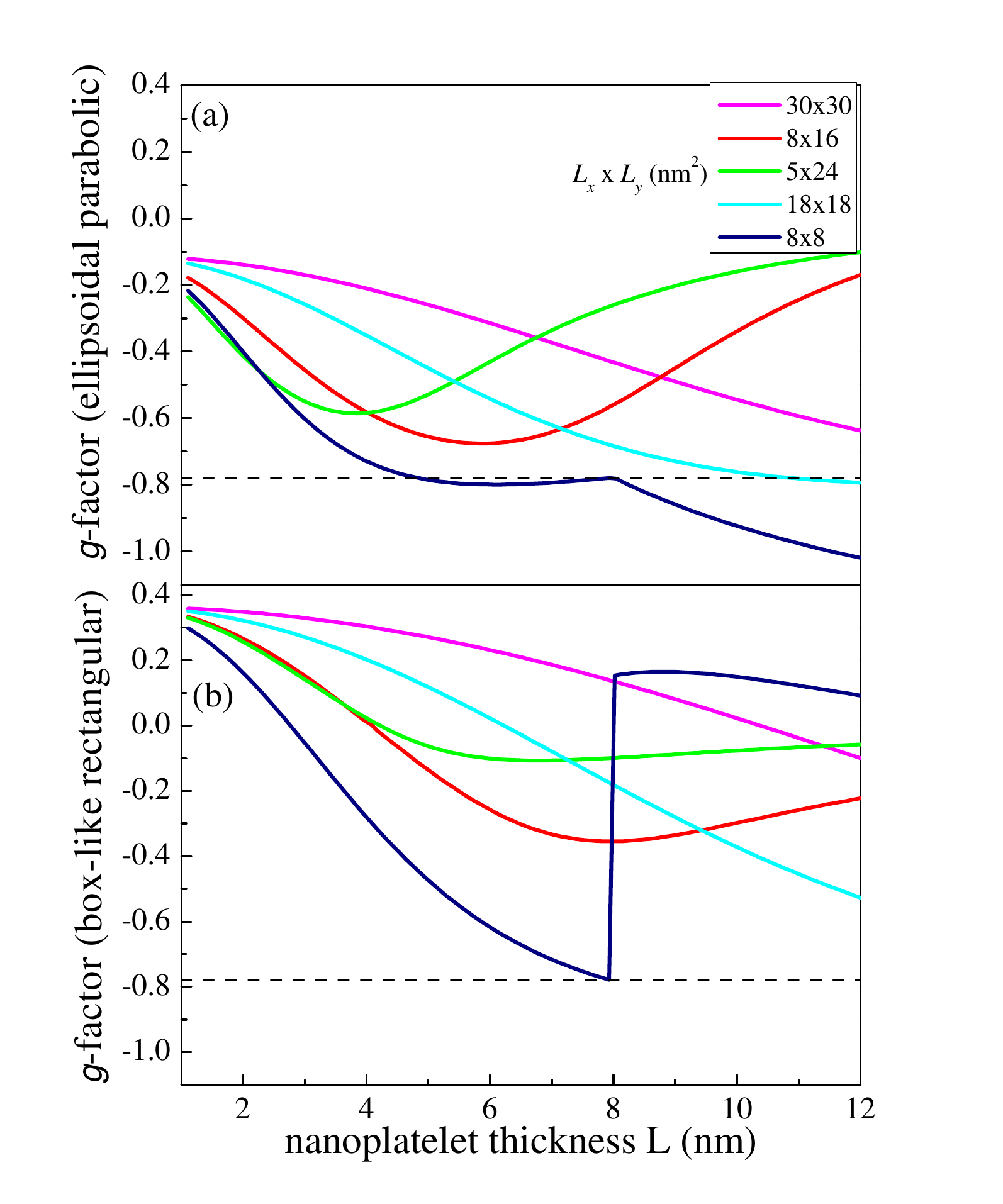}
\caption{Size dependencies of hole ground state $g$-factor  for (a) ellipsoidal nanocrystals with parabolic potential and  (b) cuboid nanocrystals with  box-like infinite potential. The nanocrystal width $L$ is the width along the nanocrystal axis $z$ being the direction of the external magnetic field and the curves of different colors correspond to different NC in-plane sizes shown as $L_x \times L_y$ in nm$^2$. The calculations are done for zb-CdSe parameters given in Table \ref{Table_g1} with $\beta = 0.22$. The dashed horizontal lines correspond the $g$-factor value in (a) spherical  and (b) cube CdSe NCs: $g_h^{\rm sph}\approx g_{h,3/2}^{\rm cube} \approx -0.78$.  
}\label{fig_width}
\end{figure}

Above, we considered the ideal model nanostructures, for example, 2D quantum well or quantum wires with an infinite in-plane area or length, respectively. For such structures as well as for spherical and cube NCs considered in the framework of Luttinger Hamiltonian, hole $g$-factors are independent of the characteristic size of the system unless the effect of the renormalization of Luttinger parameters is considered. The realistic  colloidal NPLs, however,  have a finite thickness and a finite in-plane size  \cite{Olutas2015_1,Shornikova2018,Ayari2020}.  We discuss below the effect of the finite in-plain  size of  NPL on the heavy hole $g$-factor.

 Even conventional semiconductor quantum wells, while their in-plane size could be quite large,  have a finite thickness. The finiteness of the ratio  of the NPL thickness to in-plane size results in non-zero light and heavy holes mixing even in zero magnetic field. This leads to the nonzero first order perturbation renormalization of the spin Zeeman term as well as orbital correction to the hole effective $g$-factor. As a consequence, the  hole ground state $g$-factor is somewhere in between the 2D,  $g_{h,3/2}^{2D}$, and spherical, $g_h^{\rm sph}$ limits. The in-plane anisotropy of the structure results in additional corrections to $g$-factor values as well. In Fig. \ref{fig_width} we show the dependencies of the hole ground state $g$-factor on nanocrystal width $L$ for different in-plane sizes in the case of  (a) box-like infinite and  (b) parabolic potentials. The calculations are done for the zb-CdSe parameters given in Table \ref{Table_g1} with $\beta = 0.22$. For the sake of comparison, we define that size $L_\alpha$ of the nanocrystals with parabolic potential along each direction $\alpha =x,y,z$ as $L_\alpha=4L_{h,\alpha}$, where  $L_{h,\alpha}$  is the oscillator length along respective direction calculated with the heavy hole effective mass. One can see from the Fig. \ref{fig_width}, that in the limit $L=L_z \rightarrow 0$  for both potentials all curves tend to the same limit, corresponding to the ideal 2D quantum well: $g_{h,3/2}^{2D}=-0.11$ for the parabolic potential and $g_{h,3/2}^{2D}=0.36$ for the abrupt box-like potential. Dashed line on both panels corresponds to the value of the heavy hole $g$-factors in cubic (spherical) NC which are  occasionally the same in the case of zb-CdSe: $g_h^{\rm sph} \approx g_{h,3/2}^{\rm cube} \approx -0.78$. With an increase of $L$, $g_{h,3/2}$ starts to depend on the NC  in-plane size and thickness even without account of the energy dependence of the Luttinger parameters. Overall, the smaller the in-plane cross-section, $L_x \times L_y$,  the stronger is the dependence. For the smallest cross-section, $8\times 8$ nm$^2$, at $L=8$ nm the case of cubic (spherical) NCs takes place, so at $L>8$ nm we have the prolate NC with the ground state being the light hole. It leads to the discontinuity of the ground state $g$-factor in cube NC and to the knee in $g$-factor dependence for parabolic NC (see Fig. \ref{shape_anis}).  On both panels one can see, that for realistic in-plane size of the NPL, the dependence of hole $g$-factor on its thickness could be important. Also the values of hole $g$-factors even in thin NPL can differ from  value for  2D-limit because of the strong nonparabolicity effect. 

\section{Exciton effective $g$-factor }
\label{exciton}
 \begin{figure*}[ht]
\includegraphics[width=1\textwidth]{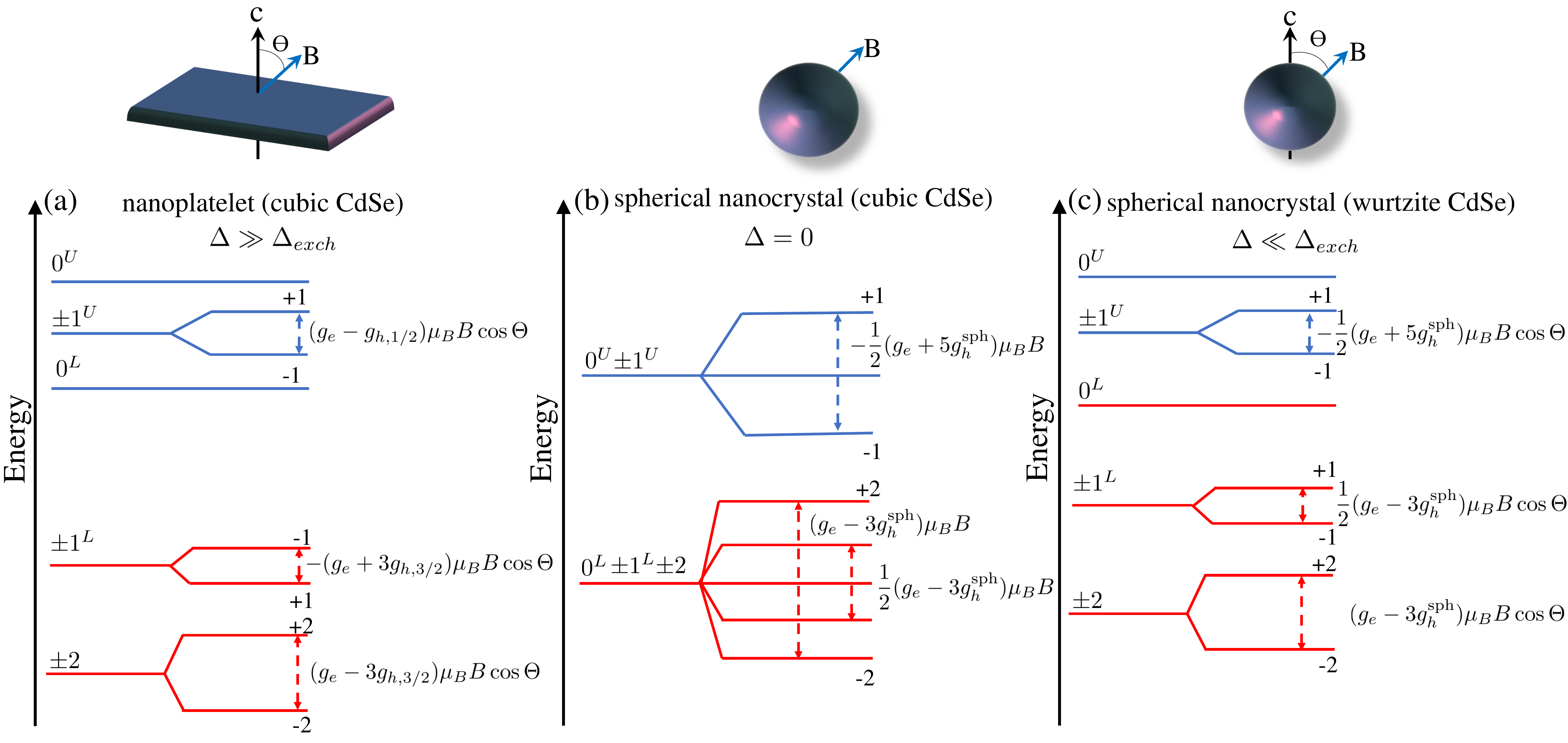}
\caption{Scheme of the exciton energy levels in magnetic field. (a) light and heavy hole splitting $\Delta$ is large as compared with exciton splitting due to exchange interaction $\Delta_{exch}$; (b) $\Delta=0$; (c) $\Delta\ll \Delta_{exch}$, all resulting $g$-factors are assumed to be positive, $\Theta$ is the angle between anisotropy axis and magnetic field. The order of energy levels is sketched for $g_e=1.7$, $g_h^{\rm sph} = -0.7$, $g_{h,3/2}=-0.1$,  $g_{h,1/2}=0.1$. } \label{scheme}
\end{figure*}
The knowledge of electron and hole effective $g$-factors in nanostructures allows one to describe energy splittings of allowed optical transitions in a low external magnetic field and to calculate respective bright exciton effective $g$-factors.  For  excitons formed with heavy holes $J_{hz}=\pm 3/2$ (or $M=\pm 3/2$ in nanocrystals), the allowed transitions in $\sigma^\pm$ polarization are:
	 $$E_{\sigma^{\pm}}= E_g +E_e+E_{h,3/2} \mp\frac{1}{2}g_e\mu_B B\mp\frac{3}{2}g_{h,3/2}\mu_B B.$$
For transitions with light hole excitons, $J_{hz}=\pm 1/2$ or $M=\pm 1/2$ one has 
$$E_{\sigma^\pm} =E_g +E_e+E_{h,1/2}\pm\frac{1}{2}g_e\mu_B B\mp\frac{1}{2}g_{h,1/2}\mu_B B \, .$$
Here we neglected both the direct Coulomb and exchange electron-hole interaction and electron and hole are treated as independent non-interacting quasi-particles.   In Fig. \ref{fig:2} the energy levels and allowed transitions are shown schematically for the 4-fold degenerate hole energy level in zero field  $E_{h,3/2}=E_{h,1/2}=E_h$ and 
$g_{h,3/2}=g_{h,1/2}=g_h$ realized in the bulk zinc blende semiconductors or spherical nanocrystals.

\begin{equation}\label{g_X}
g_{ex,1}=\frac{E_{\sigma_+}-E_{\sigma_-}}{\mu_{\rm B} B} = \frac{E_{F=+1}-E_{F=-1}}{\mu_{\rm B} B},
\end{equation}
where $F=J_{hz}+S_z=\pm 1$ or $F=M+S_z$ denotes the total exciton spin projection on magnetic field direction. 
From Eq. \eqref{g_X}, the exciton effective $g$-factors for the bright exciton with heavy, $g_{ex,1}^{hh}$, and light, $g_{ex,1}^{lh}$,  hole are:
\begin{equation} \label{gbright}
g_{ex,1}^{hh}=-(g_e+3g_{h,3/2}), \quad  g_{ex,1}^{lh}=g_e-g_{h,1/2} \, .               
\end{equation}
The dark exciton states with $F=\pm 2$   are not optically active in one-photon processes in dipole approximation. However, for such excitons one can use the definition of $g$-factor  similar to Eq. \eqref{g_X}:
	\begin{equation}\label{g_X2}
	g_{ex,2} = \frac{E_{F=+2}-E_{F=-2}}{\mu_{\rm B} B} = g_e-3g_{h,3/2} \, .
	\end{equation}
 The definition \eqref{g_X},\eqref{g_X2} is used, for example, in Refs. \cite{vanKesteren1990,Efros1996,EfrosCh3}. Note, however, that depending on the chosen  hole $g$-factor definition, the resulting expressions for the exciton $g$-factors may differ. In addition,  another definition of the exciton $g$-factors with $g_{ex,|F|}=(E_F-E_{-F})/2F\mu_{\rm B}B$ can be used. Often in the literature, for example in \cite{Shornikova2020nn,qiang2020},  the dark exciton $g$-factor is denoted as 	$g_F \equiv g_{ex,2}$.

For the correct description of exciton Zeeman splitting the fine structure of exciton energy needs to be taken into account even for excitons confined in the nanocrystals in a strong confinement regime.  First, we consider exciton fine energy structure in zero magnetic field. In the 2D case of a  nanoplatelet, the strong quantization along the symmetry axis splits the heavy and light hole states by the large energy $\Delta$. The further structure of the exciton states formed with heavy and light holes is determined by the electron-hole exchange interaction with characteristic energy   $\Delta_{exch} = 4\eta>0$ which is much smaller than $\Delta$ and results  in the energy splitting between states with different values of $|F|$ (Fig. \ref{scheme} (a)). In the notation of the exciton states, apart from the total projection of the exciton spin $F$, the letters $U$ and $L$ are used for the upper and lower levels with the same $F$. In the 2D case in Fig. \ref{scheme} (a), lower excitons are the heavy-hole excitons and the upper are formed with the light holes. The ground  exciton $\pm 2$ is dark and the heavy-hole  bright exciton  $\pm 1^{L}$ is shifted up by $\Delta E_{AF} = 3\eta \approx 5$  in 4 ml CdSe NPL \cite{Shornikova2018}.  We assume that the magnetic field energy $\mu_{\rm B}B$ is much smaller than $\Delta$ and $\Delta_{exch}$. Then the Zeeman energy splitting of the exciton states with $F\neq 0$ is highly anisotropic (depends on the angle between the anisotropic axis and external magnetic field $\Theta$) and it is controlled by   $g_{ex,1}^L = g_{1,ex}^{hh}$ and $g_{ex,1}^U = g_{1,ex}^{lh}$ from  Eqs. (\ref{gbright}) for bright and Eq. (\ref{g_X2}) for dark excitons, correspondingly. Note, that the $0^{L}$ is a dark state and the excitation probability of the $0^{U}$  is strongly dependent on light propagation direction \cite{EfrosCh3}.
	
In opposite,  for small spherical NCs of zinc blende or wurtzite semiconductors exchange interaction plays a key role in exciton energy structure. 	The electron-hole exchange interaction 	in the spherically symmetric structures of zinc-blende semiconductor split the 8-fold degenerate by the total exciton projection $F=M+S_z$ exciton ground state  into two states: 3-fold degenerate state with total momentum ${\cal F}=1$ and 5-fold degenerate state with ${\cal F}=2$ (Fig. \ref{scheme} (b)) \cite{Efros1996,EfrosCh3,Rodina2010}. The exchange energy splitting $\Delta_{exch}= E_1-E_2= 4\eta >0$ increases in small nanocrystals as $\eta \propto 1/a^3$. On the other hand, one has to consider the total anisotropic splitting $\Delta = \Delta_{cr}+\Delta_{sh}$ of the hole states caused by the crystal field in wurtzite semiconductors and shape anisotropy as discussed in Sect. \ref{anis}. As it was already mentioned, for wz-CdSe,  "quasi-spherical" nanocrystals with isotropic excitons ($\Delta=0$) shown in Fig. \ref{scheme} (b) can be realized for the prolate shape depending on the NC radius and the anisotropy parameter $\mu$ \cite{Gupta2002}.

In the general case, the exciton fine structure in zero magnetic field is caused by the joint effect of the exchange interaction and the uniaxial anisotropy. The energy splittings, including $\Delta E_{AF}$ between dark $\pm 2$ and $\pm 1^L$ bright excitons depended on both, $\Delta$ and $\eta$ energy parameters.   As $\Delta$ and $\eta$ scales with $a$ differently, the splitting between levels changes with NC size. The level structure in Fig. \ref{scheme} (c) is shown  for the case $0 < \Delta \ll \Delta_{exch}$. 
  \begin{figure}[h!]
	\includegraphics[width=0.85\linewidth]{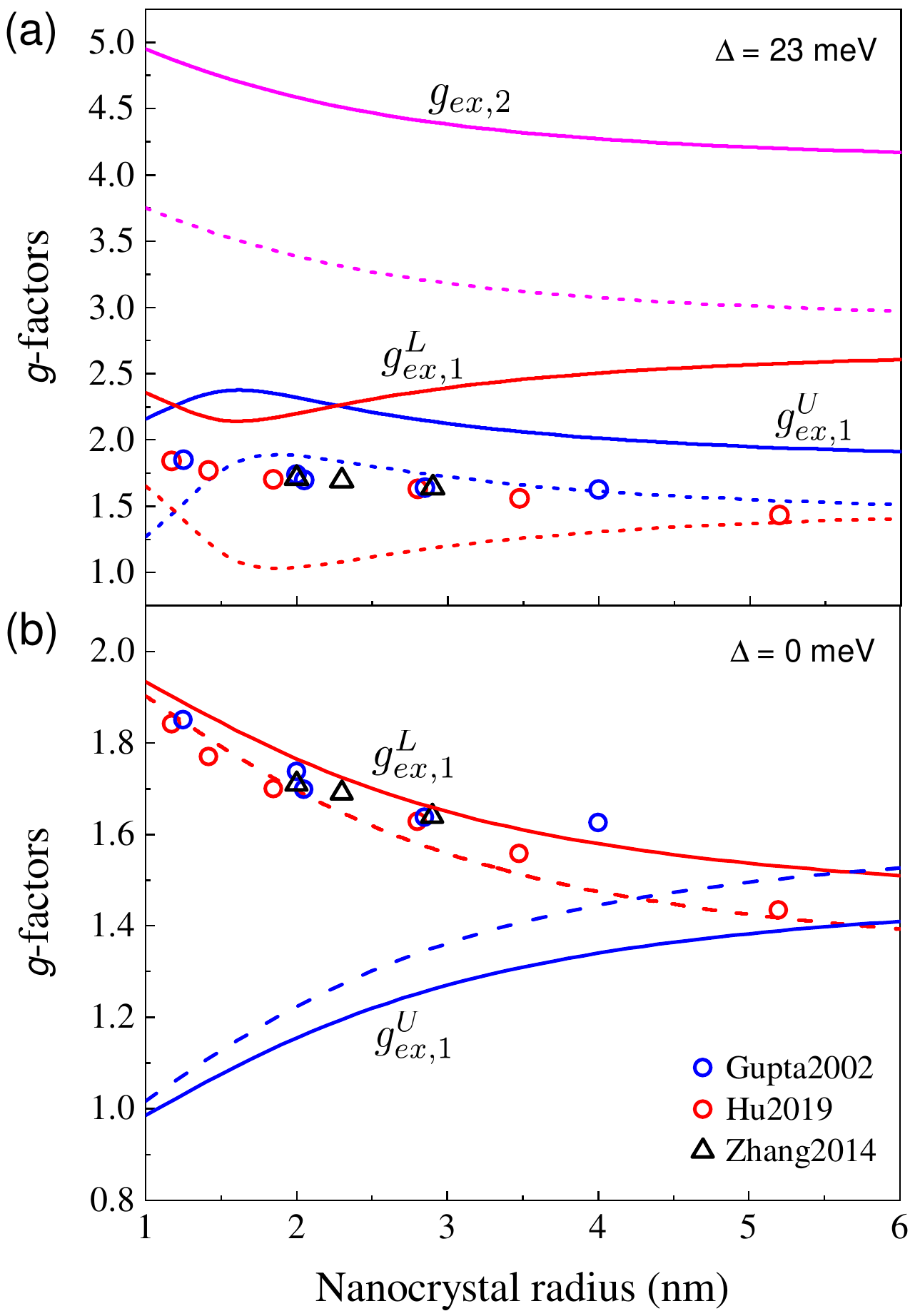}
	\caption{(a) Size dependence of $g$-factors for dark exciton (magenta line) and  upper and lower bright excitons (blue and red lines) in  spherical wz-CdSe NCs. Solid lines are calculated with $g_h^{\rm sph}=-1.1$ (see Table \ref{Table_g1} for $\beta=0.28$), and dashed lines with $g_h^{\rm sph} = -0.73$ from Ref. \cite{Gupta2002}.  Symbols show the experimental  $g_2$ values from Refs. \cite{Gupta2002,Hu2019,Zhang2014}, the same as in Fig. \ref{gefit}(a). (b) Size dependencies of $g_{ex,1}^L$ and $g_{ex,1}^U$ with  $g_h^{\rm sph} = -0.73$ calculated according to Eq. \eqref{g_small} in quasi-spherical ($\Delta=0$) wz-CdSe NCs (solid lines)  and zb-CdSe NCs (dash lines).}\label{crystal}
\end{figure}

 \begin{figure}[h!]
	\includegraphics[width=0.85\linewidth]{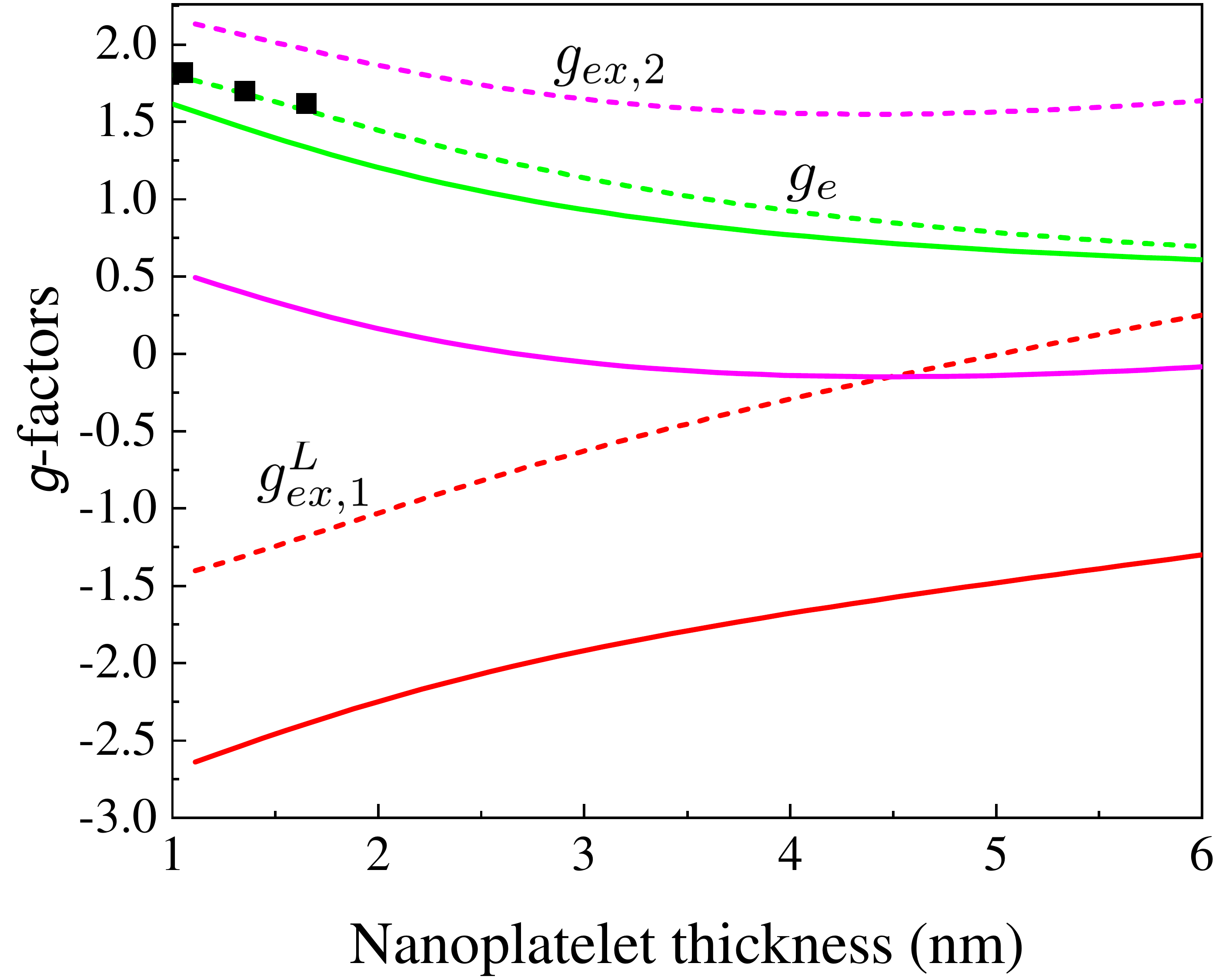}
	\caption{ Size dependence of $g$-factors for the dark exciton (magenta line) and the  lower bright exciton (red lines) in  zb-CdSe NPLs. Calculations are made for zb-CdSe (see Table \ref{Table_g1} for $\beta=0.22$)  with magnetic field directed perpendicular to the platelet  plane. Green lines show the calculated electron $g$-factor $g_e$.  Dash lines correspond  to the parabolic potential ( $g_{h,3/2} = -0.1$ in the 2D limit) and solid lines to the box potential ( $g_h=0.36$  in the 2D limit) for the in-plane size $L_x \times L_y = 30 \times 30$~nm$^2$.   Symbols show the  experimental $g_e$ data  from  Ref.~[\onlinecite{Kudlacik2019}], the same as in Fig. \ref{gefit}. }\label{crystal}
\end{figure}

In external magnetic field the Zeeman splitting of the dark exciton state with $|F|=2$ is, for any nanostructure, still given by  Eq. \eqref{g_X2} with $g_{h,3/2} = g_h^{\rm sph}$ in spherical nanocrystals. However, $g$-factors of bright exciton states with $|F|=1$ are modified due to the mixing of the light and heavy hole states and are given in general case as
\begin{eqnarray}\label{g_LU}
g_{ex,1}^{L,U}= |C^\pm|^2 g_{ex,1}^{hh}+|C^\mp|^2 g_{ex,1}^{lh} \, .
\end{eqnarray}
Here $C^\pm$ are the heavy to light hole mixing coefficients introduced in Ref. \cite{Efros1996,EfrosCh3}: 
\begin{equation}\label{Cpm}
	|C^\pm|^2 = \frac{1}{2} \left( 1 \pm \frac{\Delta - 2\eta}{\sqrt{(\Delta -2\eta)^2 +12 \eta^2}} \right) \, .
\end{equation}
Again, as $\Delta$ and $\eta$ scales with $a$ differently, the values of $|C^+|^2$ and $|C^-|^2=1-|C^+|^2$ are varying with nanocrystal size from $|C^+|^2=1$ for large nanocrystals with $\Delta >0$ similar to the QW case (or $|C^+|^2=0$ for  $\Delta <0$ similar to the nanowire case) to  $|C^+|^2=1/4$ in small nanocrystals. In the latter case and for spherical nanocrystals with $g_{h,3/2}=g_{h,1/2}=g_h^{\rm sph}$, the effective $g$-factors can be obtained from Eq. \eqref{g_LU} as \cite{EfrosCh3}:
\begin{equation}\label{g_small}
g_{ex,1}^{L}\approx \frac{g_e-3g_h^{\rm sph}}{2} \, , \quad 
g_{ex,1}^{U}\approx -\frac{g_e+5g_h^{\rm sph}}{2}. \end{equation}
The energy level splitting for this limit are shown in Fig. \ref{scheme}(b) for zb-CdSe (isotropic splitting) and Fig. \ref{scheme}(c) for wz-CdSe (anisotropic splitting depending on the angle $\Theta$). Note,  that for spheroidal nanocrystals with the shape anisotropy contribution $\Delta$, one has to use Eq. \eqref{g_LU} with $g_{h,3/2}\ne g_{h,1/2}$. 

The states with $F=0$ do not split in the magnetic field. However,  strong fields ${\bm B} \parallel {\bm c}$ may mix the $0^L$ with $0^U$ states as well as $\pm 1^L$ states with $\pm 1^U$ \cite{Rodina2016}, while the perpendicular component of the magnetic field $\propto B \sin \Theta$ mixes the state with $F$ differing by $\pm 1$ \cite{Efros1996}. Such a mixing should be taken into account in the analysis of exciton energy level structure in strong magnetic fields with $\mu_{\rm B}B$ comparable or larger than the zero-field fine structure splittings (see, for example, \cite{Brodu2019}.)

If one takes into account the cubic symmetry of the crystal lattice additional term in exchange interaction $\propto (\sigma_{e,x}J_{h,x}^3+\sigma_{e,y}J_{h,y}^3+\sigma_{e,z}J_{h,z}^3) $  appears \cite{PhysRevB.11.1512}. It splits the exciton state with ${\cal F}=2$ into 2-fold degenerate in zero field state and the 3-degenerate state.  The effect of the external field in that case would be strongly dependent on the field direction with respect to the crystallographic axes. This effect is rather weak and we neglect it. If one considers nanocrystals with cubic shape a similar term in exchange interaction will appear (as it happens for hole $g$-factors) and in that case, the effect might be strong. We will discuss this problem in a separate paper.

All above consideration of the exciton $g$-factor in NCs is valid if the magnetic field is directed along the crystal axis $c$ for wz-CdSe.  If it is not the case, effective exciton $g$-factors in Eqs. \eqref{gbright}, \eqref{g_X2}, \eqref{g_small} will obtain the factor $\cos\Theta$ as it is shown in Fig.~\ref{scheme}. In real ensemble of randomly oriented nanocrystals the observable lines would be broadened. In the following Fig. \ref{crystal} (a) it is assumed, that magnetic field is parallel to NC $c$-axis ($\cos\Theta=1$) and perpendicular to NPL plane in Fig. \ref{crystal} (b).

The dependence of exciton and electron $g$-factors on nanocrystal diameter for spherical  nanocrystals without shell and based on wurtzite CdSe are shown in Fig. \ref{crystal} (a). Hole $g$-factor is $g_h=-1.12$, see Table \ref{Table_g1} for $\beta=0.28$, (solid lines) and $g_h=-0.73$ \cite{Gupta2002} (dashed lines), crystal field splitting is $\Delta=23$ meV and exchange interaction parameter $\eta (a/a_B)^3 =0.1$ meV corresponds to the account of short-range interaction only \cite{Nirmal1995}.  One can see the significant difference between the  exciton and single electron  $g$-factors. So measuring $g$-factors in experiment can in the ideal situation help one to distinguish the state of NC being observed.   Symbols show the experimentally measured  $g_2$ values from Refs. \cite{Gupta2002,Hu2019,Zhang2014}, the same as in Fig. \ref{gefit}(a). As it was discussed in Sect. \ref{electron}, these $g_2$ values can not be  ascribed to ground state of the resident electron neither in wz-CdSe nor in zb-CdSe. As it is shown in Fig. \ref{crystal} (a),  these values indeed  could be possibly ascribed to $g_{1,ex}^L$ exciton in the ''quasi-spherical" wz-CdSe and spherical zb-CdSe with $\Delta=0$ (see dash-dotted lines) as it was suggested in Ref. [\onlinecite{Gupta2002}]. However, this good fit does not prove the origin of these $g_2$ values as they can be also fit for the electron excited states or assigned to  surface-localized  electron states as it was suggested in  Ref. [\onlinecite{Hu2019}]. Further experimental studies of the bright exciton $g$-factor size-dependence in spherical NCs by different experimental techniques, for example by the SFRS, are needed to clarify this situation.  The available  data for the dark exciton $g$-factor in CdSe NCs were obtained from the polarized photoluminesence in magnetic field \cite{JohnstonHalperin2001,Granadosdelguila2017,qiang2020} or from the Zeeman splitting in single NC \cite{Biadala2010} and correspond to the calculations with $|g_h^{\rm sph}| \le 0.73$ (dashed line in  Fig. \ref{crystal} (a)) rather than with $g_h = -1.1$ (solid line). 

 In Fig. \ref{crystal} (b) we show the size dependence of $g$-factors for the dark exciton (magenta line) and the lower bright exciton $g_{ex,1}^L=g_{ex,1}^{hh}$ (Eq. \eqref{gbright}, red lines) in  CdSe NPLs. Calculations are made for zb-CdSe (see Table \ref{Table_g1} for $\beta=0.22$). Solid lines correspond  to the parabolic potential ($g_{h,3/2} = -0.1$ in the 2D limit) and dashed lines to the box potential ( $g_h=0.36$  in the 2D limit) for the in-plane size $L_x \times L_y = 30 \times 30$~nm$^2$. The respective size dependencies of $g_{h,3/2}$ can be found in Fig. \ref{fig_width}.  Green lines show the electron $g$-factor. The comparison with the experimental data from Ref. \cite{Shornikova2020nn}, not only $g_{h,3/2}$, but also the dark exciton $g$-factor, $g_{ex,2}$ calculated for the parabolic confining potential correspond to the observed values  $g_{h,3/2} \approx -0.1$ and $g_{ex,2} \approx 2$ for 4 ml NPL. The main difference for the exciton $g$-factors calculated for two confining potentials comes from the hole $g$-factor, while the difference for the electron $g$-factor is not so significant. However, one can see in Fig. \ref{crystal} (b) that the electron $g$-factor size dependence calculated for the parabolic potential perfectly describes the experimental data for the resident electron $g$-factor measured by the SFRS in Ref. [\onlinecite{Kudlacik2019}].
 

\section{Conclusion} \label{sum}

To summarize, we have revisited the problem of definition and calculation of electron, hole, and exciton $g$-factors in II-VI and III-V-based nanocrystals. We have calculated the electron $g$-factor within 8-band  $\bm k\cdot\bm p$-model in spherical and planar bare NCs and demonstrated a good agreement with the results of the tight-binding calculations as well as with the experimental data. 

We have presented the semi-analytical methods for calculation of hole $g$-factor in NCs of different shapes and symmetry: spherical, axial (planar, spheroid and wire), and cube in a full range of heavy and light hole effective mass ratio $\beta$.  The results allow one to calculate the hole $g$-factor for all studied structures just knowing the set of Luttinger parameters. We have shown that the main contribution to the renormalization of the hole $g$-factor in nanostructure as compared with the bulk value comes from the orbital effect. It is more substantial in semiconductors with small $\beta$ and a large value of $\gamma_1$ Luttinger parameter. Our original results for cube NCs demonstrate the consequence of breaking the rotational symmetry and call for further experimental studies of cube zb-CdSe which can be probably synthesized by the cation exchange from the cube PbSe NCs. We have predicted the non-equidistant Zeeman splitting of the hole ground state which is four-fold degenerate in the isotropic cube NCs in zero magnetic field.    We have also considered the effect of the NC shape on the heavy and light hole $g$-factors and predicted their difference in the prolate and oblate NCs even if the hole ground state degenerate in zero field, for example in "quasi-spherical" wz-CdSe or "isotropic" InP NCs. 

Finally, we have investigated the size-dependences of dark and bright exciton $g$-factors in spherical NCs and planar NPLs and discussed the attribution of the experimentally observed multiply $g$-values in spherical CdSe NCs. We have shown that the experimental data for the electron, hole, and exciton $g$-factors in thin CdSe NPLs can be well described by assuming that electron and hole are confined in a smooth parabolic potential.

\section*{Acknowledgments}
We thank D.R. Yakovlev, M.M. Glazov, P. Sercel and Al.L. Efros for valuable discussions. 
This work was funded by the Russian Science Foundation (Grant No. 20-42-01008).

\appendix
\setcounter{equation}{0}
\setcounter{figure}{0}
\setcounter{table}{0}
\renewcommand{\thefigure}{A\arabic{figure}}
\renewcommand{\theequation}{A\arabic{equation}}
\renewcommand{\thetable}{A\arabic{table}}

\section{Size dependence  of the electron energy level and effective $g$-factor in the eight-band Kane model}
 \label{AA}
 \begin{table*}[ht]
 	\centering
 	\small
 	\caption {\ Parameters for calculation of electron $g$-factor according to Eq.~\eqref{eqApp2}. $E_g$ is the energy gap, $E_p$ is the interband coupling matrix element in the Kane model, $\Delta_{SO}$ is the spin orbit energy, $g_{e}$ and $m_e/m_0$ are the electron $g$-factor and   effective mass in bulk semiconductor. For a comparison with tight-binding results for size dependence of electron $g$-factor in wz-CdSe, CdTe, GaAs, InP from Ref. [\onlinecite{Tadjine2017}] we use set of parameters from Ref.[\onlinecite{Tadjine2017}], except electron effective masses. The second set of parameters for wz-CdSe is taken from Refs.~[\onlinecite{Ekimov1993,Piper1967}]. For zb-CdSe we use set of parameters from Ref. [\onlinecite{Karimov2000}].    }
 	\begin{tabular*}{1\textwidth}{@{\extracolsep{\fill}}lllllll}
 		\hline
 		Semiconductor & $E_g$, eV & $E_p$, eV & $\Delta_{SO}$, eV & $g_{e}$& $g_{e}$ (exp.)&$m_e/m_0$ \\
 		\hline
 		wz-CdSe & 1.8174 & 21.40 & 0.3871 & 0.633 &0.68[\onlinecite{Piper1967}]& 0.13 [\onlinecite{Adachi2004}] \\
 		wz-CdSe & 1.84 & 17.5 & 0.42 & 0.68  [\onlinecite{Piper1967}] & & 0.11 \\
 		zb-CdSe &  1.764 & 18.3 & 0.47 & 0.42 &0.42 [\onlinecite{Karimov2000}]& 0.13 \\
 		CdTe &  1.611 & 19.57 & 0.8221 & -1.236&-1.66 [\onlinecite{Oestreich1996}]&  0.09 [\onlinecite{Adachi2004}] \\
 		GaAs &  1.519  & 25.34 & 0.3399 & -0.065&-0.44 [\onlinecite{Weisbuch1977}]& 0.067 [\onlinecite{Adachi2005}]\\
 		InP &  1.424  & 20.45 & 0.108 & 1.22&1.2[\onlinecite{Oestreich1996}]&0.08 [\onlinecite{Adachi2005}]\\	    
 		\hline
 	\end{tabular*}
 	\label{tab:table1}
 \end{table*}
 
In the eight-band ${\bm k} \cdot {\bm p}$ model, the normalization condition for the total electron wave function $\Psi_e = \Psi_e^c+\Psi_e^v$, where $\Psi_e^c(r)$ and  $\Psi_e^v(r)$ describe the conduction and valence band contributions, respectively, reads $\int|\Psi_e({\bf r})|^2d^3{\bf r}= \int(|\Psi_e^c({\bf r})|^2+|\Psi_e^v({\bf r})|^2)d^3{\bf r}=1$. Using the expression of $\Psi_e^v({\bf r})$ via $\Psi_e^c({\bf r})$ \cite{Rodina2008,Merkulov2010book} one can rewrite the normalization condition as
\begin{eqnarray}\label{eqApp1}	
 \int |\Psi_e({\bf r})|^2d^3{\bf r} = \int|  \Psi_e^c({\bf r})|^2  {\cal A}^{-1}(E_e )d^3{\bf r}= 1 \, , \nonumber \\
  {\cal A}(E_e ) = \left[ 1+  \frac{\partial m_e(E_e)}{\partial E_e} \frac{E_e}{ m_e(E_e)} \right]^{-1} \, .
\end{eqnarray}
Here $m_e(E_e)$ is the electron effective mass at the energy $E_e$ calculated from bottom of the conduction band:
\begin{multline}\label{eqApp4}	
m_e(E_e)=m_0\Bigg[\gamma_{rb}+\frac{E_p}{3}\left(\frac{2}{\tilde E}+\frac{1}{\tilde E_g+\Delta_{SO}}\right)\Bigg]^{-1} \, ,
\end{multline} 
where $\tilde E= E_g+E_e$ and $\gamma_{rb}$ takes into account the contribution of remote bands and  $m_e \equiv m_e(E_e=0)$ is the electron effective mass at the bottom of the conduction band. Note, that in the case of semiconductor heterostructure, the electron effective mass in \eqref{eqApp1}	as well as the electron effective $g$-factor $\tilde g_e(E_e)$ in \eqref{geNC} can be different in different materials.
In bare NCs, ${\cal A}$ does not depend on the coordinate and can be directly used as the renormalization constant for the conduction band contribution. It can be also written explicitly as 
 	\begin{eqnarray} \label{AEe}
{\cal A}(E_e)=\left[1+\alpha_p(E_e)E_e m_e(E)/m_0 \right]^{-1} \, , \\
\alpha_p(E_e) = \frac{E_p}{3} \left( \frac{2}{\tilde E^2}+\frac{1}{(\tilde E_g+\Delta_{SO})^2} \right) \, . \nonumber
\end{eqnarray}
At small energies $E_e \ll E_g$, one can approximate $m_e^{-1}(E) \approx m_e^{-1} - m_0^{-1}\alpha_p|_{E_e=0}E_e$  and ${\cal A}(E_e) \approx m_e/m_e(E)$ \cite{Rodina2008}.

As it was discussed in Ref. \cite{Rodina2003}, the surface contribution $g_{\rm sur}$ to the electron $g$-factor can be nonzero even in the case of the bare semiconductor nanostructure with the infinite potential barrier at the surface. For simplicity, we consider below the case of vanishing electron conduction band wave function component at the surface $\Psi_e^c({\bm r})|_{{\bm r}={\bm s}}=0$ corresponding to $g_{\rm sur}=0$. In this case, equation \eqref{eqApp2}	gives the universal dependence of the electron effective $g$-factor on the electron energy $E_e$ (not the optical transition energy) within eight-band ${\bm k} \cdot {\bm p}$ model valid for any nanostructure potential shape. The size dependence of the electron $g$-factor can be found by establishing the correspondence between the electron quantization energy $E_e$ and the size of the nanostructure of a particular shape.

\begin{figure} 
	\includegraphics[width=0.85\linewidth]{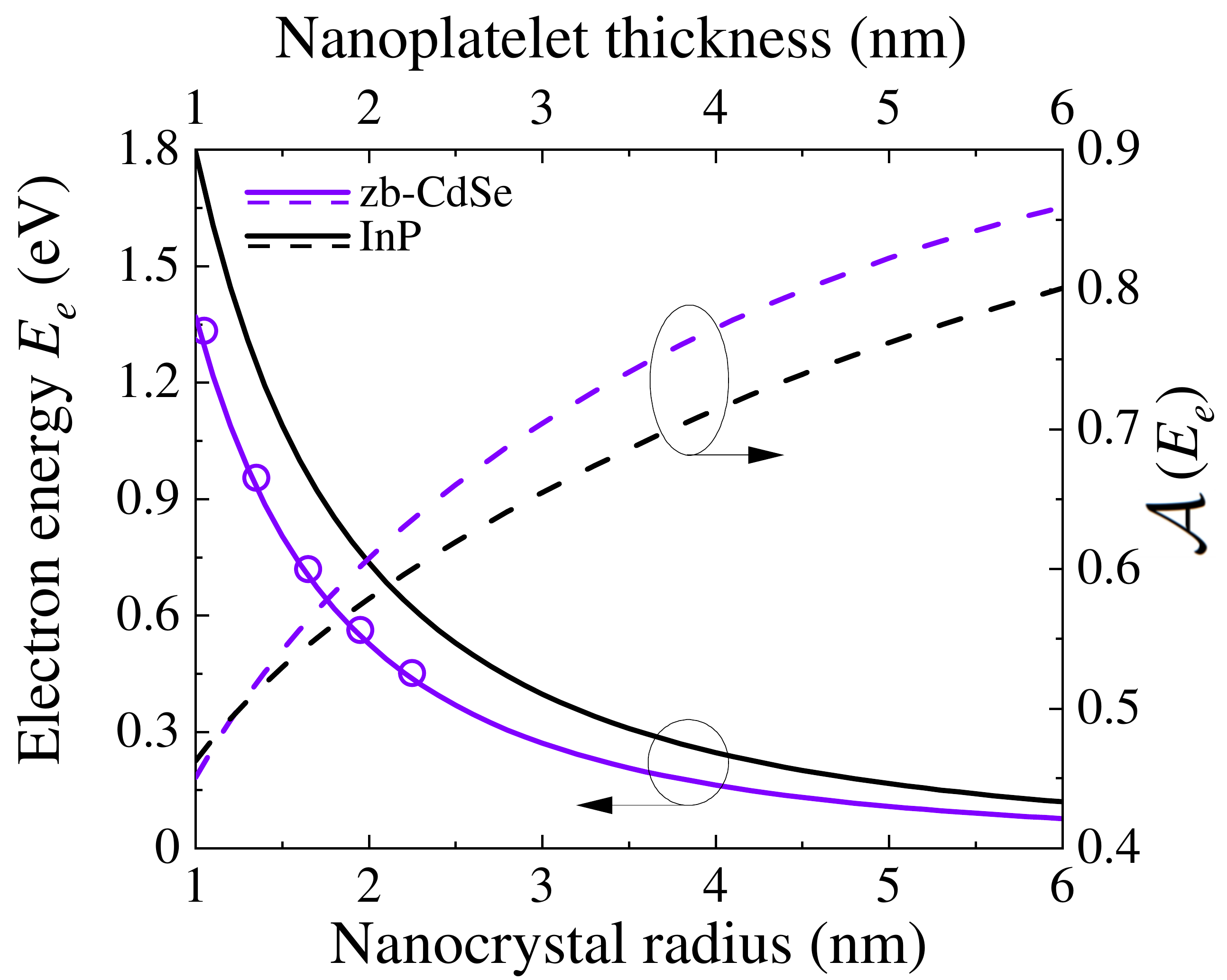}
	\caption{Size dependencies of the electron energy $E_e$ (solid lines, left axis) and renormalization constant ${\cal A}(E_e)$ (dashed lines, right axis)  in zb-CdSe and InP nanocrystals. Open circles show the results of the tight-binding calculations for zb-CdSe NPLs from \cite{Benchamekh2014}. }\label{fig:Eeme}
\end{figure}

For spherical semiconductor nanocrystals of the radius $a$, the   equation for the  ground state, $1S_{e}$, electron quantization energy under assumption of zero boundary condition $\Psi_e^c(a)=0$ reads:
\begin{eqnarray}\label{eqApp5}	
E_e=\frac{\hbar^2\pi^2}{2m_e(E_e)a^2} \, .
\end{eqnarray}
It can be solved numerically for $E_e$ at the a given $a$ to obtain $g_e(E_e(a))$. Alternatively, one can use the electron energy $E_e$ to obtain the parameterized dependence $g_e(a)$ Eqs.~\eqref{eqApp2} and \eqref{eqApp5}. 
 Fig.~\ref{fig:Eeme} shows the size dependencies of  the electron energy $E_e$ (solid lines, left axis) and renormalization constant ${\cal A}(E_e)$ (dashed lines, right axis)  in zb-CdSe and InP nanocrystals. The electron energies calculated  according to Eq.~\eqref{eqApp5} are in very good agreement with the results of the tight-binding calculations for zb-CdSe NPLs from \cite{Benchamekh2014}. One can see that in small NCs the renormalization ${\cal A}(E_e)$ can be as small as 0.5 resulting in additional corrections to the electron $g$-factor, up to 10\% as compared with Eq.~\eqref{LR}. It is also possible to calculate the electron $g$-factor for the excited $nS_{e}$ states in spherical NCs using Eq. \eqref{eqApp2} and factor $n^2$ in  Eq. \eqref{eqApp5}.   
The size dependencies of the electron energy level, effective mass and effective $g$-factor in  a QW or NPL with the thickness $L$ can be obtained  after the replacement of nanocrystal radius $a$ by the well width $L$ in Eq. \eqref{eqApp5}.

We consider also spherical NCs and 2D nanoplatelets with parabolic confining potential $V_{\rm ext}({\bf r}) = \kappa r^2/2$ and  $V_{\rm ext}({\bf r}) = \kappa z^2/2$, respectively.  For such potentials, the electron energy level can be found form: 
\begin{equation} \label{epar}
E_e=\frac{\hbar^2 n}{2\sqrt{m_e m_e(E_e)}L_e^2}.
\end{equation} 
where $n=3$ for NC and $n=1$ for NPL is the number of spatial directions and  $L_e=\hbar^{1/2}/(\kappa m_e)^{1/4}$ is the potential oscillator length with electron mass $m_e$ taken at the bottom of conduction band ($E_e=0$).  In order to calculate the size dependence of the electron $g$-factor in such NCs one needs to  define the NC radius $a$ or NPL  thickness $L$, in other words, relate it to $L_0$. For example, the choice $L=\pi L_e$ allows one to have in thick NPLs (where $E_e \ll E_g$ and the energy dependence of the electron effective mass in Eq. \eqref{epar} can be neglected)   the electron energy $E_e$, and consequently $g_e(E_e)$,   the same as  in the box-like potential. In thin NPLs, however, the electron energy level and respective electron $g$-factor in the parabolic potential becomes larger than for the box-like potential due to the weaker effect of the electron effective mass energy dependence. The size dependence of the electron $g$ factor in CdSe NPLs is plotted in Fig. \ref{crystal}(b) for  $L=4 L_e$ corresponding to 99.5\% of the electron density inside the NPL with boundaries at  $z=\pm L/2$.
\newline

\section{Magnetic field and anisotropic perturbation to the hole Hamiltonian}\label{AB}
\setcounter{figure}{0}
\renewcommand{\thefigure}{B\arabic{figure}}
\renewcommand{\theequation}{B\arabic{equation}}
The orbital Bloch functions of the top of valence band in studied semiconductors are of $p$-like symmetry and are often designated as $X$, $Y$, and $Z$, each having two possible spins, $\uparrow$ and $\downarrow$ \cite{ivchenko05a}. Taking into account spin-orbit interaction leads to the splitting of the valence band with  the following set of functions being the basis of the topmost $\Gamma_8$ subband \cite{ivchenko05a}:
\begin{equation}\label{Bloch}
|\Gamma_8,+\frac{3}{2}\rangle=-\uparrow\frac{X+\mathrm{i}Y}{\sqrt{2}}
\end{equation}
$$|\Gamma_8,+\frac{1}{2}\rangle=\sqrt{\frac{2}{3}}\uparrow Z-\downarrow\frac{X+\mathrm{i}Y}{\sqrt{6}} $$
$$ |\Gamma_8,-\frac{1}{2}\rangle=\sqrt{\frac{2}{3}}\downarrow Z+\uparrow\frac{X-\mathrm{i}Y}{\sqrt{6}}$$
$$|\Gamma_8,-\frac{3}{2}\rangle=\downarrow\frac{X-\mathrm{i}Y}{\sqrt{2}}.$$
Each of functions \ref{Bloch} have momentum $3/2$ and its projection on $z$-axis $+3/2$, $+1/2$, $-1/2$, and $-3/2$.

Luttinger Hamiltonian of Eq. (\ref{lutt})  can be written in a in matrix form in the basis \eqref{Bloch}  as follows  \cite{ivchenko05a}:
\begin{multline}\label{lutt_matrix}
\hat{H}_L=\frac{\hbar^2}{2m_0}\left(
\begin{array}{cccc}
 F & H & I &
   0 \\
 H^* & G & 0 & I \\
 I^* & 0 & G & -H \\
 0 & I^* & -H^* & F \\
\end{array}
\right),\\
F=(\gamma_1-2\gamma)k_z^2+(\gamma_1+\gamma)(k_x^2+k_y^2),\\
G=(\gamma_1+2\gamma)k_z^2+(\gamma_1-\gamma)(k_x^2+k_y^2),\\
H=-2 \sqrt{3} \gamma k_z (k_x-\mathrm i k_y),~ I=-\sqrt{3} \gamma (k_x-\mathrm i k_y)^2.
\end{multline}

 Anisotropic correction to Luttinger Hamiltonian has the form \cite{Efros1993,Semina2016}:
\begin{eqnarray}
\hat{H}_L^{an}&& =-\frac{2\mu}{3} \frac{\hbar^2 } {2m_0} \Bigg[ (\gamma_1 + \frac{5}{2}\gamma) (\hat k^2 - 3 \hat k_z^2)- \\ && 2\gamma [ (\hat {\bm k}{\bm J})^2 - 3\{ (\hat {\bm k}{\bm J}){\hat k_z}J_z \} ] \Bigg]. \nonumber
\end{eqnarray}
Note,  that here we use the opposite sign of $\mu$ as compared with Refs. [\onlinecite{Efros1993,Semina2016}]. In matrix form it can be written as:
\begin{eqnarray}
&&\hat{H}_L^{an}=-\frac{2\mu}{3} \frac{\hbar^2 } {2m_0} \left(
\begin{array}{cccc}
F^{an} &H^{an} & I^{an} & 0 \\
H^{an*} & G^{an} & 0 & I^{an} \\
 I^{an*} & 0 &  G^{an} &-H^{an} \\
 0 & I^{an*} & -H^{an*} &F^{an}  \\
\end{array}
\right), \nonumber \\
&&F^{an}=-2(\gamma_1-2\gamma)k_z^2+(\gamma_1+\gamma)(k_x^2+k_y^2),\\
&&G^{an}=-2(\gamma_1+2\gamma)k_z^2+(\gamma_1-\gamma)(k_x^2+k_y^2), \nonumber \\
&&H^{an}=\sqrt{3}\gamma k_z (k_x- \mathrm i k_y),~I^{an}=-\sqrt{3}\gamma(k_x- \mathrm i k_y)^2. \nonumber
\end{eqnarray}

In the presence of external magnetic field, the hole wave vector $\bm  k$ has to be renormalized as  
$\bm k\rightarrow \bm k-\frac{e\bm A}{c\hbar}$, where $\bm A$ is the vector potential of external magnetic field. The case $\bm B\parallel z$ in the Landau gauge correspond to $\bm A=(0,Bx,0)$. The B-linear correction to the Luttinger Hamiltonian \eqref{lutt} is \cite{Semina2015}:
\begin{eqnarray}
&&\hat{H}_B=\mu_BB\left\{-\left(2\gamma_1+5\right) 
  xk_y+ \right. \\ && \left. 2 \gamma \left[ 2J_y^2
   xk_y+\{J_yJ_z\} xk_z+\{J_xJ_y\}\left(xk_x-\frac{i}{2}\right)\right]\right\}. \nonumber
\end{eqnarray}
Here we neglect $\propto B^2$ corrections leading to diamagnetic shift as we consider a low field regime. It can  also be rewritten  in the  matrix form as B-linear corrections to the matrix Luttinger Hamiltonian \eqref{lutt_matrix}:
\begin{widetext}
\small{\begin{equation}\label{HBmatrix}
\hat{H}_B=\mu_B B \left(
\begin{array}{cccc}
 -2 x k_y(\gamma_1+\gamma) & -2 \mathrm i \sqrt{3} \gamma
    x k_z & -2\mathrm i \sqrt{3} \gamma x ( k_x - \mathrm i k_y)+\sqrt{3} \gamma & 0
   \\
 2 \mathrm i \sqrt{3} \gamma
    x k_z & -2 x k_y(\gamma_1-\gamma) & 0 & -2\mathrm i \sqrt{3} \gamma x ( k_x - \mathrm i k_y)+\sqrt{3} \gamma \\
 2\mathrm i\sqrt{3}  \gamma  x(k_x+\mathrm i k_y) +\sqrt{3}\gamma & 0 & -2 x k_y(\gamma_1-\gamma) & 2 \mathrm i \sqrt{3} \gamma
    x k_z \\
 0 &  2\mathrm i\sqrt{3}  \gamma  x(k_x+\mathrm i k_y) +\sqrt{3}\gamma & -2 \mathrm i \sqrt{3} \gamma
    x k_z & -2 x k_y(\gamma_1+\gamma) \\
\end{array}
\right).
\end{equation}}
\normalsize \end{widetext}

Now we obtain Eq. \eqref{g2d} from the main text as a first order correction from term $\hat{H}_B$ in the hole Hamiltonian \eqref{Hamilt_dot}. The orbital contribution from the magnetic field  $\hat{H}_B$ is more convenient to consider using its matrix form in Eq. \eqref{HBmatrix}. We are interested in energy splitting of heavy hole states ($J_z=\pm 3/2$) in thin quantum well-like structures, where the spitting of heavy and light holes is large, so far, their mixing is negligible and hole ground state can be considered as a pure heavy hole. Now we calculate the contributions to the energy of hole with $J_z=+3/2$ coming from  \eqref{HBmatrix}. Firstly comes the diagonal term, $-2\mu_B B(\gamma_1+\gamma)xk_y$:
\begin{equation}\label{3/2_3/2}
   E^{(1)}_{\frac{3}{2},\frac{3}{2}}= -2\mu_B B(\gamma_1+\gamma) \sum_n\langle\Psi_0|k_y|\Psi_n\rangle\langle\Psi_n|x|\Psi_0 \rangle,
\end{equation}
where $\Psi_0$ is the hole ground state wave function, $n$ denotes all possible intermediate hole states described by wave functions $\Psi_n$. We next use relations  \cite{Wang_2015}
\begin{multline}\label{pxy}
\langle\Psi_0| x|\Psi_n\rangle=\mathrm i\frac{ \hbar}{m_0}\frac{\langle\Psi_0| p_x|\Psi_n\rangle}{E_n-E_0},\\~\langle\Psi_0| y|\Psi_n\rangle =-\mathrm i\frac{ \hbar}{m_0}\frac{\langle\Psi_0| p_y|\Psi_n\rangle}{E_n-E_0},
\end{multline}
with $E_0$ being ground state energy and $E_n$ being intermediate state energy 
in order to express the coordinate matrix elements throw the  momentum operator  ${\bm p}$ matrix elements and  the relation
\begin{equation}\label{px}
\bm p=\frac{m_0}{\hbar}\frac{\partial \hat{H}_L}{\partial \bm k}
\end{equation}
in order to relate   ${\bm p}$ with the wave vector ${\bm k}$.  Note, that Luttinger Hamiltonian $\hat{H}_L$ is acting in space of 4-component wave functions and each case one has to take corresponding component of $\hat{H}_L$. Eq. (\ref{pxy}) and $p_x=\hbar(\gamma_1+\gamma)k_x$  from Eq. \eqref{px} for the  $\hat{H}_{L,\frac{3}{2},\frac{3}{2}}$ component   allow us to rewrite \eqref{3/2_3/2} as
\begin{equation}\label{3/2_3/2_1_1}
   E^{(1)}_{\frac{3}{2},\frac{3}{2}}= -\mu_B B(\gamma_1+\gamma)^2\frac{ 2 \mathrm i \hbar^2}{m_0} \sum_n\frac{\langle\Psi_0| k_y|\Psi_n\rangle \langle\Psi_n|k_x|\Psi_0 \rangle}{E_n-E_0}.
\end{equation}
The corresponding correction for hole with $J_z=-3/2$ is the same as \eqref{3/2_3/2_1_1}, so the diagonal matrix elements of $\hat{H}_B$ do not contribute to heavy hole $g$-factor.

Now we calculate the contribution from the matrix element $-2\mathrm i \sqrt{3} \mu_B B \gamma x k_z$, which mixes hole states with $J_z=+3/2$ and $J_z=+1/2$:
\begin{multline}\label{3/2_1/2}
E^{(1)}_{\frac{3}{2},\frac{1}{2}}= -2\mathrm i \sqrt{3} \mu_B B \gamma \sum_n\langle\Psi_0|k_z|\Psi_n\rangle\langle\Psi_n|x|\Psi_0 \rangle=\\=   \mu_B B \frac{6 \hbar^2}{m_0}\sum_n\frac{|\langle\Psi_0|\gamma k_z|\Psi_n\rangle |^2}{E_n-E_0} \, , 
\end{multline}
where $p_x=-\frac{\hbar\sqrt{3}}{m_0}\gamma k_z$ from \eqref{px} for the  $\hat{H}_{L,\frac{3}{2},\frac{1}{2}}$ component is used.  

Similarly, 
\begin{equation}
E^{(1)}_{-\frac{3}{2},-\frac{1}{2}}=-\mu_B B \frac{6 \hbar^2}{m_0}\sum_n\frac{|\langle\Psi_0|\gamma k_z|\Psi_n\rangle |^2}{E_n-E_0}.
\end{equation}
In the same way we obtain that corrections $E^{(1)}_{\frac{3}{2},-\frac{1}{2}}=E^{(1)}_{-\frac{3}{2},\frac{1}{2}}$
do not contribute to hole $g$-factors. Finally, the orbital contribution to heavy hole Zeeman splitting is
\begin{multline}\label{ggg}
\Delta E =E^{(1)}_{-\frac{3}{2},-\frac{1}{2}}-E^{(1)}_{\frac{3}{2},\frac{1}{2}}=\\=-\mu_B B \frac{12 \hbar^2}{m_0}\sum_n\frac{|\langle\Psi_0|\gamma k_z|\Psi_n\rangle |^2}{E_n-E_0}
\end{multline}
and the heavy hole $g$-factor  $g_{h,3/2}^{2D}$ is given by Eq. \eqref{g2d} in the main text.
As the only contribution to Eq.  \eqref{ggg} is coming from heavy and light hole mixing, $n$ in Eq. \eqref{g2d}  denotes even light hole states, which have non-zero matrix elements $\langle \Psi_0|\gamma k_z|\Psi_n\rangle$ with the ground state of the heavy hole. 

As the in-plane envelope wave functions are the complete basis and in Eq. \eqref{ggg} only operator $k_z$ is present, the summation over in-plane wave functions gives unity. Therefore, in Eq. \eqref{ggg}  as well as in Eq. \eqref{g2d} of the main text, $\Psi_0$ denotes the envelope wave function of heavy hole ground state and $\Psi_n$ are envelope wave functions of excited light hole states quantization along $z$ axis.

The function  $G(\beta)$ describing the orbital contribution to the heavy-hole $g$-factor for NPLs with box-like and  parabolic potentials can be expressed analytically
\begin{equation}\label{box_well}
G(\beta)=-\sum_{n=1}^{\infty} \frac{192 (\beta-1)^2 n^2}{\pi ^2 (\beta+1)
   \left(4 n^2-1\right)^2 \left(\beta-4
   n^2\right)} \nonumber
\end{equation}
for box-like potential and 
\begin{equation}\label{par_well}
G(\beta)=-\sum_{n=1}^{\infty} \frac{24 \sqrt[4]{\beta}
   \left(\sqrt{\beta}-1\right)^{2 n}
   \left(\sqrt{\beta}+1\right)^{1-2 n} \Gamma
   \left(n+\frac{1}{2}\right)}{\sqrt{\pi }
   (\beta+1) \left(\sqrt{\beta}-4 n+1\right)
   \Gamma (n)} \nonumber
\end{equation}
for parabolic potential with $\Gamma(x)$ being the Euler gamma function. In the limit $\beta\rightarrow 0$ the expression \eqref{box_well} can be expanded into a series and summarized analytically:
\begin{equation}\label{box_well_ser}
G(\beta)\approx 0.568-1.568 \beta+1.897
   \beta^2-1.82 \beta^3+1.839\beta^4. \nonumber
\end{equation}
The corresponding dependence is shown by thin dashed line in Fig.~\ref{gfact_well}  of the main text. It can be seen, that asymptotic \eqref{box_well_ser} demonstrates a good applicability for $\beta\lesssim 0.5$, corresponding to materials listen in the Table 1. For parabolic potential, the expression for $G(\beta)$ behaves as $\beta^{1/4}$ with $\beta\rightarrow 0$, although this asymptotic works only at non-physically small $\beta$, see the inset in Fig.~\ref{gfact_well} of the main text. For realistic values of $\beta$ the empirical asymptotic
\begin{equation}\label{par_well_ser}
G(\beta)\approx 1.1524 \beta^{-0.107}+9.1 \beta^2-5.714\beta  \nonumber
\end{equation}
can be written. It works for $10^{-4}\lesssim\beta\lesssim 0.3$ and shown in Fig.~\ref{gfact_well} of the main text by thin dash-dotted line.

The similar expression as Eq. \eqref{ggg} gives the orbital corrections to the Zeeman splitting of the hole at the  at the lowest light-hole quantum size level, $E_0=E_{lh_{1}}$, coming from the admixture of the even excited states of the heavy-hole,  $E_{hh_{2n}}$ \cite{Durnev2012}. In this case, $\Psi_0$ in Eq. \label{ggg} denotes the envelope wave function of light hole ground state and $\Psi_n$ are envelope wave functions of excited  even heavy hole states quantization along $z$ axis. The resulting expression for the light-hole $g$-factor reads
	\begin{equation}\label{ghl}
g_{h,1/2}^{2D} = 2\varkappa  -\mu_B B \frac{12 \hbar^2}{m_0}\sum_n\frac{|\langle hh_{2n}|\gamma k_z|lh_1\rangle |^2}{E_{lh_1}-E_{hh_{2n}}} \,.
\end{equation}

\section{Electron and hole energy levels in the nanocrystals with cubic shape}\label{AC}

\setcounter{figure}{0}
\renewcommand{\thefigure}{C\arabic{figure}}
\renewcommand{\theequation}{C\arabic{equation}}

\begin{figure}[h]
\includegraphics[width=0.9\columnwidth]{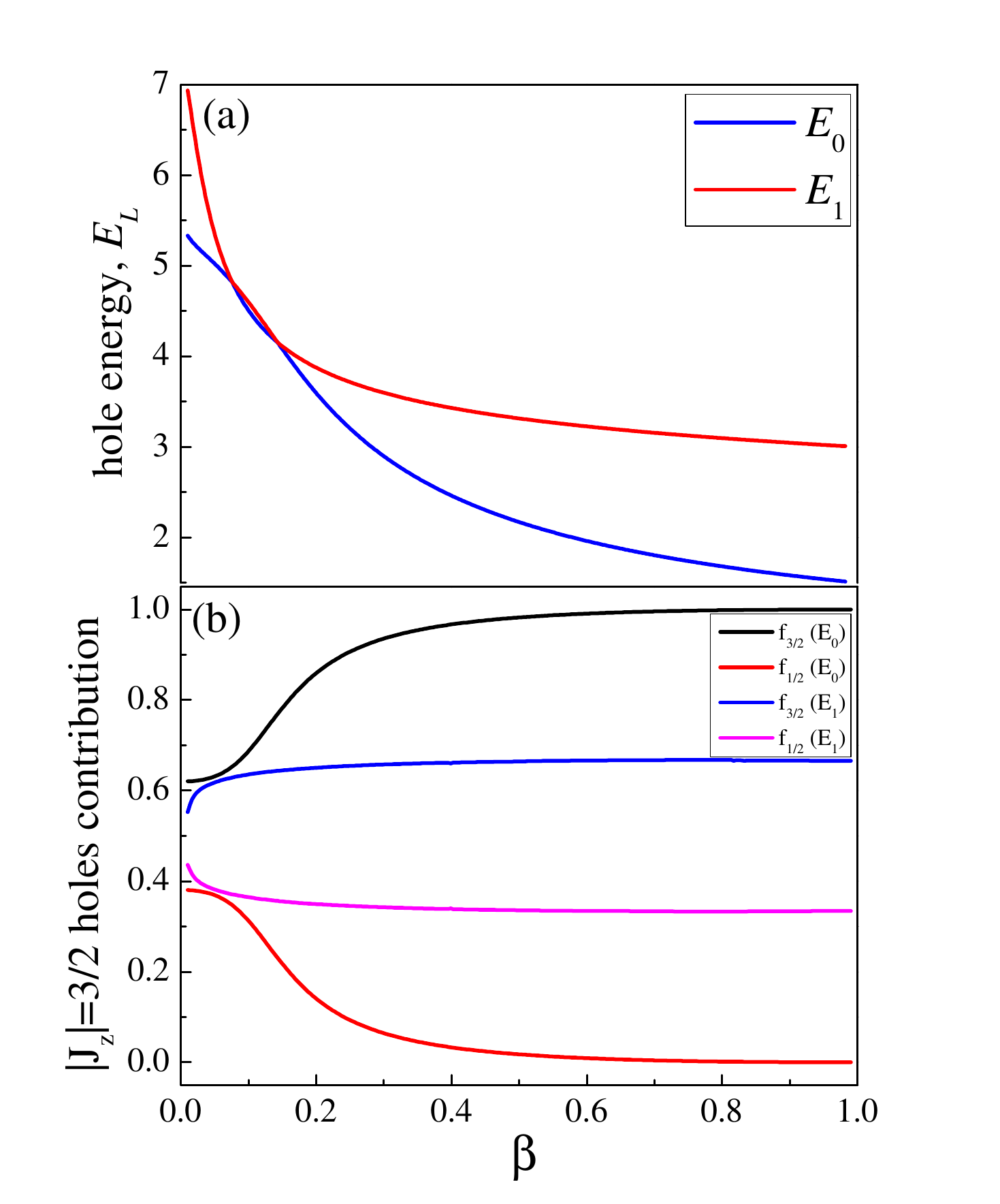}
   \caption{(a) The dependence of first two hole states energies $E_0$ and $E_1$ in cube nanulcrystals on the effective mass ratio $\beta$ in zero magnetic field. Energies are given in units $E_L=\hbar^2 \pi^2/m_{hh}L^2$, $L$ being nanocrystal size (cube width); (b) The heavy hole contributions as function of $\beta$ for the hole states originating from the ground and first excited states.}
   \label{hhcontr}
\end{figure}

{Here we consider nanocrystals with the cubic shape and edges directed along cubic axes of the crystal.  The potential of such cube nanocrystal  is described by a rectangular potential with infinite barrier:
\begin{equation}\label{cubic_pot}
V^{\text{cube}}(x,y,z)=\left\{
\begin{array}{cc}
0,&|x|,|y|,|z|\leq \frac{L}{2}\\
\infty,& |x|,|y|,|z|> \frac{L}{2}
\end{array}
\right.,
\end{equation}
where $x$, $y$, $z$ are electron or hole coordinates. The conduction band electron energy levels can be found from  \eqref{eqApp5} with $a \rightarrow L/\sqrt{3}$. With this substitution, the size dependence of the energy levels, effective mass and effective $g$-factors can be seen in Figs. \ref{fig:Eeme} and \ref{gefit}.

For holes situation is more complicated as the top of valence band is 4-fold degenerate and hole wave function is a four component one. In order to calculate hole states in cubic nanocrystal we developed numerical method. The hole kinetic energy is described by Luttinger Hamiltonian, Eq.~\eqref{lutt} and nanocrystal potential is taken from Eq.~\eqref{cubic_pot}. Potential \eqref{cubic_pot} mixes hole states with different spin projections $J_z$ and, unlike spherically symmetric case, due to lower symmetry the Schr\"odinger equation of the hole can not be simplified, compare with Ref. [\onlinecite{Gelmont1971}].   We numerically diagonalize hole Hamiltonian matrix calculated on the 4-component basis of eigenfunctions of infinite rectangular quantum well along coordinate axes:
\begin{eqnarray}\label{basis}
\Psi_{J_z}^{n_x,n_y,n_z}(x,y,z)=\phi_{J_z}^{n_x}(J_z)\phi_{J_z}^{n_y}(y)\phi_{J_z}^{n_z}(z),\\
\phi_{J_z}^{n_\alpha}(\alpha)=\sqrt{\frac{2}{L}}\sin\left[\frac{\pi n}{L}\left(\alpha+\frac{L}{2}\right)\right] \, ,  \nonumber
\end{eqnarray} 
where 
$J_z=\pm 3/2, \pm 1/2$, $n_x,n_y,n_z=1...N$, and  $\alpha=x,y,z$. All matrix elements are calculated analytically and basis size is $N=16$, which is more than enough to obtain convergence for several lowest hole states at any reasonable value of light to heavy effective mass ratio $\beta$. 

The dependencies of the energies of the two lowest hole states on $\beta$ are shown in Fig.~\ref{hhcontr}(a). The interesting point is the crossing of the levels $E_0$ and $E_1$ leading to level $E_1$ being the ground state in some range of $\beta$. This leads to the significant change of the hole ground state structure in the vicinity of the crossing point. On fig.~\ref{hhcontr} (b) we show the heavy hole ($|J_z|=3/2$) contributions to hole wave function, $f_{|M|}$, as function of $\beta$ for states, corresponding to energy levels $E_0$ and $E_1$. For rather large $\beta$ and far from crossing the ground state is $S$-like state $E_0$ and $3/2$ hole states are mostly formed from heavy holes and $1/2$ state are mostly light holes. On the opposite, for smaller $\beta$, in crossing range or near it levels $E_0$ and $E_1$ are close to each other and, consequently,  are strongly mixed. As a result, the contributions of heavy and light holes to the ground state sublevels are comparable leading to the specific features in dependencies of hole $g$-factors on $\beta$, see Fig.~\ref{cubic_gfactor} of the main text.




\begin{thebibliography}{125}%
\makeatletter
\providecommand \@ifxundefined [1]{%
 \@ifx{#1\undefined}
}%
\providecommand \@ifnum [1]{%
 \ifnum #1\expandafter \@firstoftwo
 \else \expandafter \@secondoftwo
 \fi
}%
\providecommand \@ifx [1]{%
 \ifx #1\expandafter \@firstoftwo
 \else \expandafter \@secondoftwo
 \fi
}%
\providecommand \natexlab [1]{#1}%
\providecommand \enquote  [1]{``#1''}%
\providecommand \bibnamefont  [1]{#1}%
\providecommand \bibfnamefont [1]{#1}%
\providecommand \citenamefont [1]{#1}%
\providecommand \href@noop [0]{\@secondoftwo}%
\providecommand \href [0]{\begingroup \@sanitize@url \@href}%
\providecommand \@href[1]{\@@startlink{#1}\@@href}%
\providecommand \@@href[1]{\endgroup#1\@@endlink}%
\providecommand \@sanitize@url [0]{\catcode `\\12\catcode `\$12\catcode
  `\&12\catcode `\#12\catcode `\^12\catcode `\_12\catcode `\%12\relax}%
\providecommand \@@startlink[1]{}%
\providecommand \@@endlink[0]{}%
\providecommand \url  [0]{\begingroup\@sanitize@url \@url }%
\providecommand \@url [1]{\endgroup\@href {#1}{\urlprefix }}%
\providecommand \urlprefix  [0]{URL }%
\providecommand \Eprint [0]{\href }%
\providecommand \doibase [0]{http://dx.doi.org/}%
\providecommand \selectlanguage [0]{\@gobble}%
\providecommand \bibinfo  [0]{\@secondoftwo}%
\providecommand \bibfield  [0]{\@secondoftwo}%
\providecommand \translation [1]{[#1]}%
\providecommand \BibitemOpen [0]{}%
\providecommand \bibitemStop [0]{}%
\providecommand \bibitemNoStop [0]{.\EOS\space}%
\providecommand \EOS [0]{\spacefactor3000\relax}%
\providecommand \BibitemShut  [1]{\csname bibitem#1\endcsname}%
\let\auto@bib@innerbib\@empty
\bibitem [{\citenamefont {de~Mello~Doneg{\'{a}}}(2011)}]{Doneg2011}%
  \BibitemOpen
  \bibfield  {author} {\bibinfo {author} {\bibfnamefont {Celso}\ \bibnamefont
  {de~Mello~Doneg{\'{a}}}},\ }\bibfield  {title} {\enquote {\bibinfo {title}
  {Synthesis and properties of colloidal heteronanocrystals},}\ }\href@noop {}
  {\bibfield  {journal} {\bibinfo  {journal} {Chem. Soc. Rev.}\ }\textbf
  {\bibinfo {volume} {40}},\ \bibinfo {pages} {1512} (\bibinfo {year}
  {2011})}\BibitemShut {NoStop}%
\bibitem [{\citenamefont {Ithurria}\ and\ \citenamefont
  {Dubertret}(2008)}]{Ithurria2008}%
  \BibitemOpen
  \bibfield  {author} {\bibinfo {author} {\bibfnamefont {S.}\
  \bibnamefont {Ithurria}}\ and\ \bibinfo {author} {\bibfnamefont {B.}\
  \bibnamefont {Dubertret}},\ }\bibfield  {title} {\enquote {\bibinfo {title}
  {Quasi 2d colloidal cdse platelets with thicknesses controlled at the atomic
  level},}\ } {\bibfield  {journal} {\bibinfo
   {journal} {J. Am. Chem. Soc.}\ }\textbf {\bibinfo {volume} {130}},\ \bibinfo
  {pages} {16504} (\bibinfo {year} {2008})}\BibitemShut {NoStop}%
  \bibitem [{\citenamefont {Kovalenko}\ \emph {et~al.}(2015)\citenamefont
  {Kovalenko}, \citenamefont {Manna}, \citenamefont {Cabot}, \citenamefont
  {Hens}, \citenamefont {Talapin}, \citenamefont {Kagan}, \citenamefont
  {Klimov}, \citenamefont {Rogach}, \citenamefont {Reiss}, \citenamefont
  {Milliron}, \citenamefont {Guyot-Sionnnest}, \citenamefont {Konstantatos},
  \citenamefont {Parak}, \citenamefont {Hyeon}, \citenamefont {Korgel},
  \citenamefont {Murray},\ and\ \citenamefont {Heiss}}]{Kovalenko2015}%
  \BibitemOpen
  \bibfield  {author} {\bibinfo {author} {\bibfnamefont {M.~V.}\
  \bibnamefont {Kovalenko}}, \bibinfo {author} {\bibfnamefont {L.}\
  \bibnamefont {Manna}}, \bibinfo {author} {\bibfnamefont {A.}\
  \bibnamefont {Cabot}}, \bibinfo {author} {\bibfnamefont {Z.}\ \bibnamefont
  {Hens}}, \bibinfo {author} {\bibfnamefont {D.~V.}\ \bibnamefont
  {Talapin}}, \bibinfo {author} {\bibfnamefont {C.~R.}\ \bibnamefont
  {Kagan}}, \bibinfo {author} {\bibfnamefont {V.~I.}\ \bibnamefont
  {Klimov}}, \bibinfo {author} {\bibfnamefont {A.~L.}\ \bibnamefont
  {Rogach}}, \bibinfo {author} {\bibfnamefont {P.}\ \bibnamefont {Reiss}},
  \bibinfo {author} {\bibfnamefont {D.~J.}\ \bibnamefont {Milliron}},
  \bibinfo {author} {\bibfnamefont {Ph.}\ \bibnamefont {Guyot-Sionnnest}},
  \bibinfo {author} {\bibfnamefont {G.}\ \bibnamefont {Konstantatos}},
  \bibinfo {author} {\bibfnamefont {W.~J.}\ \bibnamefont {Parak}},
  \bibinfo {author} {\bibfnamefont {T.}\ \bibnamefont {Hyeon}}, \bibinfo
  {author} {\bibfnamefont {B.~A.}\ \bibnamefont {Korgel}}, \bibinfo {author}
  {\bibfnamefont {C.~B.}\ \bibnamefont {Murray}}, \ and\ \bibinfo
  {author} {\bibfnamefont {W.}\ \bibnamefont {Heiss}},\ }\bibfield
  {title} {\enquote {\bibinfo {title} {Prospects of nanoscience with
  nanocrystals},}\ }\href@noop {} {\bibfield  {journal} {\bibinfo  {journal}
  {{ACS} Nano}\ }\textbf {\bibinfo {volume} {9}},\ \bibinfo {pages}
  {1012} (\bibinfo {year} {2015})}\BibitemShut {NoStop}%
\bibitem [{\citenamefont {Graham-Rowe}(2009)}]{GrahamRowe2009}%
  \BibitemOpen
  \bibfield  {author} {\bibinfo {author} {\bibfnamefont {D.}\ \bibnamefont
  {Graham-Rowe}},\ }\bibfield  {title} {\enquote {\bibinfo {title} {From dots
  to devices},}\ }\href@noop {} {\bibfield  {journal} {\bibinfo  {journal}
  {Nature Photonics}\ }\textbf {\bibinfo {volume} {3}},\ \bibinfo {pages}
  {307} (\bibinfo {year} {2009})}\BibitemShut {NoStop}%
\bibitem [{\citenamefont {Lohse}\ and\ \citenamefont
  {Murphy}(2012)}]{Lohse2012}%
  \BibitemOpen
  \bibfield  {author} {\bibinfo {author} {\bibfnamefont {S.~E.}\
  \bibnamefont {Lohse}}\ and\ \bibinfo {author} {\bibfnamefont {C.~J.}\
  \bibnamefont {Murphy}},\ }\bibfield  {title} {\enquote {\bibinfo {title}
  {Applications of colloidal inorganic nanoparticles: From
  medicine~to~energy},}\ }\href@noop {} {\bibfield  {journal} {\bibinfo
  {journal} {J. Am. Chem. Soc.}\ }\textbf {\bibinfo
  {volume} {134}},\ \bibinfo {pages} {15607} (\bibinfo {year}
  {2012})}\BibitemShut {NoStop}%
\bibitem [{\citenamefont {Freeman}\ and\ \citenamefont
  {Willner}(2012)}]{Freeman2012}%
  \BibitemOpen
  \bibfield  {author} {\bibinfo {author} {\bibfnamefont {R.}\ \bibnamefont
  {Freeman}}\ and\ \bibinfo {author} {\bibfnamefont {I.}\ \bibnamefont
  {Willner}},\ }\bibfield  {title} {\enquote {\bibinfo {title} {Optical
  molecular sensing with semiconductor quantum dots ({QDs})},}\ }\href@noop {}
  {\bibfield  {journal} {\bibinfo  {journal} {Chem. Soc. Rev.}\
  }\textbf {\bibinfo {volume} {41}},\ \bibinfo {pages} {4067} (\bibinfo {year}
  {2012})}\BibitemShut {NoStop}%
\bibitem [{\citenamefont {Kamat}(2012)}]{Kamat2012}%
  \BibitemOpen
  \bibfield  {author} {\bibinfo {author} {\bibfnamefont {P.~V.}\
  \bibnamefont {Kamat}},\ }\bibfield  {title} {\enquote {\bibinfo {title}
  {Boosting the efficiency of quantum dot sensitized solar cells through
  modulation of interfacial charge transfer},}\ }\href@noop {} {\bibfield
  {journal} {\bibinfo  {journal} {Acc. Chem. Res.}\ }\textbf
  {\bibinfo {volume} {45}},\ \bibinfo {pages} {1906} (\bibinfo {year}
  {2012})}\BibitemShut {NoStop}%
\bibitem [{\citenamefont {Carey}\ \emph {et~al.}(2015)\citenamefont {Carey},
  \citenamefont {Abdelhady}, \citenamefont {Ning}, \citenamefont {Thon},
  \citenamefont {Bakr},\ and\ \citenamefont {Sargent}}]{Carey2015}%
  \BibitemOpen
  \bibfield  {author} {\bibinfo {author} {\bibfnamefont {G.~H.}\
  \bibnamefont {Carey}}, \bibinfo {author} {\bibfnamefont {A.~L.}\
  \bibnamefont {Abdelhady}}, \bibinfo {author} {\bibfnamefont {Z.}\
  \bibnamefont {Ning}}, \bibinfo {author} {\bibfnamefont {S.~M.}\
  \bibnamefont {Thon}}, \bibinfo {author} {\bibfnamefont {O.~M.}\
  \bibnamefont {Bakr}}, \ and\ \bibinfo {author} {\bibfnamefont {E.~H.}\
  \bibnamefont {Sargent}},\ }\bibfield  {title} {\enquote {\bibinfo {title}
  {Colloidal quantum dot solar cells},}\ }\href@noop {} {\bibfield  {journal}
  {\bibinfo  {journal} {Chem. Rev.}\ }\textbf {\bibinfo {volume} {115}},\
  \bibinfo {pages} {12732} (\bibinfo {year} {2015})}\BibitemShut
  {NoStop}%
\bibitem [{\citenamefont {Lhuillier}\ \emph {et~al.}(2016)\citenamefont
  {Lhuillier}, \citenamefont {Scarafagio}, \citenamefont {Hease}, \citenamefont
  {Nadal}, \citenamefont {Aubin}, \citenamefont {Xu}, \citenamefont {Lequeux},
  \citenamefont {Patriarche}, \citenamefont {Ithurria},\ and\ \citenamefont
  {Dubertret}}]{Lhuillier2016}%
  \BibitemOpen
  \bibfield  {author} {\bibinfo {author} {\bibfnamefont {E.}\
  \bibnamefont {Lhuillier}}, \bibinfo {author} {\bibfnamefont {M.}\
  \bibnamefont {Scarafagio}}, \bibinfo {author} {\bibfnamefont {P.}\
  \bibnamefont {Hease}}, \bibinfo {author} {\bibfnamefont {B.}\ \bibnamefont
  {Nadal}}, \bibinfo {author} {\bibfnamefont {H.}\ \bibnamefont
  {Aubin}}, \bibinfo {author} {\bibfnamefont {X.~Z.}\ \bibnamefont {Xu}},
  \bibinfo {author} {\bibfnamefont {N.}\ \bibnamefont {Lequeux}}, \bibinfo
  {author} {\bibfnamefont {G.}\ \bibnamefont {Patriarche}}, \bibinfo
  {author} {\bibfnamefont {S.}\ \bibnamefont {Ithurria}}, \ and\ \bibinfo
  {author} {\bibfnamefont {B.}\ \bibnamefont {Dubertret}},\ }\bibfield
  {title} {\enquote {\bibinfo {title} {Infrared photodetection based on
  colloidal quantum-dot films with high mobility and optical absorption up to
  {THz}},}\ }\href@noop {} {\bibfield  {journal} {\bibinfo  {journal} {Nano
  Lett.}\ }\textbf {\bibinfo {volume} {16}},\ \bibinfo {pages} {1282.}
  (\bibinfo {year} {2016})}\BibitemShut {NoStop}%
\bibitem [{\citenamefont {Wang}\ \emph {et~al.}(2016)\citenamefont {Wang},
  \citenamefont {Shang}, \citenamefont {Kanjanaboos}, \citenamefont {Zhou},
  \citenamefont {Ning},\ and\ \citenamefont {Sargent}}]{Wang2016}%
  \BibitemOpen
  \bibfield  {author} {\bibinfo {author} {\bibfnamefont {R.}\ \bibnamefont
  {Wang}}, \bibinfo {author} {\bibfnamefont {Y.}\ \bibnamefont {Shang}},
  \bibinfo {author} {\bibfnamefont {P.}\ \bibnamefont {Kanjanaboos}},
  \bibinfo {author} {\bibfnamefont {W.}\ \bibnamefont {Zhou}}, \bibinfo
  {author} {\bibfnamefont {Z.}\ \bibnamefont {Ning}}, \ and\ \bibinfo
  {author} {\bibfnamefont {E.~H.}\ \bibnamefont {Sargent}},\ }\bibfield
  {title} {\enquote {\bibinfo {title} {Colloidal quantum dot ligand engineering
  for high performance solar cells},}\ }\href@noop {} {\bibfield  {journal}
  {\bibinfo  {journal} {Energy Environ. Sci.}\ }\textbf {\bibinfo
  {volume} {9}},\ \bibinfo {pages} {1130} (\bibinfo {year}
  {2016})}\BibitemShut {NoStop}%
\bibitem [{\citenamefont {Loss}\ and\ \citenamefont
  {DiVincenzo}(1997)}]{Loss1997}%
  \BibitemOpen
  \bibfield  {author} {\bibinfo {author} {\bibfnamefont {D.}\bibnamefont
  {Loss}}\ and\ \bibinfo {author} {\bibfnamefont {D.P.}\ \bibnamefont
  {DiVincenzo}},\ }\bibfield  {title} {\enquote {\bibinfo {title} {{Quantum
  computation with quantum dots}},}\ } {\bibfield  {journal} {\bibinfo
  {journal} {Phys. Rev. A}\ }\textbf {\bibinfo {volume} {57}},\ \bibinfo
  {pages} {120} (\bibinfo {year} {1998})}\BibitemShut {NoStop}%
\bibitem [{\citenamefont {Imamoglu}\ \emph {et~al.}(1999)\citenamefont
  {Imamoglu}, \citenamefont {Awschalom}, \citenamefont {Burkard}, \citenamefont
  {DiVincenzo}, \citenamefont {Loss}, \citenamefont {Sherwin},\ and\
  \citenamefont {Small}}]{Imamoglu1999}%
  \BibitemOpen
  \bibfield  {author} {\bibinfo {author} {\bibfnamefont {A.}~\bibnamefont
  {Imamoglu}}, \bibinfo {author} {\bibfnamefont {D.~D.}\ \bibnamefont
  {Awschalom}}, \bibinfo {author} {\bibfnamefont {G.}~\bibnamefont {Burkard}},
  \bibinfo {author} {\bibfnamefont {D.~P.}\ \bibnamefont {DiVincenzo}},
  \bibinfo {author} {\bibfnamefont {D.}~\bibnamefont {Loss}}, \bibinfo {author}
  {\bibfnamefont {M.}~\bibnamefont {Sherwin}}, \ and\ \bibinfo {author}
  {\bibfnamefont {A.}~\bibnamefont {Small}},\ }\bibfield  {title} {\enquote
  {\bibinfo {title} {Quantum information processing using quantum dot spins and
  cavity {QED}},}\ }\href@noop {} {\bibfield  {journal} {\bibinfo  {journal}
  {Phys. Rev. Let.}\ }\textbf {\bibinfo {volume} {83}},\ \bibinfo
  {pages} {4204} (\bibinfo {year} {1999})}\BibitemShut {NoStop}%
\bibitem [{\citenamefont {Nadj-Perge}\ \emph {et~al.}(2010)\citenamefont
  {Nadj-Perge}, \citenamefont {Frolov}, \citenamefont {Bakkers},\ and\
  \citenamefont {Kouwenhoven}}]{NadjPerge2010}%
  \BibitemOpen
  \bibfield  {author} {\bibinfo {author} {\bibfnamefont {S.}~\bibnamefont
  {Nadj-Perge}}, \bibinfo {author} {\bibfnamefont {S.~M.}\ \bibnamefont
  {Frolov}}, \bibinfo {author} {\bibfnamefont {E.~P. A.~M.}\ \bibnamefont
  {Bakkers}}, \ and\ \bibinfo {author} {\bibfnamefont {L.~P.}\ \bibnamefont
  {Kouwenhoven}},\ }\bibfield  {title} {\enquote {\bibinfo {title} {Spin-orbit
  qubit in a semiconductor nanowire},}\ }\href@noop {} {\bibfield  {journal}
  {\bibinfo  {journal} {Nature}\ }\textbf {\bibinfo {volume} {468}},\ \bibinfo
  {pages} {1084} (\bibinfo {year} {2010})}\BibitemShut {NoStop}%
\bibitem [{\citenamefont {Warburton}(2013)}]{Warburton2013}%
  \BibitemOpen
  \bibfield  {author} {\bibinfo {author} {\bibfnamefont {Richard~J.}\
  \bibnamefont {Warburton}},\ }\bibfield  {title} {\enquote {\bibinfo {title}
  {Single spins in self-assembled quantum dots},}\ }\href@noop {} {\bibfield
  {journal} {\bibinfo  {journal} {Nat. Mater.}\ }\textbf {\bibinfo
  {volume} {12}},\ \bibinfo {pages} {483} (\bibinfo {year}
  {2013})}\BibitemShut {NoStop}%
\bibitem [{\citenamefont {Cao}\ \emph {et~al.}(2016)\citenamefont {Cao},
  \citenamefont {Li}, \citenamefont {Yu}, \citenamefont {Wang}, \citenamefont
  {Chen}, \citenamefont {Song}, \citenamefont {Xiao}, \citenamefont {Guo},
  \citenamefont {Jiang}, \citenamefont {Hu},\ and\ \citenamefont
  {Guo}}]{Cao2016}%
  \BibitemOpen
  \bibfield  {author} {\bibinfo {author} {\bibfnamefont {G.}\ \bibnamefont
  {Cao}}, \bibinfo {author} {\bibfnamefont {H.-O.}\ \bibnamefont {Li}},
  \bibinfo {author} {\bibfnamefont {G.-D.}\ \bibnamefont {Yu}}, \bibinfo
  {author} {\bibfnamefont {B.-Ch.}\ \bibnamefont {Wang}}, \bibinfo {author}
  {\bibfnamefont {B.-B.}\ \bibnamefont {Chen}}, \bibinfo {author}
  {\bibfnamefont {X.-X.}\ \bibnamefont {Song}}, \bibinfo {author}
  {\bibfnamefont {M.}\ \bibnamefont {Xiao}}, \bibinfo {author} {\bibfnamefont
  {G.-C.}\ \bibnamefont {Guo}}, \bibinfo {author} {\bibfnamefont
  {H.-W.}\ \bibnamefont {Jiang}}, \bibinfo {author} {\bibfnamefont
  {X.}\ \bibnamefont {Hu}}, \ and\ \bibinfo {author} {\bibfnamefont
  {Guo-Ping}\ \bibnamefont {Guo}},\ }\bibfield  {title} {\enquote {\bibinfo
  {title} {Tunable hybrid qubit in a {GaAs} double quantum dot},}\ }\href@noop
  {} {\bibfield  {journal} {\bibinfo  {journal} {Phys. Rev. Lett.}\
  }\textbf {\bibinfo {volume} {116}},\ \bibinfo {pages} {086801} (\bibinfo {year} {2016})}\BibitemShut
  {NoStop}%
\bibitem [{\citenamefont {Kuno}\ \emph {et~al.}(1998)\citenamefont {Kuno},
  \citenamefont {Nirmal}, \citenamefont {Bawendi}, \citenamefont {Efros},\ and\
  \citenamefont {Rosen}}]{Kuno1998}%
  \BibitemOpen
  \bibfield  {author} {\bibinfo {author} {\bibfnamefont {M.}~\bibnamefont
  {Kuno}}, \bibinfo {author} {\bibfnamefont {M.}~\bibnamefont {Nirmal}},
  \bibinfo {author} {\bibfnamefont {M.~G.}\ \bibnamefont {Bawendi}}, \bibinfo
  {author} {\bibfnamefont {A.}\ \bibnamefont {Efros}}, \ and\ \bibinfo
  {author} {\bibfnamefont {M.}\ \bibnamefont {Rosen}},\ }\bibfield
  {title} {\enquote {\bibinfo {title} {Magnetic circular dichroism study of
  {CdSe} quantum dots},}\ } {\bibfield
  {journal} {\bibinfo  {journal} {J. Chem. Phys.}\ }\textbf
  {\bibinfo {volume} {108}},\ \bibinfo {pages} {4242} (\bibinfo {year}
  {1998})}\BibitemShut {NoStop}%
\bibitem [{\citenamefont {Gupta}\ \emph {et~al.}(2002)\citenamefont {Gupta},
  \citenamefont {Awschalom}, \citenamefont {Efros},\ and\ \citenamefont
  {Rodina}}]{Gupta2002}%
  \BibitemOpen
  \bibfield  {author} {\bibinfo {author} {\bibfnamefont {J.~A.}\ \bibnamefont
  {Gupta}}, \bibinfo {author} {\bibfnamefont {D.~D.}\ \bibnamefont
  {Awschalom}}, \bibinfo {author} {\bibfnamefont {{Al.}~L.}\ \bibnamefont
  {Efros}}, \ and\ \bibinfo {author} {\bibfnamefont {A.~V.}\ \bibnamefont
  {Rodina}},\ }\bibfield  {title} {\enquote {\bibinfo {title} {{Spin dynamics
  in semiconductor nanocrystals}},}\ } {\bibfield  {journal} {\bibinfo  {journal} {Phys.
  Rev. B}\ }\textbf {\bibinfo {volume} {66}},\ \bibinfo {pages} {125307}
  (\bibinfo {year} {2002})}\BibitemShut {NoStop}%
\bibitem [{\citenamefont {Htoon}\ \emph {et~al.}(2009)\citenamefont {Htoon},
  \citenamefont {Crooker}, \citenamefont {Furis}, \citenamefont {Jeong},
  \citenamefont {Efros},\ and\ \citenamefont {Klimov}}]{Htoon2009}%
  \BibitemOpen
  \bibfield  {author} {\bibinfo {author} {\bibfnamefont {H.}~\bibnamefont
  {Htoon}}, \bibinfo {author} {\bibfnamefont {S.A.}~\bibnamefont {Crooker}},
  \bibinfo {author} {\bibfnamefont {M.}~\bibnamefont {Furis}}, \bibinfo
  {author} {\bibfnamefont {S.}~\bibnamefont {Jeong}}, \bibinfo {author}
  {\bibfnamefont {{Al.}~L.}\ \bibnamefont {Efros}}, \ and\ \bibinfo {author}
  {\bibfnamefont {V.I.}~\bibnamefont {Klimov}},\ }\bibfield  {title} {\enquote
  {\bibinfo {title} {{Anomalous circular polarization of photoluminescence
  spectra of individual CdSe nanocrystals in an applied magnetic field}},}\
  } {\bibfield  {journal}
  {\bibinfo  {journal} {Phys. Rev. Lett.}\ }\textbf {\bibinfo {volume} {102}},\
  \bibinfo {pages} {017402} (\bibinfo {year} {2009})}\BibitemShut {NoStop}%
\bibitem [{\citenamefont {Biadala}\ \emph {et~al.}(2010)\citenamefont
  {Biadala}, \citenamefont {Louyer}, \citenamefont {Tamarat},\ and\
  \citenamefont {Lounis}}]{Biadala2010}%
  \BibitemOpen
  \bibfield  {author} {\bibinfo {author} {\bibfnamefont {L.}~\bibnamefont
  {Biadala}}, \bibinfo {author} {\bibfnamefont {Y.}~\bibnamefont {Louyer}},
  \bibinfo {author} {\bibfnamefont {Ph.}\ \bibnamefont {Tamarat}}, \ and\
  \bibinfo {author} {\bibfnamefont {B.}~\bibnamefont {Lounis}},\ }\bibfield
  {title} {\enquote {\bibinfo {title} {{Band-edge exciton fine structure of
  single CdSe/ZnS nanocrystals in external magnetic fields}},}\ } {\bibfield  {journal} {\bibinfo
  {journal} {Phys. Rev. Lett.}\ }\textbf {\bibinfo {volume} {105}},\ \bibinfo
  {pages} {157402} (\bibinfo {year} {2010})}\BibitemShut {NoStop}%
\bibitem [{\citenamefont {Fern\'{e}e}\ \emph {et~al.}(2012)\citenamefont
  {Fern\'{e}e}, \citenamefont {Sinito}, \citenamefont {Louyer}, \citenamefont
  {Potzner}, \citenamefont {Nguyen}, \citenamefont {Mulvaney}, \citenamefont
  {Tamarat},\ and\ \citenamefont {Lounis}}]{Fernee2012nc}%
  \BibitemOpen
  \bibfield  {author} {\bibinfo {author} {\bibfnamefont {M.~J.}\ \bibnamefont
  {Fern\'{e}e}}, \bibinfo {author} {\bibfnamefont {C.}~\bibnamefont {Sinito}},
  \bibinfo {author} {\bibfnamefont {Y.}~\bibnamefont {Louyer}}, \bibinfo
  {author} {\bibfnamefont {C.}~\bibnamefont {Potzner}}, \bibinfo {author}
  {\bibfnamefont {T.~L.}\ \bibnamefont {Nguyen}}, \bibinfo {author}
  {\bibfnamefont {P.}~\bibnamefont {Mulvaney}}, \bibinfo {author}
  {\bibfnamefont {P.}~\bibnamefont {Tamarat}}, \ and\ \bibinfo {author}
  {\bibfnamefont {B.}~\bibnamefont {Lounis}},\ }\bibfield  {title} {\enquote
  {\bibinfo {title} {{Magneto-optical properties of trions in non-blinking
  charged nanocrystals reveal an acoustic phonon bottleneck}},}\ }{\bibfield  {journal} {\bibinfo  {journal}
  {Nature Commun.}\ }\textbf {\bibinfo {volume} {3}},\ \bibinfo {pages} {1287}
  (\bibinfo {year} {2012})}\BibitemShut {NoStop}%
\bibitem [{\citenamefont {Liu}\ \emph {et~al.}(2013)\citenamefont {Liu},
  \citenamefont {Biadala}, \citenamefont {Rodina}, \citenamefont {Yakovlev},
  \citenamefont {Dunker}, \citenamefont {Javaux}, \citenamefont {Hermier},
  \citenamefont {Efros}, \citenamefont {Dubertret},\ and\ \citenamefont
  {Bayer}}]{Liu2013}%
  \BibitemOpen
  \bibfield  {author} {\bibinfo {author} {\bibfnamefont {F.}~\bibnamefont
  {Liu}}, \bibinfo {author} {\bibfnamefont {L.}~\bibnamefont {Biadala}},
  \bibinfo {author} {\bibfnamefont {A.~V.}\ \bibnamefont {Rodina}}, \bibinfo
  {author} {\bibfnamefont {D.~R.}\ \bibnamefont {Yakovlev}}, \bibinfo {author}
  {\bibfnamefont {D.}~\bibnamefont {Dunker}}, \bibinfo {author} {\bibfnamefont
  {C.}~\bibnamefont {Javaux}}, \bibinfo {author} {\bibfnamefont {J.~P.}\
  \bibnamefont {Hermier}}, \bibinfo {author} {\bibfnamefont {{Al.}~L.}\
  \bibnamefont {Efros}}, \bibinfo {author} {\bibfnamefont {B.}~\bibnamefont
  {Dubertret}}, \ and\ \bibinfo {author} {\bibfnamefont {M.}~\bibnamefont
  {Bayer}},\ }\bibfield  {title} {\enquote {\bibinfo {title} {{Spin dynamics of
  negatively charged excitons in CdSe/CdS colloidal nanocrystals}},}\ } {\bibfield  {journal} {\bibinfo
  {journal} {Phys. Rev. B}\ }\textbf {\bibinfo {volume} {88}},\ \bibinfo
  {pages} {035302} (\bibinfo {year} {2013})}\BibitemShut {NoStop}%
\bibitem [{\citenamefont {Ivchenko}\ and\ \citenamefont
  {Kiselev}(1992)}]{Kiselev1992}%
  \BibitemOpen
  \bibfield  {author} {\bibinfo {author} {\bibfnamefont {E.~L.}\ \bibnamefont
  {Ivchenko}}\ and\ \bibinfo {author} {\bibfnamefont {A.~A.}\ \bibnamefont
  {Kiselev}},\ }\href@noop {} {\bibfield  {journal} {\bibinfo  {journal} {Sov.
  Phys. Semicond.}\ }\textbf {\bibinfo {volume} {26}},\ \bibinfo {pages} {827}
  (\bibinfo {year} {1992})}\BibitemShut {NoStop}%
\bibitem [{\citenamefont {Kiselev}\ \emph
  {et~al.}(1998{\natexlab{a}})\citenamefont {Kiselev}, \citenamefont
  {Ivchenko},\ and\ \citenamefont {R\"{o}ssler}}]{Kiselev1998}%
  \BibitemOpen
  \bibfield  {author} {\bibinfo {author} {\bibfnamefont {A.~A.}\ \bibnamefont
  {Kiselev}}, \bibinfo {author} {\bibfnamefont {E.~L.}\ \bibnamefont
  {Ivchenko}}, \ and\ \bibinfo {author} {\bibfnamefont {U.}~\bibnamefont
  {R\"{o}ssler}},\ }\bibfield  {title} {\enquote {\bibinfo {title} {{Electron g
  factor in one- and zero-dimensional semiconductor nanostructures}},}\ } {\bibfield  {journal} {\bibinfo
  {journal} {Phys. Rev. B}\ }\textbf {\bibinfo {volume} {58}},\ \bibinfo
  {pages} {16353} (\bibinfo {year} {1998}{\natexlab{a}})}\BibitemShut {NoStop}%
\bibitem [{\citenamefont {Rodina}\ \emph {et~al.}(2003)\citenamefont {Rodina},
  \citenamefont {Efros},\ and\ \citenamefont {Alekseev}}]{Rodina2003}%
  \BibitemOpen
  \bibfield  {author} {\bibinfo {author} {\bibfnamefont {A.~V.}\ \bibnamefont
  {Rodina}}, \bibinfo {author} {\bibfnamefont {{Al.}~L.}\ \bibnamefont
  {Efros}}, \ and\ \bibinfo {author} {\bibfnamefont {A.~Yu.}\ \bibnamefont
  {Alekseev}},\ }\bibfield  {title} {\enquote {\bibinfo {title} {{Effect of the
  surface on the electron quantum size levels and electron g factor in
  spherical semiconductor nanocrystals}},}\ } {\bibfield  {journal} {\bibinfo  {journal} {Phys.
  Rev. B}\ }\textbf {\bibinfo {volume} {67}},\ \bibinfo {pages} {155312}
  (\bibinfo {year} {2003})}\BibitemShut {NoStop}%
\bibitem [{\citenamefont {Yugova}\ \emph {et~al.}(2007)\citenamefont {Yugova},
  \citenamefont {Greilich}, \citenamefont {Yakovlev}, \citenamefont {Kiselev},
  \citenamefont {Bayer}, \citenamefont {Petrov}, \citenamefont {Dolgikh},
  \citenamefont {Reuter},\ and\ \citenamefont {Wieck}}]{Yugova2007}%
  \BibitemOpen
  \bibfield  {author} {\bibinfo {author} {\bibfnamefont {I.~A.}\ \bibnamefont
  {Yugova}}, \bibinfo {author} {\bibfnamefont {A.}~\bibnamefont {Greilich}},
  \bibinfo {author} {\bibfnamefont {D.~R.}\ \bibnamefont {Yakovlev}}, \bibinfo
  {author} {\bibfnamefont {A.~A.}\ \bibnamefont {Kiselev}}, \bibinfo {author}
  {\bibfnamefont {M.}~\bibnamefont {Bayer}}, \bibinfo {author} {\bibfnamefont
  {V.~V.}\ \bibnamefont {Petrov}}, \bibinfo {author} {\bibfnamefont {Yu.~K.}\
  \bibnamefont {Dolgikh}}, \bibinfo {author} {\bibfnamefont {D.}~\bibnamefont
  {Reuter}}, \ and\ \bibinfo {author} {\bibfnamefont {A.~D.}\ \bibnamefont
  {Wieck}},\ }\bibfield  {title} {\enquote {\bibinfo {title} {Universal
  behavior of the electron $g$-factor in GaAsAl$_x$Ga$_{1-x}$As
  quantum wells},}\ }{\bibfield
  {journal} {\bibinfo  {journal} {Phys. Rev. B}\ }\textbf {\bibinfo {volume}
  {75}},\ \bibinfo {pages} {245302} (\bibinfo {year} {2007})}\BibitemShut
  {NoStop}%
\bibitem [{\citenamefont {Schrier}\ and\ \citenamefont
  {Whaley}(2003)}]{Schrier2003}%
  \BibitemOpen
  \bibfield  {author} {\bibinfo {author} {\bibfnamefont {Joshua}\ \bibnamefont
  {Schrier}}\ and\ \bibinfo {author} {\bibfnamefont {K.~Birgitta}\ \bibnamefont
  {Whaley}},\ }\bibfield  {title} {\enquote {\bibinfo {title}
  {Tight-bindingg-factor calculations of {CdSe} nanostructures},}\ } {\bibfield  {journal} {\bibinfo
  {journal} {Phys. Rev. B}\ }\textbf {\bibinfo {volume} {67}} (\bibinfo
  {year} {2003})}\BibitemShut {NoStop}%
\bibitem [{\citenamefont {Tadjine}\ \emph {et~al.}(2017)\citenamefont
  {Tadjine}, \citenamefont {Niquet},\ and\ \citenamefont
  {Delerue}}]{Tadjine2017}%
  \BibitemOpen
  \bibfield  {author} {\bibinfo {author} {\bibfnamefont {A.}\ \bibnamefont
  {Tadjine}}, \bibinfo {author} {\bibfnamefont {Y.-M.}\ \bibnamefont
  {Niquet}}, \ and\ \bibinfo {author} {\bibfnamefont {C.}\ \bibnamefont
  {Delerue}},\ }\bibfield  {title} {\enquote {\bibinfo {title} {Universal
  behavior of electron $g$-factors in semiconductor nanostructures},}\ }\ {\bibfield  {journal} {\bibinfo
  {journal} {Phys. Rev. B}\ }\textbf {\bibinfo {volume} {95}},\ \bibinfo
  {pages} {235437} (\bibinfo {year} {2017})}\BibitemShut {NoStop}%
\bibitem [{\citenamefont {Chen}\ and\ \citenamefont {Whaley}(2004)}]{Chen2004}%
  \BibitemOpen
  \bibfield  {author} {\bibinfo {author} {\bibfnamefont {P.}\ \bibnamefont
  {Chen}}\ and\ \bibinfo {author} {\bibfnamefont {K.~B.}\ \bibnamefont
  {Whaley}},\ }\bibfield  {title} {\enquote {\bibinfo {title} {Magneto-optical
  response of {CdSe} nanostructures},}\ } {\bibfield  {journal} {\bibinfo  {journal}
  {Physical Review B}\ }\textbf {\bibinfo {volume} {70}} (\bibinfo {year}
  {2004})}\BibitemShut {NoStop}%
\bibitem [{\citenamefont {Csontos}\ \emph {et~al.}(2009)\citenamefont
  {Csontos}, \citenamefont {Brusheim}, \citenamefont {Z\"{u}licke},\ and\
  \citenamefont {Xu}}]{Csontos2009}%
  \BibitemOpen
  \bibfield  {author} {\bibinfo {author} {\bibfnamefont {D.}~\bibnamefont
  {Csontos}}, \bibinfo {author} {\bibfnamefont {P.}~\bibnamefont {Brusheim}},
  \bibinfo {author} {\bibfnamefont {U.}~\bibnamefont {Z\"{u}licke}}, \ and\
  \bibinfo {author} {\bibfnamefont {H.~Q.}\ \bibnamefont {Xu}},\ }\bibfield
  {title} {\enquote {\bibinfo {title} {Spin-3/2 physics of semiconductor hole
  nanowires: Valence-band mixing and tunable interplay between bulk-material
  and orbital bound-state spin splittings},}\ } {\bibfield  {journal} {\bibinfo  {journal}
  {Physical Review B}\ }\textbf {\bibinfo {volume} {79}} (\bibinfo {year}
  {2009})}\BibitemShut {NoStop}%
\bibitem [{\citenamefont {Roth}\ \emph {et~al.}(1959)\citenamefont {Roth},
  \citenamefont {Lax},\ and\ \citenamefont {Zwerdling}}]{Roth1959}%
  \BibitemOpen
  \bibfield  {author} {\bibinfo {author} {\bibfnamefont {L.~M.}\ \bibnamefont
  {Roth}}, \bibinfo {author} {\bibfnamefont {B.}~\bibnamefont {Lax}}, \ and\
  \bibinfo {author} {\bibfnamefont {S.}~\bibnamefont {Zwerdling}},\ }\bibfield
  {title} {\enquote {\bibinfo {title} {{Theory of optical magneto-absorption
  effects in semiconductors}},}\ }
  {\bibfield  {journal} {\bibinfo  {journal} {Phys. Rev.}\ }\textbf {\bibinfo
  {volume} {114}},\ \bibinfo {pages} {90} (\bibinfo {year}
  {1959})}\BibitemShut {NoStop}%
\bibitem [{\citenamefont {Weisbuch}\ and\ \citenamefont
  {Hermann}(1977)}]{Weisbuch1977}%
  \BibitemOpen
  \bibfield  {author} {\bibinfo {author} {\bibfnamefont {C.}\ \bibnamefont
  {Weisbuch}}\ and\ \bibinfo {author} {\bibfnamefont {C.}\ \bibnamefont
  {Hermann}},\ }\bibfield  {title} {\enquote {\bibinfo {title} {Optical
  detection of conduction-electron spin resonance in GaAs, Ga$_{1-x}$In$_x$As,
  and Ga$_{1-x}$Al$_x$As},}\ }{\bibfield
  {journal} {\bibinfo  {journal} {Phys. Rev. B}\ }\textbf {\bibinfo
  {volume} {15}},\ \bibinfo {pages} {816} (\bibinfo {year}
  {1977})}\BibitemShut {NoStop}%
\bibitem [{\citenamefont {Karimov}\ \emph {et~al.}(2000)\citenamefont
  {Karimov}, \citenamefont {Wolverson}, \citenamefont {Davies}, \citenamefont
  {Stepanov}, \citenamefont {Ruf}, \citenamefont {Ivanov}, \citenamefont
  {Sorokin}, \citenamefont {O'Donnell},\ and\ \citenamefont
  {Prior}}]{Karimov2000}%
  \BibitemOpen
  \bibfield  {author} {\bibinfo {author} {\bibfnamefont {O.~Z.}\ \bibnamefont
  {Karimov}}, \bibinfo {author} {\bibfnamefont {D.}~\bibnamefont {Wolverson}},
  \bibinfo {author} {\bibfnamefont {J.~J.}\ \bibnamefont {Davies}}, \bibinfo
  {author} {\bibfnamefont {S.~I.}\ \bibnamefont {Stepanov}}, \bibinfo {author}
  {\bibfnamefont {T.}~\bibnamefont {Ruf}}, \bibinfo {author} {\bibfnamefont
  {S.~V.}\ \bibnamefont {Ivanov}}, \bibinfo {author} {\bibfnamefont {S.~V.}\
  \bibnamefont {Sorokin}}, \bibinfo {author} {\bibfnamefont {C.~B.}\
  \bibnamefont {O'Donnell}}, \ and\ \bibinfo {author} {\bibfnamefont {K.~A.}\
  \bibnamefont {Prior}},\ }\bibfield  {title} {\enquote {\bibinfo {title}
  {Electrong-factor for cubic Zn$_{1-x}$Cd$_x$Se determined by spin-flip Raman
  scattering},}\ } {\bibfield
  {journal} {\bibinfo  {journal} {Phys. Rev. B}\ }\textbf {\bibinfo
  {volume} {62}},\ \bibinfo {pages} {16582} (\bibinfo {year}
  {2000})}\BibitemShut {NoStop}%
\bibitem [{\citenamefont {Piper}(1967)}]{Piper1967}%
  \BibitemOpen
  \bibfield  {author} {\bibinfo {author} {\bibfnamefont {W~W}\ \bibnamefont
  {Piper}},\ }\bibfield  {title} {\enquote {\bibinfo {title} {II-VI
  semiconducting compounds},}\ }in\ \href@noop {} {\emph {\bibinfo {booktitle}
  {International Conference on II-VI Semiconducting Compounds (1967: Brown
  University)}}},\ \bibinfo {editor} {edited by\ \bibinfo {editor}
  {\bibfnamefont {D.~G.}\ \bibnamefont {Thomas}}}\ (\bibinfo
  {organization} {WA Benjamin},\ \bibinfo {address} {New York},\ \bibinfo
  {year} {1967})\BibitemShut {NoStop}%
\bibitem [{\citenamefont {Oestreich}\ \emph {et~al.}(1996)\citenamefont
  {Oestreich}, \citenamefont {Hallstein}, \citenamefont {Heberle},
  \citenamefont {Eberl}, \citenamefont {Bauser},\ and\ \citenamefont
  {R\"{u}hle}}]{Oestreich1996}%
  \BibitemOpen
  \bibfield  {author} {\bibinfo {author} {\bibfnamefont {M.}~\bibnamefont
  {Oestreich}}, \bibinfo {author} {\bibfnamefont {S.}~\bibnamefont
  {Hallstein}}, \bibinfo {author} {\bibfnamefont {A.~P.}\ \bibnamefont
  {Heberle}}, \bibinfo {author} {\bibfnamefont {K.}~\bibnamefont {Eberl}},
  \bibinfo {author} {\bibfnamefont {E.}~\bibnamefont {Bauser}}, \ and\ \bibinfo
  {author} {\bibfnamefont {W.~W.}\ \bibnamefont {R\"{u}hle}},\ }\bibfield
  {title} {\enquote {\bibinfo {title} {Temperature and density dependence of
  the electron land{\'{e}}gfactor in semiconductors},}\ } {\bibfield  {journal} {\bibinfo  {journal}
  {Phys. Rev. B}\ }\textbf {\bibinfo {volume} {53}},\ \bibinfo {pages}
  {7911} (\bibinfo {year} {1996})}\BibitemShut {NoStop}%
\bibitem [{\citenamefont {Kalevich}\ and\ \citenamefont
  {Korenev}(1992)}]{Kalevich1992}%
  \BibitemOpen
  \bibfield  {author} {\bibinfo {author} {\bibfnamefont {V.K.}\ \bibnamefont
  {Kalevich}}\ and\ \bibinfo {author} {\bibfnamefont {V.L.}\ \bibnamefont
  {Korenev}},\ }\bibfield  {title} {\enquote {\bibinfo {title} {Electron
  g-factor anisotropy in asymmetric GaAs/AlGaAs quantum well},}\ }\href@noop {}
  {\bibfield  {journal} {\bibinfo  {journal} {JETP Lett.}\ }\textbf {\bibinfo
  {volume} {56}},\ \bibinfo {pages} {253} (\bibinfo {year} {1992})}\BibitemShut
  {NoStop}%
\bibitem [{\citenamefont {Sirenko}\ \emph {et~al.}(1997)\citenamefont
  {Sirenko}, \citenamefont {Ruf}, \citenamefont {Cardona}, \citenamefont
  {Yakovlev}, \citenamefont {Ossau}, \citenamefont {Waag},\ and\ \citenamefont
  {Landwehr}}]{Sirenko1997}%
  \BibitemOpen
  \bibfield  {author} {\bibinfo {author} {\bibfnamefont {A.~A.}\ \bibnamefont
  {Sirenko}}, \bibinfo {author} {\bibfnamefont {T.}~\bibnamefont {Ruf}},
  \bibinfo {author} {\bibfnamefont {M.}~\bibnamefont {Cardona}}, \bibinfo
  {author} {\bibfnamefont {D.~R.}\ \bibnamefont {Yakovlev}}, \bibinfo {author}
  {\bibfnamefont {W.}~\bibnamefont {Ossau}}, \bibinfo {author} {\bibfnamefont
  {A.}~\bibnamefont {Waag}}, \ and\ \bibinfo {author} {\bibfnamefont
  {G.}~\bibnamefont {Landwehr}},\ }\bibfield  {title} {\enquote {\bibinfo
  {title} {Electron and hole $g$-factors measured by spin-flip Raman scattering
  in CdTe/Cd$_{1-x}$Mg$_{x}$Te single quantum wells},}\ } {\bibfield  {journal} {\bibinfo  {journal} {Phys.
  Rev. B}\ }\textbf {\bibinfo {volume} {56}},\ \bibinfo {pages} {2114}
  (\bibinfo {year} {1997})}\BibitemShut {NoStop}%
\bibitem [{\citenamefont {Hermann}\ and\ \citenamefont
  {Weisbuch}(1977)}]{Hermann1977}%
  \BibitemOpen
  \bibfield  {author} {\bibinfo {author} {\bibfnamefont {C.}~\bibnamefont
  {Hermann}}\ and\ \bibinfo {author} {\bibfnamefont {C.}~\bibnamefont
  {Weisbuch}},\ }\bibfield  {title} {\enquote {\bibinfo {title} {$\bm k\cdot \bm p$
  perturbation theory in III-V compounds and alloys: a reexamination},}\ } {\bibfield  {journal} {\bibinfo  {journal}
  {Phys. Rev. B}\ }\textbf {\bibinfo {volume} {15}},\ \bibinfo {pages} {823}
  (\bibinfo {year} {1977})}\BibitemShut {NoStop}%
\bibitem [{\citenamefont {Ivchenko}\ \emph {et~al.}(1996)\citenamefont
  {Ivchenko}, \citenamefont {Kiselev},\ and\ \citenamefont
  {Willander}}]{ivch_kis_will96}%
  \BibitemOpen
  \bibfield  {author} {\bibinfo {author} {\bibfnamefont {E.L.}\ \bibnamefont
  {Ivchenko}}, \bibinfo {author} {\bibfnamefont {A.A.}\ \bibnamefont
  {Kiselev}}, \ and\ \bibinfo {author} {\bibfnamefont {M.}~\bibnamefont
  {Willander}},\ }\bibfield  {title} {\enquote {\bibinfo {title} {Electronic g
  factor in biased quantum wells},}\ }\href@noop {} {\bibfield  {journal}
  {\bibinfo  {journal} {Solid State Comm.}\ }\textbf {\bibinfo
  {volume} {102}},\ \bibinfo {pages} {375} (\bibinfo {year}
  {1996})}\BibitemShut {NoStop}%
\bibitem [{\citenamefont {Kiselev}\ \emph
  {et~al.}(1998{\natexlab{b}})\citenamefont {Kiselev}, \citenamefont
  {Ivchenko},\ and\ \citenamefont {R\"ossler}}]{Kiselev98}%
  \BibitemOpen
  \bibfield  {author} {\bibinfo {author} {\bibfnamefont {A.~A.}\ \bibnamefont
  {Kiselev}}, \bibinfo {author} {\bibfnamefont {E.~L.}\ \bibnamefont
  {Ivchenko}}, \ and\ \bibinfo {author} {\bibfnamefont {U.}~\bibnamefont
  {R\"ossler}},\ }\bibfield  {title} {\enquote {\bibinfo {title} {Electron g
  factor in one- and zero-dimensional semiconductor nanostructures},}\ } {\bibfield  {journal} {\bibinfo
  {journal} {Phys. Rev. B}\ }\textbf {\bibinfo {volume} {58}},\ \bibinfo
  {pages} {16353} (\bibinfo {year} {1998}{\natexlab{b}})}\BibitemShut
  {NoStop}%
\bibitem [{\citenamefont {Merkulov}\ and\ \citenamefont
  {Rodina}(2010)}]{Merkulov2010book}%
  \BibitemOpen
  \bibfield  {author} {\bibinfo {author} {\bibfnamefont {I.~A.}\ \bibnamefont
  {Merkulov}}\ and\ \bibinfo {author} {\bibfnamefont {A.~V.}\ \bibnamefont
  {Rodina}},\ }\bibfield  {title} {\enquote {\bibinfo {title} {{Exchange
  interaction between carriers and magnetic ions in quantum size
  heterostructures}},}\ }in\ \href@noop {} {\emph {\bibinfo {booktitle}
  {Introduction to the physics of diluted magnetic semiconductors}}},\ \bibinfo
  {editor} {edited by\ \bibinfo {editor} {\bibfnamefont {J.}~\bibnamefont
  {Kossut}}\ and\ \bibinfo {editor} {\bibfnamefont {J.~A.}\ \bibnamefont
  {Gaj}}}\ (\bibinfo  {publisher} {Springer},\ \bibinfo {year} {2010})\
  Chap.~\bibinfo {chapter} {3}, pp.\ \bibinfo {pages} {65--101}\BibitemShut
  {NoStop}%
\bibitem [{\citenamefont {Rodina}\ and\ \citenamefont
  {Alekseev}(2008)}]{Rodina2008}%
  \BibitemOpen
  \bibfield  {author} {\bibinfo {author} {\bibfnamefont {A.~V.}\ \bibnamefont
  {Rodina}}\ and\ \bibinfo {author} {\bibfnamefont {A.~Yu.}\ \bibnamefont
  {Alekseev}},\ }\bibfield  {title} {\enquote {\bibinfo {title} {{Theory of
  intrinsic electric polarization and spin Hall current in spin-orbit-coupled
  semiconductor heterostructures}},}\ } {\bibfield  {journal} {\bibinfo  {journal} {Phys.
  Rev. B}\ }\textbf {\bibinfo {volume} {78}},\ \bibinfo {pages} {115304}
  (\bibinfo {year} {2008})}\BibitemShut {NoStop}%
\bibitem [{\citenamefont {Ekimov}\ \emph {et~al.}(1993)\citenamefont {Ekimov},
  \citenamefont {Hache}, \citenamefont {Schanne-Klein}, \citenamefont {Ricard},
  \citenamefont {Flytzanis}, \citenamefont {Kudryavtsev}, \citenamefont
  {Yazeva}, \citenamefont {Rodina},\ and\ \citenamefont {Efros}}]{Ekimov1993}%
  \BibitemOpen
  \bibfield  {author} {\bibinfo {author} {\bibfnamefont {A.~I.}\ \bibnamefont
  {Ekimov}}, \bibinfo {author} {\bibfnamefont {F.}~\bibnamefont {Hache}},
  \bibinfo {author} {\bibfnamefont {M.~C.}\ \bibnamefont {Schanne-Klein}},
  \bibinfo {author} {\bibfnamefont {D.}~\bibnamefont {Ricard}}, \bibinfo
  {author} {\bibfnamefont {Ch.}\ \bibnamefont {Flytzanis}}, \bibinfo {author}
  {\bibfnamefont {I.~A.}\ \bibnamefont {Kudryavtsev}}, \bibinfo {author}
  {\bibfnamefont {T.~V.}\ \bibnamefont {Yazeva}}, \bibinfo {author}
  {\bibfnamefont {A.~V.}\ \bibnamefont {Rodina}}, \ and\ \bibinfo {author}
  {\bibfnamefont {{Al.}~L.}\ \bibnamefont {Efros}},\ }\bibfield  {title}
  {\enquote {\bibinfo {title} {{Absorption and intensity-dependent
  photoluminescence measurements on CdSe quantum dots: assignment of the first
  electronic transitions}},}\ }
  {\bibfield  {journal} {\bibinfo  {journal} {JOSA B}\ }\textbf {\bibinfo
  {volume} {10}},\ \bibinfo {pages} {100} (\bibinfo {year}
  {1993})}\BibitemShut {NoStop}%
\bibitem [{\citenamefont {Hu}\ \emph {et~al.}(2019)\citenamefont {Hu},
  \citenamefont {Yakovlev}, \citenamefont {Liang}, \citenamefont {Qiang},
  \citenamefont {Chen}, \citenamefont {Jia}, \citenamefont {Sun}, \citenamefont
  {Bayer},\ and\ \citenamefont {Feng}}]{Hu2019}%
  \BibitemOpen
  \bibfield  {author} {\bibinfo {author} {\bibfnamefont {R.}\
  \bibnamefont {Hu}}, \bibinfo {author} {\bibfnamefont {D.~R.}\
  \bibnamefont {Yakovlev}}, \bibinfo {author} {\bibfnamefont {P.}\
  \bibnamefont {Liang}}, \bibinfo {author} {\bibfnamefont {G.}\ \bibnamefont
  {Qiang}}, \bibinfo {author} {\bibfnamefont {C.}\ \bibnamefont {Chen}},
  \bibinfo {author} {\bibfnamefont {T.}\ \bibnamefont {Jia}}, \bibinfo
  {author} {\bibfnamefont {Zh.}\ \bibnamefont {Sun}}, \bibinfo {author}
  {\bibfnamefont {M.}\ \bibnamefont {Bayer}}, \ and\ \bibinfo {author}
  {\bibfnamefont {D.}\ \bibnamefont {Feng}},\ }\bibfield  {title}
  {\enquote {\bibinfo {title} {Origin of two larmor frequencies in the coherent
  spin dynamics of colloidal {CdSe} quantum dots revealed by controlled
  charging},}\ }\ {\bibfield
  {journal} {\bibinfo  {journal} {J. Phys. Chem. Lett.}\
  }\textbf {\bibinfo {volume} {10}},\ \bibinfo {pages} {3681} (\bibinfo
  {year} {2019})}\BibitemShut {NoStop}%
\bibitem [{\citenamefont {Zhang}\ \emph {et~al.}(2014)\citenamefont {Zhang},
  \citenamefont {Jin}, \citenamefont {Ma}, \citenamefont {Xu}, \citenamefont
  {Lin}, \citenamefont {Ma},\ and\ \citenamefont {Sun}}]{Zhang2014}%
  \BibitemOpen
  \bibfield  {author} {\bibinfo {author} {\bibfnamefont {Zh.}\
  \bibnamefont {Zhang}}, \bibinfo {author} {\bibfnamefont {Z.}\
  \bibnamefont {Jin}}, \bibinfo {author} {\bibfnamefont {H.}\ \bibnamefont
  {Ma}}, \bibinfo {author} {\bibfnamefont {Y.}\ \bibnamefont {Xu}}, \bibinfo
  {author} {\bibfnamefont {X.}\ \bibnamefont {Lin}}, \bibinfo {author}
  {\bibfnamefont {G.}\ \bibnamefont {Ma}}, \ and\ \bibinfo {author}
  {\bibfnamefont {X.}\ \bibnamefont {Sun}},\ }\bibfield  {title} {\enquote
  {\bibinfo {title} {Room-temperature spin coherence in zinc blende {CdSe}
  quantum dots studied by time-resolved faraday ellipticity},}\ } {\bibfield  {journal} {\bibinfo
  {journal} {Physica E}\ }\textbf
  {\bibinfo {volume} {56}},\ \bibinfo {pages} {85--89} (\bibinfo {year}
  {2014})}\BibitemShut {NoStop}%
\bibitem [{\citenamefont {Kudlacik}\ \emph {et~al.}(2019)\citenamefont
  {Kudlacik}, \citenamefont {Sapega}, \citenamefont {Yakovlev}, \citenamefont
  {Kalitukha}, \citenamefont {Shornikova}, \citenamefont {Rodina},
  \citenamefont {Ivchenko}, \citenamefont {Dimitriev}, \citenamefont
  {Nasilowski}, \citenamefont {Dubertret},\ and\ \citenamefont
  {Bayer}}]{Kudlacik2019}%
  \BibitemOpen
  \bibfield  {author} {\bibinfo {author} {\bibfnamefont {D.}\ \bibnamefont
  {Kudlacik}}, \bibinfo {author} {\bibfnamefont {V.~F.}\ \bibnamefont
  {Sapega}}, \bibinfo {author} {\bibfnamefont {D.~R.}\ \bibnamefont
  {Yakovlev}}, \bibinfo {author} {\bibfnamefont {I.~V.}\ \bibnamefont
  {Kalitukha}}, \bibinfo {author} {\bibfnamefont {E.~V.}\ \bibnamefont
  {Shornikova}}, \bibinfo {author} {\bibfnamefont {A.~V.}\ \bibnamefont
  {Rodina}}, \bibinfo {author} {\bibfnamefont {E.~L.}\ \bibnamefont
  {Ivchenko}}, \bibinfo {author} {\bibfnamefont {G.~S.}\ \bibnamefont
  {Dimitriev}}, \bibinfo {author} {\bibfnamefont {M.}\ \bibnamefont
  {Nasilowski}}, \bibinfo {author} {\bibfnamefont {B.}\ \bibnamefont
  {Dubertret}}, \ and\ \bibinfo {author} {\bibfnamefont {M.}\ \bibnamefont
  {Bayer}},\ }\bibfield  {title} {\enquote {\bibinfo {title} {Single and double
  electron spin-flip raman scattering in {CdSe} colloidal nanoplatelets},}\
  } {\bibfield  {journal}
  {\bibinfo  {journal} {Nano Lett.}\ }\textbf {\bibinfo {volume} {20}},\
  \bibinfo {pages} {517} (\bibinfo {year} {2019})}\BibitemShut {NoStop}%
\bibitem [{\citenamefont {Rodina}\ and\ \citenamefont
  {Ivchenko}(2020)}]{Rodina2020}%
  \BibitemOpen
  \bibfield  {author} {\bibinfo {author} {\bibfnamefont {A.~V.}\ \bibnamefont
  {Rodina}}\ and\ \bibinfo {author} {\bibfnamefont {E.~L.}\ \bibnamefont
  {Ivchenko}},\ }\bibfield  {title} {\enquote {\bibinfo {title} {Theory of
  single and double electron spin-flip raman scattering in semiconductor
  nanoplatelets},}\ }\href@noop {} {\bibfield  {journal} {\bibinfo  {journal}
  {arXiv:2010.10385}\ } (\bibinfo {year}
  {2020})}\BibitemShut {NoStop}%
\bibitem [{\citenamefont {van Bree}\ \emph {et~al.}(2014)\citenamefont {van
  Bree}, \citenamefont {Silov}, \citenamefont {Koenraad},\ and\ \citenamefont
  {Flatt{\'{e}}}}]{vanBree2014}%
  \BibitemOpen
  \bibfield  {author} {\bibinfo {author} {\bibfnamefont {J.}~\bibnamefont {van
  Bree}}, \bibinfo {author} {\bibfnamefont {A.{\hspace{0.167em}}Yu.}\
  \bibnamefont {Silov}}, \bibinfo {author} {\bibfnamefont
  {P.{\hspace{0.167em}}M.}\ \bibnamefont {Koenraad}}, \ and\ \bibinfo {author}
  {\bibfnamefont {M.{\hspace{0.167em}}E.}\ \bibnamefont {Flatt{\'{e}}}},\
  }\bibfield  {title} {\enquote {\bibinfo {title} {Spin-orbit-induced
  circulating currents in a semiconductor nanostructure},}\ } {\bibfield  {journal} {\bibinfo  {journal}
  {Phys. Rev. Lett.}\ }\textbf {\bibinfo {volume} {112}} (\bibinfo
  {year} {2014})}\BibitemShut {NoStop}%
\bibitem [{\citenamefont {Luttinger}(1956)}]{Luttinger1956}%
  \BibitemOpen
  \bibfield  {author} {\bibinfo {author} {\bibfnamefont {J.~M.}\ \bibnamefont
  {Luttinger}},\ }\bibfield  {title} {\enquote {\bibinfo {title} {{Quantum
  theory of cyclotron resonance in semiconductors: general theory}},}\ } {\bibfield  {journal} {\bibinfo
  {journal} {Phys. Rev.}\ }\textbf {\bibinfo {volume} {102}},\ \bibinfo {pages}
  {1030} (\bibinfo {year} {1956})}\BibitemShut {NoStop}%
\bibitem [{\citenamefont {Gel'mont}\ and\ \citenamefont
  {D'yakonov}(1971)}]{Gelmont1971}%
  \BibitemOpen
  \bibfield  {author} {\bibinfo {author} {\bibfnamefont {B.~L.}\ \bibnamefont
  {Gel'mont}}\ and\ \bibinfo {author} {\bibfnamefont {M.~I.}\ \bibnamefont
  {D'yakonov}},\ }\bibfield  {title} {\enquote {\bibinfo {title} {{Acceptor
  levels in diamond-type semiconductors}},}\ }\href@noop {} {\bibfield
  {journal} {\bibinfo  {journal} {Soviet Physics. Semiconductors}\ }\textbf
  {\bibinfo {volume} {5}},\ \bibinfo {pages} {2191} (\bibinfo {year}
  {1971})}\BibitemShut {NoStop}%
\bibitem [{\citenamefont {Semina}\ and\ \citenamefont
  {Suris}(2015)}]{Semina2015}%
  \BibitemOpen
  \bibfield  {author} {\bibinfo {author} {\bibfnamefont {M.~A.}\ \bibnamefont
  {Semina}}\ and\ \bibinfo {author} {\bibfnamefont {R.~A.}\ \bibnamefont
  {Suris}},\ }\bibfield  {title} {\enquote {\bibinfo {title} {{Holes localized
  in nanostructures in an external magnetic field: g-factor and mixing of
  states}},}\ } {\bibfield  {journal}
  {\bibinfo  {journal} {Semiconductors}\ }\textbf {\bibinfo {volume} {49}},\
  \bibinfo {pages} {797} (\bibinfo {year} {2015})}\BibitemShut {NoStop}%
\bibitem [{\citenamefont {Marie}\ \emph {et~al.}(1999)\citenamefont {Marie},
  \citenamefont {Amand}, \citenamefont {Jeune}, \citenamefont {Paillard},
  \citenamefont {Renucci}, \citenamefont {Golub}, \citenamefont {Dymnikov},\
  and\ \citenamefont {Ivchenko}}]{Marie1999}%
  \BibitemOpen
  \bibfield  {author} {\bibinfo {author} {\bibfnamefont {X.}~\bibnamefont
  {Marie}}, \bibinfo {author} {\bibfnamefont {T.}~\bibnamefont {Amand}},
  \bibinfo {author} {\bibfnamefont {P.~Le}\ \bibnamefont {Jeune}}, \bibinfo
  {author} {\bibfnamefont {M.}~\bibnamefont {Paillard}}, \bibinfo {author}
  {\bibfnamefont {P.}~\bibnamefont {Renucci}}, \bibinfo {author} {\bibfnamefont
  {L.~E.}\ \bibnamefont {Golub}}, \bibinfo {author} {\bibfnamefont {V.~D.}\
  \bibnamefont {Dymnikov}}, \ and\ \bibinfo {author} {\bibfnamefont {E.~L.}\
  \bibnamefont {Ivchenko}},\ }\bibfield  {title} {\enquote {\bibinfo {title}
  {Hole spin quantum beats in quantum-well structures},}\ } {\bibfield  {journal} {\bibinfo  {journal}
  {Phys. Rev. B}\ }\textbf {\bibinfo {volume} {60}},\ \bibinfo {pages}
  {5811} (\bibinfo {year} {1999})}\BibitemShut {NoStop}%
\bibitem [{\citenamefont {Gel'mont}\ and\ \citenamefont
  {D'yakonov}(1973)}]{Gelmont1973}%
  \BibitemOpen
  \bibfield  {author} {\bibinfo {author} {\bibfnamefont {B.~L.}\ \bibnamefont
  {Gel'mont}}\ and\ \bibinfo {author} {\bibfnamefont {M.~I.}\ \bibnamefont
  {D'yakonov}},\ }\bibfield  {title} {\enquote {\bibinfo {title} {{g-factor of
  acceptors in semiconductors with the diamond structure}},}\ }\href@noop {}
  {\bibfield  {journal} {\bibinfo  {journal} {Soviet Physics. Semiconductors}\
  }\textbf {\bibinfo {volume} {7}},\ \bibinfo {pages} {2013} (\bibinfo
  {year} {1973})}\BibitemShut {NoStop}%
\bibitem [{\citenamefont {Efros}\ \emph {et~al.}(1996)\citenamefont {Efros},
  \citenamefont {Rosen}, \citenamefont {Kuno}, \citenamefont {Nirmal},
  \citenamefont {Norris},\ and\ \citenamefont {Bawendi}}]{Efros1996}%
  \BibitemOpen
  \bibfield  {author} {\bibinfo {author} {\bibfnamefont {{Al.}~L.}\
  \bibnamefont {Efros}}, \bibinfo {author} {\bibfnamefont {M.}~\bibnamefont
  {Rosen}}, \bibinfo {author} {\bibfnamefont {M.}~\bibnamefont {Kuno}},
  \bibinfo {author} {\bibfnamefont {M.}~\bibnamefont {Nirmal}}, \bibinfo
  {author} {\bibfnamefont {D.J.}~\bibnamefont {Norris}}, \ and\ \bibinfo {author}
  {\bibfnamefont {M.}~\bibnamefont {Bawendi}},\ }\bibfield  {title} {\enquote
  {\bibinfo {title} {{Band-edge exciton in quantum dots of semiconductors with
  a degenerate valence band: Dark and bright exciton states}},}\ } {\bibfield  {journal} {\bibinfo
  {journal} {Phys. Rev. B}\ }\textbf {\bibinfo {volume} {54}},\ \bibinfo
  {pages} {4843} (\bibinfo {year} {1996})}\BibitemShut {NoStop}%
\bibitem [{\citenamefont {Efros}(2003)}]{EfrosCh3}%
  \BibitemOpen
  \bibfield  {author} {\bibinfo {author} {\bibfnamefont {{Al.}~L.}\
  \bibnamefont {Efros}},\ }\bibfield  {title} {\enquote {\bibinfo {title}
  {{Fine Structure and Polarization Properties of Band-Edge Excitons in
  Semiconductor Nanocrystals; Chapter 3}},}\ }in\ \href@noop {} {\emph
  {\bibinfo {booktitle} {Semiconductor and Metal Nanocrystals: Synthesis and
  Electronic and Optical Properties}}},\ \bibinfo {editor} {edited by\ \bibinfo
  {editor} {\bibnamefont {{V. I. Klimov and M. Dekker}}}}\ (\bibinfo
  {publisher} {New York},\ \bibinfo {year} {2003})\ pp.\ \bibinfo {pages}
  {103--141}\BibitemShut {NoStop}%
\bibitem [{\citenamefont {Shornikova}\ \emph
  {et~al.}(2020{\natexlab{a}})\citenamefont {Shornikova}, \citenamefont
  {Golovatenko}, \citenamefont {Yakovlev}, \citenamefont {Rodina},
  \citenamefont {Biadala}, \citenamefont {Qiang}, \citenamefont {Kuntzmann},
  \citenamefont {Nasilowski}, \citenamefont {Dubertret}, \citenamefont
  {Polovitsyn}, \citenamefont {Moreels},\ and\ \citenamefont
  {Bayer}}]{Shornikova2020nn}%
  \BibitemOpen
\bibfield  {journal} {  }\bibfield  {author} {\bibinfo {author} {\bibfnamefont
  {E.~V.}\ \bibnamefont {Shornikova}}, \bibinfo {author} {\bibfnamefont
  {A.~A.}\ \bibnamefont {Golovatenko}}, \bibinfo {author} {\bibfnamefont
  {D.~R.}\ \bibnamefont {Yakovlev}}, \bibinfo {author} {\bibfnamefont
  {A.~V.}\ \bibnamefont {Rodina}}, \bibinfo {author} {\bibfnamefont {L.}\
  \bibnamefont {Biadala}}, \bibinfo {author} {\bibfnamefont {G.}\
  \bibnamefont {Qiang}}, \bibinfo {author} {\bibfnamefont {A.}\
  \bibnamefont {Kuntzmann}}, \bibinfo {author} {\bibfnamefont {M.}\
  \bibnamefont {Nasilowski}}, \bibinfo {author} {\bibfnamefont {B.}\
  \bibnamefont {Dubertret}}, \bibinfo {author} {\bibfnamefont {A.}\
  \bibnamefont {P.}}, \bibinfo {author} {\bibfnamefont {I.}\
  \bibnamefont {Moreels}}, \ and\ \bibinfo {author} {\bibfnamefont {M.}\
  \bibnamefont {Bayer}},\ }\bibfield  {title} {\enquote {\bibinfo {title}
  {Surface spin magnetism controls the polarized exciton emission from {CdSe}
  nanoplatelets},}\ } {\bibfield
  {journal} {\bibinfo  {journal} {Nature Nanotechnology}\ }\textbf {\bibinfo
  {volume} {15}},\ \bibinfo {pages} {277} (\bibinfo {year}
  {2020}{\natexlab{a}})}\BibitemShut {NoStop}%
\bibitem [{\citenamefont {Shornikova}\ \emph
  {et~al.}(2020{\natexlab{b}})\citenamefont {Shornikova}, \citenamefont
  {Yakovlev}, \citenamefont {Biadala}, \citenamefont {Crooker}, \citenamefont
  {Belykh}, \citenamefont {Kochiev}, \citenamefont {Kuntzmann}, \citenamefont
  {Nasilowski}, \citenamefont {Dubertret},\ and\ \citenamefont
  {Bayer}}]{Shornikova2020nl}%
  \BibitemOpen
  \bibfield  {author} {\bibinfo {author} {\bibfnamefont {E.}\ \bibnamefont
  {Shornikova}}, \bibinfo {author} {\bibfnamefont {D.}\ \bibnamefont
  {Yakovlev}}, \bibinfo {author} {\bibfnamefont {L.}\ \bibnamefont
  {Biadala}}, \bibinfo {author} {\bibfnamefont {S.}\ \bibnamefont
  {Crooker}}, \bibinfo {author} {\bibfnamefont {V.}\ \bibnamefont
  {Belykh}}, \bibinfo {author} {\bibfnamefont {M.}\ \bibnamefont
  {Kochiev}}, \bibinfo {author} {\bibfnamefont {A.}\ \bibnamefont
  {Kuntzmann}}, \bibinfo {author} {\bibfnamefont {M.}\ \bibnamefont
  {Nasilowski}}, \bibinfo {author} {\bibfnamefont {B.}\ \bibnamefont
  {Dubertret}}, \ and\ \bibinfo {author} {\bibfnamefont {M.}\ \bibnamefont
  {Bayer}},\ }\bibfield  {title} {\enquote {\bibinfo {title} {Negatively
  charged excitons in cdse nanoplatelets},}\ } {\bibfield  {journal} {\bibinfo  {journal}
  {Nano Let.}\ }\textbf {\bibinfo {volume} {20}},\ \bibinfo {pages}
  {1370} (\bibinfo {year} {2020}{\natexlab{b}})}\BibitemShut {NoStop}%
\bibitem [{\citenamefont {Shornikova}\ \emph
  {et~al.}(2020{\natexlab{c}})\citenamefont {Shornikova}, \citenamefont
  {Yakovlev}, \citenamefont {Tolmachev}, \citenamefont {Ivanov}, \citenamefont
  {Kalitukha}, \citenamefont {Sapega}, \citenamefont {Kudlacik}, \citenamefont
  {Kusrayev}, \citenamefont {Golovatenko}, \citenamefont {Shendre},
  \citenamefont {Delikanli}, \citenamefont {Demir},\ and\ \citenamefont
  {Bayer}}]{Shornikova2020acs}%
  \BibitemOpen
  \bibfield  {author} {\bibinfo {author} {\bibfnamefont {E.}\ \bibnamefont
  {Shornikova}}, \bibinfo {author} {\bibfnamefont {D.}\ \bibnamefont
  {Yakovlev}}, \bibinfo {author} {\bibfnamefont {D.}\ \bibnamefont
  {Tolmachev}}, \bibinfo {author} {\bibfnamefont {V.}\ \bibnamefont
  {Ivanov}}, \bibinfo {author} {\bibfnamefont {I.}\ \bibnamefont {Kalitukha}},
  \bibinfo {author} {\bibfnamefont {Victor}\ \bibnamefont {Sapega}}, \bibinfo
  {author} {\bibfnamefont {D.}\ \bibnamefont {Kudlacik}}, \bibinfo {author}
  {\bibfnamefont {Y.}\ \bibnamefont {Kusrayev}}, \bibinfo {author}
  {\bibfnamefont {A.}\ \bibnamefont {Golovatenko}}, \bibinfo {author}
  {\bibfnamefont {S.}\ \bibnamefont {Shendre}}, \bibinfo {author}
  {\bibfnamefont {S.}\ \bibnamefont {Delikanli}}, \bibinfo {author}
  {\bibfnamefont {H.}\ \bibnamefont {Demir}}, \ and\ \bibinfo {author}
  {\bibfnamefont {M.}\ \bibnamefont {Bayer}},\ }\bibfield  {title}
  {\enquote {\bibinfo {title} {Magneto-optics of excitons interacting with
  magnetic ions in cdse/cdmns colloidal nanoplatelets},}\ } {\bibfield  {journal} {\bibinfo  {journal} {ACS
  Nano}\ }\textbf {\bibinfo {volume} {14}},\ \bibinfo {pages}
  {9032}  (\bibinfo {year}
  {2020}{\natexlab{c}})}\BibitemShut {NoStop}%
\bibitem [{\citenamefont {Kiselev}\ and\ \citenamefont
  {Moiseev}(1996)}]{Kiselev1996}%
  \BibitemOpen
  \bibfield  {author} {\bibinfo {author} {\bibfnamefont {A.~A.}\ \bibnamefont
  {Kiselev}}\ and\ \bibinfo {author} {\bibfnamefont {L.V.}\ \bibnamefont
  {Moiseev}},\ }\href@noop {} {\bibfield  {journal} {\bibinfo  {journal}
  {Physics of the Solid State}\ }\textbf {\bibinfo {volume} {38}},\ \bibinfo
  {pages} {866} (\bibinfo {year} {1996})}\BibitemShut {NoStop}%
\bibitem [{\citenamefont {Durnev}\ \emph {et~al.}(2012)\citenamefont {Durnev},
  \citenamefont {Glazov},\ and\ \citenamefont {Ivchenko}}]{Durnev2012}%
  \BibitemOpen
  \bibfield  {author} {\bibinfo {author} {\bibfnamefont {M.~V.}\ \bibnamefont
  {Durnev}}, \bibinfo {author} {\bibfnamefont {M.~M.}\ \bibnamefont {Glazov}},
  \ and\ \bibinfo {author} {\bibfnamefont {E.~L.}\ \bibnamefont {Ivchenko}},\
  }\bibfield  {title} {\enquote {\bibinfo {title} {Giant zeeman splitting of
  light holes in {GaAs}/{AlGaAs} quantum wells},}\ }\href@noop {} {\bibfield
  {journal} {\bibinfo  {journal} {Physica E}\ }\textbf {\bibinfo {volume}
  {44}},\ \bibinfo {pages} {797} (\bibinfo {year} {2012})}\BibitemShut
  {NoStop}%
\bibitem [{\citenamefont {van Kesteren}\ \emph
  {et~al.}(1990{\natexlab{b}})\citenamefont {van Kesteren}, \citenamefont
  {Cosman}, \citenamefont {van~der Poel},\ and\ \citenamefont
  {Foxon}}]{vanKesteren1990}%
  \BibitemOpen
  \bibfield  {author} {\bibinfo {author} {\bibfnamefont {H.~W.}\ \bibnamefont
  {van Kesteren}}, \bibinfo {author} {\bibfnamefont {E.~C.}\ \bibnamefont
  {Cosman}}, \bibinfo {author} {\bibfnamefont {W.~A. J.~A.}\ \bibnamefont
  {van~der Poel}}, \ and\ \bibinfo {author} {\bibfnamefont {C.~T.}\
  \bibnamefont {Foxon}},\ }\bibfield  {title} {\enquote {\bibinfo {title} {Fine
  structure of excitons in type-{II} {GaAs}/{AlAs} quantum wells},}\
  }\href@noop {} {\bibfield  {journal} {\bibinfo  {journal} {Physical Review
  B}\ }\textbf {\bibinfo {volume} {41}},\ \bibinfo {pages} {5283}
  (\bibinfo {year} {1990}{\natexlab{b}})}\BibitemShut {NoStop}%
\bibitem [{\citenamefont {Rodina}\ \emph {et~al.}(2001)\citenamefont {Rodina},
  \citenamefont {Dietrich}, \citenamefont {G{\"o}ldner}, \citenamefont {Eckey},
  \citenamefont {Hoffmann}, \citenamefont {Efros}, \citenamefont {Rosen},\ and\
  \citenamefont {Meyer}}]{Rodina2001f}%
  \BibitemOpen
  \bibfield  {author} {\bibinfo {author} {\bibfnamefont {A.~V.}\ \bibnamefont
  {Rodina}}, \bibinfo {author} {\bibfnamefont {M.}~\bibnamefont {Dietrich}},
  \bibinfo {author} {\bibfnamefont {A.}~\bibnamefont {G{\"o}ldner}}, \bibinfo
  {author} {\bibfnamefont {L.}~\bibnamefont {Eckey}}, \bibinfo {author}
  {\bibfnamefont {A.}~\bibnamefont {Hoffmann}}, \bibinfo {author}
  {\bibfnamefont {{Al.}~L.}\ \bibnamefont {Efros}}, \bibinfo {author}
  {\bibfnamefont {M.}~\bibnamefont {Rosen}}, \ and\ \bibinfo {author}
  {\bibfnamefont {B.~K.}\ \bibnamefont {Meyer}},\ }\bibfield  {title} {\enquote
  {\bibinfo {title} {{Free excitons in wurtzite GaN}},}\ } {\bibfield  {journal} {\bibinfo  {journal} {Phys.
  Rev. B}\ }\textbf {\bibinfo {volume} {64}},\ \bibinfo {pages} {115204}
  (\bibinfo {year} {2001})}\BibitemShut {NoStop}%
\bibitem [{\citenamefont {Broido}\ and\ \citenamefont
  {Sham}(1985{\natexlab{a}})}]{Broido1985}%
  \BibitemOpen
  \bibfield  {author} {\bibinfo {author} {\bibfnamefont {D.~A.}\ \bibnamefont
  {Broido}}\ and\ \bibinfo {author} {\bibfnamefont {L.~J.}\ \bibnamefont
  {Sham}},\ }\bibfield  {title} {\enquote {\bibinfo {title} {Effective masses
  of holes at gaas-algaas heterojunctions},}\ } {\bibfield  {journal} {\bibinfo  {journal} {Phys.
  Rev. B}\ }\textbf {\bibinfo {volume} {31}},\ \bibinfo {pages} {888}
  (\bibinfo {year} {1985}{\natexlab{a}})}\BibitemShut {NoStop}%
\bibitem [{\citenamefont {Broido}\ and\ \citenamefont
  {Sham}(1985{\natexlab{b}})}]{Broido1985err}%
  \BibitemOpen
  \bibfield  {author} {\bibinfo {author} {\bibfnamefont {D.~A.}\ \bibnamefont
  {Broido}}\ and\ \bibinfo {author} {\bibfnamefont {L.~J.}\ \bibnamefont
  {Sham}},\ }\bibfield  {title} {\enquote {\bibinfo {title} {Erratum: Effective
  masses of holes at gaas-algaas heterojunctions},}\ } {\bibfield  {journal} {\bibinfo  {journal} {Phys.
  Rev. B}\ }\textbf {\bibinfo {volume} {31}},\ \bibinfo {pages} {6831(E)}
  (\bibinfo {year} {1985}{\natexlab{b}})}\BibitemShut {NoStop}%
\bibitem [{\citenamefont {Vahala}\ and\ \citenamefont
  {Sercel}(1990)}]{SercelPRL90}%
  \BibitemOpen
  \bibfield  {author} {\bibinfo {author} {\bibfnamefont {K.~J.}\
  \bibnamefont {Vahala}}\ and\ \bibinfo {author} {\bibfnamefont {P.~C.}\
  \bibnamefont {Sercel}},\ }\bibfield  {title} {\enquote {\bibinfo {title}
  {Application of a total-angular-momentum basis to quantum-dot band
  structure},}\ } {\bibfield
  {journal} {\bibinfo  {journal} {Phys. Rev. Lett.}\ }\textbf {\bibinfo
  {volume} {65}},\ \bibinfo {pages} {239} (\bibinfo {year}
  {1990})}\BibitemShut {NoStop}%
\bibitem [{\citenamefont {Sercel}\ and\ \citenamefont
  {Vahala}(1990)}]{SercelPRB90}%
  \BibitemOpen
  \bibfield  {author} {\bibinfo {author} {\bibfnamefont {P.~C.}\
  \bibnamefont {Sercel}}\ and\ \bibinfo {author} {\bibfnamefont {K.~J.}\
  \bibnamefont {Vahala}},\ }\bibfield  {title} {\enquote {\bibinfo {title}
  {Analytical formalism for determining quantum-wire and quantum-dot band
  structure in the multiband envelope-function approximation},}\ } {\bibfield  {journal} {\bibinfo
  {journal} {Phys. Rev. B}\ }\textbf {\bibinfo {volume} {42}},\ \bibinfo
  {pages} {3690} (\bibinfo {year} {1990})}\BibitemShut {NoStop}%
\bibitem [{\citenamefont {Rego}\ \emph {et~al.}(1997)\citenamefont {Rego},
  \citenamefont {Hawrylak}, \citenamefont {Brum},\ and\ \citenamefont
  {Wojs}}]{Rego1997}%
  \BibitemOpen
  \bibfield  {author} {\bibinfo {author} {\bibfnamefont {L.~G.~C.}\
  \bibnamefont {Rego}}, \bibinfo {author} {\bibfnamefont {P.}~\bibnamefont
  {Hawrylak}}, \bibinfo {author} {\bibfnamefont {J.~A.}\ \bibnamefont {Brum}},
  \ and\ \bibinfo {author} {\bibfnamefont {A.}~\bibnamefont {Wojs}},\
  }\bibfield  {title} {\enquote {\bibinfo {title} {{Interacting valence holes
  in p-type SiGe quantum disks in a magnetic field}},}\ } {\bibfield  {journal} {\bibinfo  {journal} {Phys.
  Rev. B}\ }\textbf {\bibinfo {volume} {55}},\ \bibinfo {pages} {15694}
  (\bibinfo {year} {1997})}\BibitemShut {NoStop}%
\bibitem [{\citenamefont {Baldereschi}\ and\ \citenamefont
  {Lipari}(1973)}]{Baldereschi1973}%
  \BibitemOpen
  \bibfield  {author} {\bibinfo {author} {\bibfnamefont {A.}~\bibnamefont
  {Baldereschi}}\ and\ \bibinfo {author} {\bibfnamefont {N.~O.}\ \bibnamefont
  {Lipari}},\ }\bibfield  {title} {\enquote {\bibinfo {title} {{Spherical model
  of shallow acceptor states in semiconductors}},}\ } {\bibfield  {journal} {\bibinfo  {journal} {Phys.
  Rev. B}\ }\textbf {\bibinfo {volume} {8}},\ \bibinfo {pages} {2697}
  (\bibinfo {year} {1973})}\BibitemShut {NoStop}%
\bibitem [{\citenamefont {Adachi}(2004)}]{Adachi2004}%
  \BibitemOpen
  \bibfield  {author} {\bibinfo {author} {\bibfnamefont {Sadao}\ \bibnamefont
  {Adachi}},\ }\href@noop {} {\emph {\bibinfo {title} {{Handbook on physical
  properties of semiconductors}}}}\ (\bibinfo  {publisher} {Springer US},\
  \bibinfo {year} {2004})\BibitemShut {NoStop}%
\bibitem [{\citenamefont {Karazhanov}(2005)}]{Karazhanov2005}%
  \BibitemOpen
  \bibfield  {author} {\bibinfo {author} {\bibfnamefont {S.~Zh.}\ \bibnamefont
  {Karazhanov}},\ }\bibfield  {title} {\enquote {\bibinfo {title} {Ab initio
  studies of the band parameters of {III}{\textendash}v and
  {II}{\textendash}{VI} zinc-blende semiconductors},}\ }\href@noop {}
  {\bibfield  {journal} {\bibinfo  {journal} {Semiconductors}\ }\textbf
  {\bibinfo {volume} {39}},\ \bibinfo {pages} {161} (\bibinfo {year}
  {2005})}\BibitemShut {NoStop}%
\bibitem [{\citenamefont {Horodysk{\'{a}}}\ \emph {et~al.}(2010)\citenamefont
  {Horodysk{\'{a}}}, \citenamefont {N{\v{e}}mec}, \citenamefont {Sprinzl},
  \citenamefont {Mal{\'{y}}}, \citenamefont {Gladilin},\ and\ \citenamefont
  {Devreese}}]{Horodysk2010}%
  \BibitemOpen
  \bibfield  {author} {\bibinfo {author} {\bibfnamefont {P.}~\bibnamefont
  {Horodysk{\'{a}}}}, \bibinfo {author} {\bibfnamefont {P.}~\bibnamefont
  {N{\v{e}}mec}}, \bibinfo {author} {\bibfnamefont {D.}~\bibnamefont
  {Sprinzl}}, \bibinfo {author} {\bibfnamefont {P.}~\bibnamefont {Mal{\'{y}}}},
  \bibinfo {author} {\bibfnamefont {V.~N.}\ \bibnamefont {Gladilin}}, \ and\
  \bibinfo {author} {\bibfnamefont {J.~T.}\ \bibnamefont {Devreese}},\
  }\bibfield  {title} {\enquote {\bibinfo {title} {Exciton spin dynamics in
  spherical {CdS} quantum dots},}\ }\href@noop {} {\bibfield  {journal}
  {\bibinfo  {journal} {Physical Review B}\ }\textbf {\bibinfo {volume} {81}}
  (\bibinfo {year} {2010})}\BibitemShut {NoStop}%
\bibitem [{\citenamefont {Edmonds}(1957)}]{Edmonds}%
  \BibitemOpen
  \bibfield  {author} {\bibinfo {author} {\bibfnamefont {A.~R.}\ \bibnamefont
  {Edmonds}},\ }\href@noop {} {\emph {\bibinfo {title} {{Angular momentum in
  Quantum mechanics}}}}\ (\bibinfo  {publisher} {Princenton University Press},\
  \bibinfo {address} {Princenton},\ \bibinfo {year} {1957})\BibitemShut
  {NoStop}%
\bibitem [{\citenamefont {Ivchenko}(2005)}]{ivchenko05a}%
  \BibitemOpen
  \bibfield  {author} {\bibinfo {author} {\bibfnamefont {E.~L.}\ \bibnamefont
  {Ivchenko}},\ }\href@noop {} {\emph {\bibinfo {title} {Optical Spectroscopy
  of Semiconductor Nanostructures}}}\ (\bibinfo  {publisher} {Alpha Science,
  Harrow UK},\ \bibinfo {year} {2005})\BibitemShut {NoStop}%
\bibitem [{\citenamefont {Efros}(1992)}]{Efros92}%
  \BibitemOpen
  \bibfield  {author} {\bibinfo {author} {\bibfnamefont {Al.~L.}\ \bibnamefont
  {Efros}},\ }\bibfield  {title} {\enquote {\bibinfo {title} {Luminescence
  polarization of cdse microcrystals},}\ } {\bibfield  {journal} {\bibinfo  {journal} {Phys.
  Rev. B}\ }\textbf {\bibinfo {volume} {46}},\ \bibinfo {pages} {7448}
  (\bibinfo {year} {1992})}\BibitemShut {NoStop}%
\bibitem [{\citenamefont {Efros}\ and\ \citenamefont
  {Rodina}(1993{\natexlab{a}})}]{PhysRevB.47.10005}%
  \BibitemOpen
  \bibfield  {author} {\bibinfo {author} {\bibfnamefont {Al.~L.}\ \bibnamefont
  {Efros}}\ and\ \bibinfo {author} {\bibfnamefont {A.~V.}\ \bibnamefont
  {Rodina}},\ }\bibfield  {title} {\enquote {\bibinfo {title} {Band-edge
  absorption and luminescence of nonspherical nanometer-size crystals},}\
  } {\bibfield  {journal} {\bibinfo
  {journal} {Phys. Rev. B}\ }\textbf {\bibinfo {volume} {47}},\ \bibinfo
  {pages} {10005} (\bibinfo {year} {1993}{\natexlab{a}})}\BibitemShut
  {NoStop}%
\bibitem [{\citenamefont {Semina}\ \emph {et~al.}(2016)\citenamefont {Semina},
  \citenamefont {Golovatenko},\ and\ \citenamefont {Rodina}}]{Semina2016}%
  \BibitemOpen
  \bibfield  {author} {\bibinfo {author} {\bibfnamefont {M.~A.}\ \bibnamefont
  {Semina}}, \bibinfo {author} {\bibfnamefont {A.~A.}\ \bibnamefont
  {Golovatenko}}, \ and\ \bibinfo {author} {\bibfnamefont {A.~V.}\ \bibnamefont
  {Rodina}},\ }\bibfield  {title} {\enquote {\bibinfo {title} {{Ground state of
  the holes localized in II-VI quantum dots with Gaussian potential
  profiles}},}\ } {\bibfield
  {journal} {\bibinfo  {journal} {Phys. Rev. B}\ }\textbf {\bibinfo {volume}
  {93}},\ \bibinfo {pages} {045409} (\bibinfo {year} {2016})}\BibitemShut
  {NoStop}%
\bibitem [{\citenamefont {Rodina}(1993)}]{Rodina1993}%
  \BibitemOpen
  \bibfield  {author} {\bibinfo {author} {\bibfnamefont {A.~V.}\ \bibnamefont
  {Rodina}},\ }\bibfield  {title} {\enquote {\bibinfo {title} {{$A^+$-center
  and exciton bound to neutral acceptor in diamond-like semiconductors}},}\
  } {\bibfield  {journal}
  {\bibinfo  {journal} {Solid State Commun.}\ }\textbf {\bibinfo {volume}
  {85}},\ \bibinfo {pages} {23} (\bibinfo {year} {1993})}\BibitemShut
  {NoStop}%
\bibitem [{\citenamefont {Efros}\ and\ \citenamefont
  {Rodina}(1993{\natexlab{b}})}]{Efros1993}%
  \BibitemOpen
  \bibfield  {author} {\bibinfo {author} {\bibfnamefont {{Al.}~L.}\
  \bibnamefont {Efros}}\ and\ \bibinfo {author} {\bibfnamefont {A.~V.}\
  \bibnamefont {Rodina}},\ }\bibfield  {title} {\enquote {\bibinfo {title}
  {{Band-edge absorption and luminescence of nonspherical nanometer-size
  crystals}},}\ } {\bibfield
  {journal} {\bibinfo  {journal} {Phys. Rev. B}\ }\textbf {\bibinfo {volume}
  {47}},\ \bibinfo {pages} {10005} (\bibinfo {year}
  {1993}{\natexlab{b}})}\BibitemShut {NoStop}%
\bibitem [{\citenamefont {Rodina}\ and\ \citenamefont
  {Efros}(2010)}]{Rodina2010}%
  \BibitemOpen
  \bibfield  {author} {\bibinfo {author} {\bibfnamefont {A.~V.}\ \bibnamefont
  {Rodina}}\ and\ \bibinfo {author} {\bibfnamefont {{Al.}~L.}\ \bibnamefont
  {Efros}},\ }\bibfield  {title} {\enquote {\bibinfo {title} {{Band-edge
  biexciton in nanocrystals of semiconductors with a degenerate valence
  band}},}\ } {\bibfield  {journal}
  {\bibinfo  {journal} {Phys. Rev. B}\ }\textbf {\bibinfo {volume} {82}},\
  \bibinfo {pages} {125324} (\bibinfo {year} {2010})}\BibitemShut {NoStop}%
\bibitem [{\citenamefont {Baldereschi}\ and\ \citenamefont
  {Lipari}(1974)}]{Baldereschi1974}%
  \BibitemOpen
  \bibfield  {author} {\bibinfo {author} {\bibfnamefont {A.}~\bibnamefont
  {Baldereschi}}\ and\ \bibinfo {author} {\bibfnamefont {Nunzio~O.}\
  \bibnamefont {Lipari}},\ }\bibfield  {title} {\enquote {\bibinfo {title}
  {Cubic contributions to the spherical model of shallow acceptor states},}\
  } {\bibfield  {journal} {\bibinfo
  {journal} {Phys. Rev. B}\ }\textbf {\bibinfo {volume} {9}},\ \bibinfo {pages}
  {1525} (\bibinfo {year} {1974})}\BibitemShut {NoStop}%
\bibitem [{\citenamefont {Golovatenko}\ \emph {et~al.}(2018)\citenamefont
  {Golovatenko}, \citenamefont {Semina}, \citenamefont {Rodina},\ and\
  \citenamefont {Shubina}}]{Golovatenko2018}%
  \BibitemOpen
  \bibfield  {author} {\bibinfo {author} {\bibfnamefont {A.~A.}\ \bibnamefont
  {Golovatenko}}, \bibinfo {author} {\bibfnamefont {M.~A.}\ \bibnamefont
  {Semina}}, \bibinfo {author} {\bibfnamefont {A.~V.}\ \bibnamefont {Rodina}},
  \ and\ \bibinfo {author} {\bibfnamefont {T.~V.}\ \bibnamefont {Shubina}},\
  }\bibfield  {title} {\enquote {\bibinfo {title} {Excitons and biexcitons in
  spheroidal quantum dots A$_2$B$_6$},}\ }
  {\bibfield  {journal} {\bibinfo  {journal} {Phys. of  Sol. State}\
  }\textbf {\bibinfo {volume} {60}},\ \bibinfo {pages} {1510} (\bibinfo
  {year} {2018})}\BibitemShut {NoStop}%
\bibitem [{\citenamefont {Semina}\ and\ \citenamefont
  {Suris}(2011)}]{Semina2011}%
  \BibitemOpen
  \bibfield  {author} {\bibinfo {author} {\bibfnamefont {M.~A.}\ \bibnamefont
  {Semina}}\ and\ \bibinfo {author} {\bibfnamefont {R.~A.}\ \bibnamefont
  {Suris}},\ }\bibfield  {title} {\enquote {\bibinfo {title} {Effect of
  localization in quantum wells and quantum wires on heavy-light hole mixing
  and acceptor binding energy},}\ }\href@noop {} {\bibfield  {journal}
  {\bibinfo  {journal} {Semiconductors}\ }\textbf {\bibinfo {volume} {45}},\
  \bibinfo {pages} {917} (\bibinfo {year} {2011})}\BibitemShut {NoStop}%
\bibitem [{\citenamefont {Shornikova}\ \emph
  {et~al.}(2018{\natexlab{a}})\citenamefont {Shornikova}, \citenamefont
  {Biadala}, \citenamefont {Yakovlev}, \citenamefont {Sapega}, \citenamefont
  {Kusrayev}, \citenamefont {Mitioglu}, \citenamefont {Ballottin},
  \citenamefont {Christianen}, \citenamefont {Belykh}, \citenamefont {Kochiev},
  \citenamefont {Sibeldin}, \citenamefont {Golovatenko}, \citenamefont
  {Rodina}, \citenamefont {Gippius}, \citenamefont {Kuntzmann}, \citenamefont
  {Jiang}, \citenamefont {Nasilowski}, \citenamefont {Dubertret},\ and\
  \citenamefont {Bayer}}]{Shornikova2018}%
  \BibitemOpen
  \bibfield  {author} {\bibinfo {author} {\bibfnamefont {E.~V.}\
  \bibnamefont {Shornikova}}, \bibinfo {author} {\bibfnamefont {L.}\
  \bibnamefont {Biadala}}, \bibinfo {author} {\bibfnamefont {D.~R.}\
  \bibnamefont {Yakovlev}}, \bibinfo {author} {\bibfnamefont {V.~F.}\
  \bibnamefont {Sapega}}, \bibinfo {author} {\bibfnamefont {Y.~G.}\
  \bibnamefont {Kusrayev}}, \bibinfo {author} {\bibfnamefont {A.~A.}\
  \bibnamefont {Mitioglu}}, \bibinfo {author} {\bibfnamefont {M.~V.}\
  \bibnamefont {Ballottin}}, \bibinfo {author} {\bibfnamefont {P. C.~M.}\
  \bibnamefont {Christianen}}, \bibinfo {author} {\bibfnamefont {V.~V.}\
  \bibnamefont {Belykh}}, \bibinfo {author} {\bibfnamefont {M.~V.}\
  \bibnamefont {Kochiev}}, \bibinfo {author} {\bibfnamefont {N.~N.}\
  \bibnamefont {Sibeldin}}, \bibinfo {author} {\bibfnamefont {A.~A.}\
  \bibnamefont {Golovatenko}}, \bibinfo {author} {\bibfnamefont {A.~V.}\
  \bibnamefont {Rodina}}, \bibinfo {author} {\bibfnamefont {N.~A.}\
  \bibnamefont {Gippius}}, \bibinfo {author} {\bibfnamefont {A.}\
  \bibnamefont {Kuntzmann}}, \bibinfo {author} {\bibfnamefont {Y.}~\bibnamefont
  {Jiang}}, \bibinfo {author} {\bibfnamefont {M.}\ \bibnamefont
  {Nasilowski}}, \bibinfo {author} {\bibfnamefont {B.}\ \bibnamefont
  {Dubertret}}, \ and\ \bibinfo {author} {\bibfnamefont {M.}\ \bibnamefont
  {Bayer}},\ }\bibfield  {title} {\enquote {\bibinfo {title} {Addressing the
  exciton fine structure in colloidal nanocrystals: the case of {CdSe}
  nanoplatelets},}\ } {\bibfield  {journal}
  {\bibinfo  {journal} {Nanoscale}\ }\textbf {\bibinfo {volume} {10}},\
  \bibinfo {pages} {646} (\bibinfo {year}
  {2018}{\natexlab{a}})}\BibitemShut {NoStop}%
\bibitem [{\citenamefont {Wimbauer}\ \emph {et~al.}(1994)\citenamefont
  {Wimbauer}, \citenamefont {Oettinger}, \citenamefont {Efros}, \citenamefont
  {Meyer},\ and\ \citenamefont {Brugger}}]{Wimbauer1994}%
  \BibitemOpen
  \bibfield  {author} {\bibinfo {author} {\bibfnamefont {Th.}\ \bibnamefont
  {Wimbauer}}, \bibinfo {author} {\bibfnamefont {K.}~\bibnamefont {Oettinger}},
  \bibinfo {author} {\bibfnamefont {Al.~L.}\ \bibnamefont {Efros}}, \bibinfo
  {author} {\bibfnamefont {B.~K.}\ \bibnamefont {Meyer}}, \ and\ \bibinfo
  {author} {\bibfnamefont {H.}~\bibnamefont {Brugger}},\ }\bibfield  {title}
  {\enquote {\bibinfo {title} {Zeeman splitting of the excitonic recombination
  in In$_x$Ga$_{1-x}$As/GaAs single quantum wells},}\ }\href@noop {} {\bibfield
  {journal} {\bibinfo  {journal} {Phys. Rev. B}\ }\textbf {\bibinfo
  {volume} {50}},\ \bibinfo {pages} {8889} (\bibinfo {year}
  {1994})}\BibitemShut {NoStop}%
\bibitem [{\citenamefont {Kubisa}\ \emph {et~al.}(2011)\citenamefont {Kubisa},
  \citenamefont {Ryczko},\ and\ \citenamefont {Misiewicz}}]{Kubisa2011}%
  \BibitemOpen
  \bibfield  {author} {\bibinfo {author} {\bibfnamefont {M.}~\bibnamefont
  {Kubisa}}, \bibinfo {author} {\bibfnamefont {K.}~\bibnamefont {Ryczko}}, \
  and\ \bibinfo {author} {\bibfnamefont {J.}~\bibnamefont {Misiewicz}},\
  }\bibfield  {title} {\enquote {\bibinfo {title} {{Spin splitting of holes in
  symmetric GaAs/Ga${}_{x}$Al${}_{1-x}$As quantum wells}},}\ } {\bibfield  {journal} {\bibinfo  {journal} {Phys.
  Rev. B}\ }\textbf {\bibinfo {volume} {83}},\ \bibinfo {pages} {195324}
  (\bibinfo {year} {2011})}\BibitemShut {NoStop}%
\bibitem [{\citenamefont {Shornikova}\ \emph
  {et~al.}(2018{\natexlab{b}})\citenamefont {Shornikova}, \citenamefont
  {Biadala}, \citenamefont {Yakovlev}, \citenamefont {Feng}, \citenamefont
  {Sapega}, \citenamefont {Flipo}, \citenamefont {Golovatenko}, \citenamefont
  {Semina}, \citenamefont {Rodina}, \citenamefont {Mitioglu}, \citenamefont
  {Ballottin}, \citenamefont {Christianen}, \citenamefont {Kusrayev},
  \citenamefont {Nasilowski}, \citenamefont {Dubertret},\ and\ \citenamefont
  {Bayer}}]{Shornikova2017nl}%
  \BibitemOpen
  \bibfield  {author} {\bibinfo {author} {\bibfnamefont {E.~V.}\
  \bibnamefont {Shornikova}}, \bibinfo {author} {\bibfnamefont {L.}\
  \bibnamefont {Biadala}}, \bibinfo {author} {\bibfnamefont {D.~R.}\
  \bibnamefont {Yakovlev}}, \bibinfo {author} {\bibfnamefont {D.}\
  \bibnamefont {Feng}}, \bibinfo {author} {\bibfnamefont {V.~F.}\
  \bibnamefont {Sapega}}, \bibinfo {author} {\bibfnamefont {N.}\
  \bibnamefont {Flipo}}, \bibinfo {author} {\bibfnamefont {A.~A.}\
  \bibnamefont {Golovatenko}}, \bibinfo {author} {\bibfnamefont {M.~A.}\
  \bibnamefont {Semina}}, \bibinfo {author} {\bibfnamefont {A.~V.}\
  \bibnamefont {Rodina}}, \bibinfo {author} {\bibfnamefont {A.~A.}\
  \bibnamefont {Mitioglu}}, \bibinfo {author} {\bibfnamefont {M.~V.}\
  \bibnamefont {Ballottin}}, \bibinfo {author} {\bibfnamefont {P. C.~M.}\
  \bibnamefont {Christianen}}, \bibinfo {author} {\bibfnamefont {Y.~G.}\
  \bibnamefont {Kusrayev}}, \bibinfo {author} {\bibfnamefont {M.}\
  \bibnamefont {Nasilowski}}, \bibinfo {author} {\bibfnamefont {B.}\
  \bibnamefont {Dubertret}}, \ and\ \bibinfo {author} {\bibfnamefont {M.}\
  \bibnamefont {Bayer}},\ }\bibfield  {title} {\enquote {\bibinfo {title}
  {Electron and hole g-factors and spin dynamics of negatively charged excitons
  in cdse/cds colloidal nanoplatelets with thick shells},}\ } {\bibfield  {journal} {\bibinfo  {journal}
  {Nano Lett.}\ }\textbf {\bibinfo {volume} {18}},\ \bibinfo {pages}
  {373} (\bibinfo {year} {2018}{\natexlab{b}})}\BibitemShut {NoStop}%
\bibitem [{\citenamefont {Norris}\ \emph {et~al.}(1996)\citenamefont {Norris},
  \citenamefont {Efros}, \citenamefont {Rosen},\ and\ \citenamefont
  {Bawendi}}]{Norris1996}%
  \BibitemOpen
  \bibfield  {author} {\bibinfo {author} {\bibfnamefont {D.J.}~\bibnamefont
  {Norris}}, \bibinfo {author} {\bibfnamefont {{Al.}~L.}\ \bibnamefont
  {Efros}}, \bibinfo {author} {\bibfnamefont {M.}~\bibnamefont {Rosen}}, \ and\
  \bibinfo {author} {\bibfnamefont {M.G.}~\bibnamefont {Bawendi}},\ }\bibfield
  {title} {\enquote {\bibinfo {title} {{Size dependence of exciton fine
  structure in CdSe quantum dots}},}\ } {\bibfield  {journal} {\bibinfo  {journal} {Phys.
  Rev. B}\ }\textbf {\bibinfo {volume} {53}},\ \bibinfo {pages} {16347}
  (\bibinfo {year} {1996})}\BibitemShut {NoStop}%
\bibitem [{\citenamefont {Fu}\ \emph {et~al.}(1998)\citenamefont {Fu},
  \citenamefont {Wang},\ and\ \citenamefont {Zunger}}]{Fu1998n2}%
  \BibitemOpen
  \bibfield  {author} {\bibinfo {author} {\bibfnamefont {H.}\
  \bibnamefont {Fu}}, \bibinfo {author} {\bibfnamefont {L.-W.}\ \bibnamefont
  {Wang}}, \ and\ \bibinfo {author} {\bibfnamefont {A.}\ \bibnamefont
  {Zunger}},\ }\bibfield  {title} {\enquote {\bibinfo {title} {Applicability of
  thek$\cdot$pmethod to the electronic structure of quantum dots},}\
  } {\bibfield  {journal} {\bibinfo  {journal} {Physical Review
  B}\ }\textbf {\bibinfo {volume} {57}},\ \bibinfo {pages} {9971}
  (\bibinfo {year} {1998})}\BibitemShut {NoStop}%
\bibitem [{\citenamefont {Kapustina}\ \emph {et~al.}(2000)\citenamefont
  {Kapustina}, \citenamefont {Petrov}, \citenamefont {Rodina},\ and\
  \citenamefont {Seisyan}}]{Kapustina2000}%
  \BibitemOpen
  \bibfield  {author} {\bibinfo {author} {\bibfnamefont {A.~B.}\ \bibnamefont
  {Kapustina}}, \bibinfo {author} {\bibfnamefont {B.~V.}\ \bibnamefont
  {Petrov}}, \bibinfo {author} {\bibfnamefont {A.~V.}\ \bibnamefont {Rodina}},
  \ and\ \bibinfo {author} {\bibfnamefont {R.~P.}\ \bibnamefont {Seisyan}},\
  }\bibfield  {title} {\enquote {\bibinfo {title} {{Magnetic absorption of
  hexagonal crystals CdSe in strong and weak fields: Quasi-cubic
  approximation}},}\ } {\bibfield  {journal}
  {\bibinfo  {journal} {Phys. of Sol. State}\ }\textbf {\bibinfo {volume}
  {42}},\ \bibinfo {pages} {1242} (\bibinfo {year} {2000})}\BibitemShut
  {NoStop}%
\bibitem [{\citenamefont {Berezovsky}\ \emph {et~al.}(2005)\citenamefont
  {Berezovsky}, \citenamefont {Ouyang}, \citenamefont {Meier}, \citenamefont
  {Awschalom}, \citenamefont {Battaglia},\ and\ \citenamefont
  {Peng}}]{Berezovsky2005}%
  \BibitemOpen
  \bibfield  {author} {\bibinfo {author} {\bibfnamefont {J.}\ \bibnamefont
  {Berezovsky}}, \bibinfo {author} {\bibfnamefont {M.}\ \bibnamefont
  {Ouyang}}, \bibinfo {author} {\bibfnamefont {F.}\ \bibnamefont {Meier}},
  \bibinfo {author} {\bibfnamefont {D.~D.}\ \bibnamefont {Awschalom}},
  \bibinfo {author} {\bibfnamefont {D.}\ \bibnamefont {Battaglia}}, \ and\
  \bibinfo {author} {\bibfnamefont {X.}\ \bibnamefont {Peng}},\
  }\bibfield  {title} {\enquote {\bibinfo {title} {Spin dynamics and level
  structure of quantum-dot quantum wells},}\ }\href@noop {} {\bibfield
  {journal} {\bibinfo  {journal} {Phys. Rev. B}\ }\textbf {\bibinfo
  {volume} {71}} (\bibinfo {year} {2005})}\BibitemShut {NoStop}%
\bibitem [{\citenamefont {Lawaetz}(1971)}]{Lawaetz1971}%
  \BibitemOpen
  \bibfield  {author} {\bibinfo {author} {\bibfnamefont {P.}~\bibnamefont
  {Lawaetz}},\ }\bibfield  {title} {\enquote {\bibinfo {title} {Valence-band
  parameters in cubic semiconductors},}\ }\href@noop {} {\bibfield  {journal}
  {\bibinfo  {journal} {Physical Review B}\ }\textbf {\bibinfo {volume} {4}},\
  \bibinfo {pages} {3460} (\bibinfo {year} {1971})}\BibitemShut {NoStop}%
\bibitem [{\citenamefont {Stradling}(1968)}]{Stradling1968}%
  \BibitemOpen
  \bibfield  {author} {\bibinfo {author} {\bibfnamefont {R.A.}\ \bibnamefont
  {Stradling}},\ }\bibfield  {title} {\enquote {\bibinfo {title} {Cyclotron
  resonance from thermally excited holes in {ZnTe}},}\ }\href@noop {}
  {\bibfield  {journal} {\bibinfo  {journal} {Solid State Comm.}\
  }\textbf {\bibinfo {volume} {6}},\ \bibinfo {pages} {665} (\bibinfo
  {year} {1968})}\BibitemShut {NoStop}%
\bibitem [{\citenamefont {Said}\ and\ \citenamefont
  {Kanehisa}(1990)}]{Said1990}%
  \BibitemOpen
  \bibfield  {author} {\bibinfo {author} {\bibfnamefont {M.}~\bibnamefont
  {Said}}\ and\ \bibinfo {author} {\bibfnamefont {M.~A.}\ \bibnamefont
  {Kanehisa}},\ }\bibfield  {title} {\enquote {\bibinfo {title} {Excited
  acceptor states in {II}{\textendash}{VI} and {III}{\textendash}V
  semiconductors},}\ }\href@noop {} {\bibfield  {journal} {\bibinfo  {journal}
  {phys. stat. sol. (b)}\ }\textbf {\bibinfo {volume} {157}},\ \bibinfo
  {pages} {311} (\bibinfo {year} {1990})}\BibitemShut {NoStop}%
\bibitem [{\citenamefont {Oka}\ and\ \citenamefont {Cardona}(1981)}]{Oka1981}%
  \BibitemOpen
  \bibfield  {author} {\bibinfo {author} {\bibfnamefont {Y.}\ \bibnamefont
  {Oka}}\ and\ \bibinfo {author} {\bibfnamefont {M.}\ \bibnamefont
  {Cardona}},\ }\bibfield  {title} {\enquote {\bibinfo {title} {Resonant
  spin-flip raman scattering on donor and acceptor states in {ZnTe}},}\
  }\href@noop {} {\bibfield  {journal} {\bibinfo  {journal} {Phys. Rev.
  B}\ }\textbf {\bibinfo {volume} {23}},\ \bibinfo {pages} {4129}
  (\bibinfo {year} {1981})}\BibitemShut {NoStop}%
\bibitem [{\citenamefont {Wagner}\ \emph {et~al.}(1992)\citenamefont {Wagner},
  \citenamefont {Lankes}, \citenamefont {Wolf}, \citenamefont {Kuhn},
  \citenamefont {Link},\ and\ \citenamefont {Gebhardt}}]{Wagner1992}%
  \BibitemOpen
  \bibfield  {author} {\bibinfo {author} {\bibfnamefont {H.P.}\ \bibnamefont
  {Wagner}}, \bibinfo {author} {\bibfnamefont {S.}~\bibnamefont {Lankes}},
  \bibinfo {author} {\bibfnamefont {K.}~\bibnamefont {Wolf}}, \bibinfo {author}
  {\bibfnamefont {W.}~\bibnamefont {Kuhn}}, \bibinfo {author} {\bibfnamefont
  {P.}~\bibnamefont {Link}}, \ and\ \bibinfo {author} {\bibfnamefont
  {W.}~\bibnamefont {Gebhardt}},\ }\bibfield  {title} {\enquote {\bibinfo
  {title} {Spectroscopic investigations of donor and acceptor states in n- and
  p-doped {ZnTe} epilayers},}\ }\href@noop {} {\bibfield  {journal} {\bibinfo
  {journal} {Journal of Crystal Growth}\ }\textbf {\bibinfo {volume} {117}},\
  \bibinfo {pages} {303} (\bibinfo {year} {1992})}\BibitemShut {NoStop}%
\bibitem [{\citenamefont {Friedrich}\ \emph {et~al.}(1994)\citenamefont
  {Friedrich}, \citenamefont {Kraus}, \citenamefont {Meininger}, \citenamefont
  {Schaack},\ and\ \citenamefont {Schmitt}}]{Friedrich1994}%
  \BibitemOpen
  \bibfield  {author} {\bibinfo {author} {\bibfnamefont {T.}~\bibnamefont
  {Friedrich}}, \bibinfo {author} {\bibfnamefont {J.}~\bibnamefont {Kraus}},
  \bibinfo {author} {\bibfnamefont {M.}~\bibnamefont {Meininger}}, \bibinfo
  {author} {\bibfnamefont {G.}~\bibnamefont {Schaack}}, \ and\ \bibinfo {author}
  {\bibfnamefont {W.~O.~G.}\ \bibnamefont {Schmitt}},\ }\bibfield  {title}
  {\enquote {\bibinfo {title} {Zeeman levels of the shallow lithium acceptor
  and band parameters in cadmium telluride},}\ }\href@noop {} {\bibfield
  {journal} {\bibinfo  {journal} {J. Phys.: Condens. Matter}\
  }\textbf {\bibinfo {volume} {6}},\ \bibinfo {pages} {4307} (\bibinfo
  {year} {1994})}\BibitemShut {NoStop}%
\bibitem [{\citenamefont {Neumann}\ \emph {et~al.}(1988)\citenamefont
  {Neumann}, \citenamefont {N\"{o}the},\ and\ \citenamefont
  {Lipari}}]{Neumann1988}%
  \BibitemOpen
  \bibfield  {author} {\bibinfo {author} {\bibfnamefont {Ch.}\ \bibnamefont
  {Neumann}}, \bibinfo {author} {\bibfnamefont {A.}~\bibnamefont {N\"{o}the}},
  \ and\ \bibinfo {author} {\bibfnamefont {N.~O.}\ \bibnamefont {Lipari}},\
  }\bibfield  {title} {\enquote {\bibinfo {title} {Two-photon magnetoabsorption
  of {ZnTe}, {CdTe}, and {GaAs}},}\ }\href@noop {} {\bibfield  {journal}
  {\bibinfo  {journal} {Physical Review B}\ }\textbf {\bibinfo {volume} {37}},\
  \bibinfo {pages} {922} (\bibinfo {year} {1988})}\BibitemShut {NoStop}%
\bibitem [{\citenamefont {Skolnick}\ \emph {et~al.}(1976)\citenamefont
  {Skolnick}, \citenamefont {Jain}, \citenamefont {Stradling}, \citenamefont
  {Leotin},\ and\ \citenamefont {Ousset}}]{Skolnick1976}%
  \BibitemOpen
  \bibfield  {author} {\bibinfo {author} {\bibfnamefont {M~S}\ \bibnamefont
  {Skolnick}}, \bibinfo {author} {\bibfnamefont {A~K}\ \bibnamefont {Jain}},
  \bibinfo {author} {\bibfnamefont {R~A}\ \bibnamefont {Stradling}}, \bibinfo
  {author} {\bibfnamefont {J}~\bibnamefont {Leotin}}, \ and\ \bibinfo {author}
  {\bibfnamefont {J~C}\ \bibnamefont {Ousset}},\ }\bibfield  {title} {\enquote
  {\bibinfo {title} {An investigation of the anisotropy of the valence band of
  {GaAs} by cyclotron resonance},}\ } {\bibfield  {journal} {\bibinfo  {journal}
  {J. Phys. C}\ }\textbf {\bibinfo {volume}
  {9}},\ \bibinfo {pages} {2809} (\bibinfo {year} {1976})}\BibitemShut
  {NoStop}%
\bibitem [{Lan(1982)}]{Landoldt22a}%
  \BibitemOpen
  \bibfield  {title} {\enquote {\bibinfo {title} {Semiconductors. group IV
  elements, IV-IV and III-V compounds. lattice properties},}\ }in\ \href@noop
  {} {\emph {\bibinfo {booktitle} {Landoldt-B\"{o}rnstein Tables}}},\ Vol.\
  \bibinfo {volume} {New Series, Group III, 22a},\ \bibinfo {editor} {edited
  by\ \bibinfo {editor} {\bibfnamefont {O.}~\bibnamefont {Madelung}}}\
  (\bibinfo  {publisher} {Springer},\ \bibinfo {address} {Berlin},\ \bibinfo
  {year} {1982})\BibitemShut {NoStop}%
\bibitem [{\citenamefont {Molenkamp}\ \emph {et~al.}(1988)\citenamefont
  {Molenkamp}, \citenamefont {Eppenga}, \citenamefont {'t~Hooft}, \citenamefont
  {Dawson}, \citenamefont {Foxon},\ and\ \citenamefont
  {Moore}}]{Molenkamp1988}%
  \BibitemOpen
  \bibfield  {author} {\bibinfo {author} {\bibfnamefont {L.~W.}\ \bibnamefont
  {Molenkamp}}, \bibinfo {author} {\bibfnamefont {R.}~\bibnamefont {Eppenga}},
  \bibinfo {author} {\bibfnamefont {G.~W.}\ \bibnamefont {'t~Hooft}}, \bibinfo
  {author} {\bibfnamefont {P.}~\bibnamefont {Dawson}}, \bibinfo {author}
  {\bibfnamefont {C.~T.}\ \bibnamefont {Foxon}}, \ and\ \bibinfo {author}
  {\bibfnamefont {K.~J.}\ \bibnamefont {Moore}},\ }\bibfield  {title} {\enquote
  {\bibinfo {title} {Determination of valence-band effective-mass anisotropy in
  gaas quantum wells by optical spectroscopy},}\ } {\bibfield  {journal} {\bibinfo  {journal} {Phys.
  Rev. B}\ }\textbf {\bibinfo {volume} {38}},\ \bibinfo {pages} {4314}
  (\bibinfo {year} {1988})}\BibitemShut {NoStop}%
\bibitem [{\citenamefont {Shanabrook}\ \emph {et~al.}(1989)\citenamefont
  {Shanabrook}, \citenamefont {Glembocki}, \citenamefont {Broido},\ and\
  \citenamefont {Wang}}]{Shanabrook1989}%
  \BibitemOpen
  \bibfield  {author} {\bibinfo {author} {\bibfnamefont {B.~V.}\ \bibnamefont
  {Shanabrook}}, \bibinfo {author} {\bibfnamefont {O.~J.}\ \bibnamefont
  {Glembocki}}, \bibinfo {author} {\bibfnamefont {D.~A.}\ \bibnamefont
  {Broido}}, \ and\ \bibinfo {author} {\bibfnamefont {W.~I.}\ \bibnamefont
  {Wang}},\ }\bibfield  {title} {\enquote {\bibinfo {title} {Luttinger
  parameters for gaas determined from the intersubband transitions in
  GaAs}/Al$_{x}$Ga$_{1-x}$As
  multiple quantum wells}\ }
  {\bibfield  {journal} {\bibinfo  {journal} {Phys. Rev. B}\ }\textbf {\bibinfo
  {volume} {39}},\ \bibinfo {pages} {3411} (\bibinfo {year}
  {1989})}\BibitemShut {NoStop}%
\bibitem [{\citenamefont {Binggeli}\ and\ \citenamefont
  {Baldereschi}(1991)}]{Bingelli1991}%
  \BibitemOpen
  \bibfield  {author} {\bibinfo {author} {\bibfnamefont {N.}~\bibnamefont
  {Binggeli}}\ and\ \bibinfo {author} {\bibfnamefont {A.}~\bibnamefont
  {Baldereschi}},\ }\bibfield  {title} {\enquote {\bibinfo {title}
  {Determination of the hole effective masses in GaAs from acceptor spectra},}\
  } {\bibfield  {journal} {\bibinfo
  {journal} {Phys. Rev. B}\ }\textbf {\bibinfo {volume} {43}},\ \bibinfo
  {pages} {14734} (\bibinfo {year} {1991})}\BibitemShut {NoStop}%
\bibitem [{\citenamefont {Hackenberg}\ \emph {et~al.}(1994)\citenamefont
  {Hackenberg}, \citenamefont {Phillips},\ and\ \citenamefont
  {Hughes}}]{PhysRevB.50.10598}%
  \BibitemOpen
  \bibfield  {author} {\bibinfo {author} {\bibfnamefont {W.}~\bibnamefont
  {Hackenberg}}, \bibinfo {author} {\bibfnamefont {R.~T.}\ \bibnamefont
  {Phillips}}, \ and\ \bibinfo {author} {\bibfnamefont {H.~P.}\ \bibnamefont
  {Hughes}},\ }\bibfield  {title} {\enquote {\bibinfo {title} {Investigation of
  the luttinger parameters for inp using hot-electron luminescence},}\ } {\bibfield  {journal} {\bibinfo
  {journal} {Phys. Rev. B}\ }\textbf {\bibinfo {volume} {50}},\ \bibinfo
  {pages} {10598} (\bibinfo {year} {1994})}\BibitemShut {NoStop}%
\bibitem [{\citenamefont {Lipari}\ and\ \citenamefont
  {Baldereschi}(1970)}]{Baldereschi1970}%
  \BibitemOpen
  \bibfield  {author} {\bibinfo {author} {\bibfnamefont {N.~O.}\ \bibnamefont
  {Lipari}}\ and\ \bibinfo {author} {\bibfnamefont {A.}~\bibnamefont
  {Baldereschi}},\ }\bibfield  {title} {\enquote {\bibinfo {title} {{Angular
  momentum theory and localized states in solids. Investigation of shallow
  acceptor states in semiconductors}},}\ } {\bibfield  {journal} {\bibinfo  {journal}
  {Phys. Rev. Lett.}\ }\textbf {\bibinfo {volume} {25}},\ \bibinfo {pages}
  {1660} (\bibinfo {year} {1970})}\BibitemShut {NoStop}%
\bibitem [{\citenamefont {Kotlyar}\ \emph {et~al.}(2001)\citenamefont
  {Kotlyar}, \citenamefont {Reinecke}, \citenamefont {Bayer},\ and\
  \citenamefont {Forchel}}]{Kotlyar2001}%
  \BibitemOpen
  \bibfield  {author} {\bibinfo {author} {\bibfnamefont {R.}~\bibnamefont
  {Kotlyar}}, \bibinfo {author} {\bibfnamefont {T.~L.}\ \bibnamefont
  {Reinecke}}, \bibinfo {author} {\bibfnamefont {M.}~\bibnamefont {Bayer}}, \
  and\ \bibinfo {author} {\bibfnamefont {A.}~\bibnamefont {Forchel}},\
  }\bibfield  {title} {\enquote {\bibinfo {title} {Zeeman spin splittings in
  semiconductor nanostructures},}\ }
  {\bibfield  {journal} {\bibinfo  {journal} {Phys. Rev. B}\ }\textbf {\bibinfo
  {volume} {63}},\ \bibinfo {pages} {085310} (\bibinfo {year}
  {2001})}\BibitemShut {NoStop}%
\bibitem [{\citenamefont {Kapoor}\ \emph {et~al.}(2010)\citenamefont {Kapoor},
  \citenamefont {Kumar},\ and\ \citenamefont {Sen}}]{Kapoor2010}%
  \BibitemOpen
  \bibfield  {author} {\bibinfo {author} {\bibfnamefont {S.}~\bibnamefont
  {Kapoor}}, \bibinfo {author} {\bibfnamefont {J.}~\bibnamefont {Kumar}}, \
  and\ \bibinfo {author} {\bibfnamefont {P.K.}\ \bibnamefont {Sen}},\
  }\bibfield  {title} {\enquote {\bibinfo {title} {Magneto-optical analysis of
  anisotropic {CdZnSe} quantum dots},}\ }\href@noop {} {\bibfield  {journal}
  {\bibinfo  {journal} {Physica E}\
  }\textbf {\bibinfo {volume} {42}},\ \bibinfo {pages} {2380} (\bibinfo
  {year} {2010})}\BibitemShut {NoStop}%
\bibitem [{\citenamefont {Traynor}\ \emph {et~al.}(1995)\citenamefont
  {Traynor}, \citenamefont {Harley},\ and\ \citenamefont
  {Warburton}}]{Traynor1995}%
  \BibitemOpen
  \bibfield  {author} {\bibinfo {author} {\bibfnamefont {N.~J.}\ \bibnamefont
  {Traynor}}, \bibinfo {author} {\bibfnamefont {R.~T.}\ \bibnamefont {Harley}},
  \ and\ \bibinfo {author} {\bibfnamefont {R.~J.}\ \bibnamefont {Warburton}},\
  }\bibfield  {title} {\enquote {\bibinfo {title} {Zeeman splitting and g
  factor of heavy-hole excitons in
  In$_x$Ga$_{1-x}$As/ GaAs
  quantum wells},}\ } {\bibfield
  {journal} {\bibinfo  {journal} {Phys. Rev. B}\ }\textbf {\bibinfo {volume}
  {51}},\ \bibinfo {pages} {7361} (\bibinfo {year} {1995})}\BibitemShut
  {NoStop}%
\bibitem [{\citenamefont {Jadczak}\ \emph {et~al.}(2012)\citenamefont
  {Jadczak}, \citenamefont {Kubisa}, \citenamefont {Ryczko}, \citenamefont
  {Bryja},\ and\ \citenamefont {Potemski}}]{Potemski2012}%
  \BibitemOpen
  \bibfield  {author} {\bibinfo {author} {\bibfnamefont {J.}~\bibnamefont
  {Jadczak}}, \bibinfo {author} {\bibfnamefont {M.}~\bibnamefont {Kubisa}},
  \bibinfo {author} {\bibfnamefont {K.}~\bibnamefont {Ryczko}}, \bibinfo
  {author} {\bibfnamefont {L.}~\bibnamefont {Bryja}}, \ and\ \bibinfo {author}
  {\bibfnamefont {M.}~\bibnamefont {Potemski}},\ }\bibfield  {title} {\enquote
  {\bibinfo {title} {High magnetic field spin splitting of excitons in
  asymmetric GaAs quantum wells},}\ } {\bibfield  {journal} {\bibinfo  {journal} {Phys.
  Rev. B}\ }\textbf {\bibinfo {volume} {86}},\ \bibinfo {pages} {245401}
  (\bibinfo {year} {2012})}\BibitemShut {NoStop}%
\bibitem [{\citenamefont {Pfeffer}\ and\ \citenamefont
  {Zawadzki}(1996)}]{Pfeffer1996}%
  \BibitemOpen
  \bibfield  {author} {\bibinfo {author} {\bibfnamefont {P.}~\bibnamefont
  {Pfeffer}}\ and\ \bibinfo {author} {\bibfnamefont {W.}~\bibnamefont
  {Zawadzki}},\ }\bibfield  {title} {\enquote {\bibinfo {title}
  {Five-levelk$\cdot$pmodel for the conduction and valence bands of {GaAs} and
  {InP}},}\ } {\bibfield  {journal}
  {\bibinfo  {journal} {Phys. Rev. B}\ }\textbf {\bibinfo {volume} {53}},\
  \bibinfo {pages} {12813} (\bibinfo {year} {1996})}\BibitemShut
  {NoStop}%
\bibitem [{\citenamefont {Efros}\ and\ \citenamefont
  {Rosen}(1998)}]{Efros1998}%
  \BibitemOpen
  \bibfield  {author} {\bibinfo {author} {\bibfnamefont {{Al.}~L.}\
  \bibnamefont {Efros}}\ and\ \bibinfo {author} {\bibfnamefont
  {M.}~\bibnamefont {Rosen}},\ }\bibfield  {title} {\enquote {\bibinfo {title}
  {{Quantum size level structure of narrow-gap semiconductor nanocrystals:
  Effect of band coupling}},}\ }
  {\bibfield  {journal} {\bibinfo  {journal} {Phys. Rev. B}\ }\textbf {\bibinfo
  {volume} {58}},\ \bibinfo {pages} {7120} (\bibinfo {year}
  {1998})}\BibitemShut {NoStop}%
\bibitem [{\citenamefont {Snelling}\ \emph {et~al.}(1992)\citenamefont
  {Snelling}, \citenamefont {Blackwood}, \citenamefont {McDonagh},
  \citenamefont {Harley},\ and\ \citenamefont {Foxon}}]{Snelling1992}%
  \BibitemOpen
  \bibfield  {author} {\bibinfo {author} {\bibfnamefont {M.~J.}\ \bibnamefont
  {Snelling}}, \bibinfo {author} {\bibfnamefont {E.}~\bibnamefont {Blackwood}},
  \bibinfo {author} {\bibfnamefont {C.~J.}\ \bibnamefont {McDonagh}}, \bibinfo
  {author} {\bibfnamefont {R.~T.}\ \bibnamefont {Harley}}, \ and\ \bibinfo
  {author} {\bibfnamefont {C.~T.~B.}\ \bibnamefont {Foxon}},\ }\bibfield
  {title} {\enquote {\bibinfo {title} {Exciton, heavy-hole, and electron g
  factors in type-I
  GaAs/Al$_x$Ga$_{1-x}$As
  quantum wells},}\ } {\bibfield
  {journal} {\bibinfo  {journal} {Phys. Rev. B}\ }\textbf {\bibinfo {volume}
  {45}},\ \bibinfo {pages} {3922} (\bibinfo {year} {1992})}\BibitemShut
  {NoStop}%
\bibitem [{\citenamefont {Bayer}\ \emph {et~al.}(1995)\citenamefont {Bayer},
  \citenamefont {Timofeev}, \citenamefont {Gutbrod}, \citenamefont {Forchel},
  \citenamefont {Steffen},\ and\ \citenamefont {Oshinowo}}]{Bayer1995}%
  \BibitemOpen
  \bibfield  {author} {\bibinfo {author} {\bibfnamefont {M.}~\bibnamefont
  {Bayer}}, \bibinfo {author} {\bibfnamefont {V.~B.}\ \bibnamefont {Timofeev}},
  \bibinfo {author} {\bibfnamefont {T.}~\bibnamefont {Gutbrod}}, \bibinfo
  {author} {\bibfnamefont {A.}~\bibnamefont {Forchel}}, \bibinfo {author}
  {\bibfnamefont {R.}~\bibnamefont {Steffen}}, \ and\ \bibinfo {author}
  {\bibfnamefont {J.}~\bibnamefont {Oshinowo}},\ }\bibfield  {title} {\enquote
  {\bibinfo {title} {Enhancement of spin splitting due to spatial confinement
  in
  In$_{x}$Ga$_{1-x}$As
  quantum dots},}\ }{\bibfield
  {journal} {\bibinfo  {journal} {Phys. Rev. B}\ }\textbf {\bibinfo {volume}
  {52}},\ \bibinfo {pages} {R11623} (\bibinfo {year}
  {1995})}\BibitemShut {NoStop}%
\bibitem [{\citenamefont {Brodu}\ \emph {et~al.}(2019)\citenamefont {Brodu},
  \citenamefont {Chandrasekaran}, \citenamefont {Scarpelli}, \citenamefont
  {Buhot}, \citenamefont {Masia}, \citenamefont {Ballottin}, \citenamefont
  {Severijnen}, \citenamefont {Tessier}, \citenamefont {Dupont}, \citenamefont
  {Rabouw}, \citenamefont {Christianen}, \citenamefont {de~Mello~Donega},
  \citenamefont {Vanmaekelbergh}, \citenamefont {Langbein},\ and\ \citenamefont
  {Hens}}]{Brodu2019}%
  \BibitemOpen
  \bibfield  {author} {\bibinfo {author} {\bibfnamefont {A.}\
  \bibnamefont {Brodu}}, \bibinfo {author} {\bibfnamefont {V.}\
  \bibnamefont {Chandrasekaran}}, \bibinfo {author} {\bibfnamefont {L.}\
  \bibnamefont {Scarpelli}}, \bibinfo {author} {\bibfnamefont {J.}\
  \bibnamefont {Buhot}}, \bibinfo {author} {\bibfnamefont {F.}\
  \bibnamefont {Masia}}, \bibinfo {author} {\bibfnamefont {M.~V.}\
  \bibnamefont {Ballottin}}, \bibinfo {author} {\bibfnamefont {M.}\
  \bibnamefont {Severijnen}}, \bibinfo {author} {\bibfnamefont {M.~D.}\
  \bibnamefont {Tessier}}, \bibinfo {author} {\bibfnamefont {D.}\
  \bibnamefont {Dupont}}, \bibinfo {author} {\bibfnamefont {F.~T.}\
  \bibnamefont {Rabouw}}, \bibinfo {author} {\bibfnamefont {P. C.~M.}\
  \bibnamefont {Christianen}}, \bibinfo {author} {\bibfnamefont {C.}\
  \bibnamefont {de~Mello~Donega}}, \bibinfo {author} {\bibfnamefont {D.}\
  \bibnamefont {Vanmaekelbergh}}, \bibinfo {author} {\bibfnamefont {W.}\
  \bibnamefont {Langbein}}, \ and\ \bibinfo {author} {\bibfnamefont {Z.}\
  \bibnamefont {Hens}},\ }\bibfield  {title} {\enquote {\bibinfo {title} {Fine
  structure of nearly isotropic bright excitons in {InP}/{ZnSe} colloidal
  quantum dots},}\ } {\bibfield
  {journal} {\bibinfo  {journal} {J. Phys. Chem. Lett.}\
  }\textbf {\bibinfo {volume} {10}},\ \bibinfo {pages} {5468} (\bibinfo
  {year} {2019})}\BibitemShut {NoStop}%
\bibitem [{\citenamefont {Olutas}\ \emph {et~al.}(2015)\citenamefont {Olutas},
  \citenamefont {Guzelturk}, \citenamefont {Kelestemur}, \citenamefont
  {Yeltik}, \citenamefont {Delikanli},\ and\ \citenamefont
  {Demir}}]{Olutas2015_1}%
  \BibitemOpen
  \bibfield  {author} {\bibinfo {author} {\bibfnamefont {M.}\ \bibnamefont
  {Olutas}}, \bibinfo {author} {\bibfnamefont {B.k}\ \bibnamefont
  {Guzelturk}}, \bibinfo {author} {\bibfnamefont {Y.}\ \bibnamefont
  {Kelestemur}}, \bibinfo {author} {\bibfnamefont {A.}\ \bibnamefont
  {Yeltik}}, \bibinfo {author} {\bibfnamefont {S.}\ \bibnamefont
  {Delikanli}}, \ and\ \bibinfo {author} {\bibfnamefont {H.~V.}\
  \bibnamefont {Demir}},\ }\bibfield  {title} {\enquote {\bibinfo {title}
  {Lateral size-dependent spontaneous and stimulated emission properties in
  colloidal cdse nanoplatelets},}\ }
  {\bibfield  {journal} {\bibinfo  {journal} {ACS Nano}\ }\textbf {\bibinfo
  {volume} {9}},\ \bibinfo {pages} {5041} (\bibinfo {year} {2015})}
   \BibitemShut {NoStop}%
\bibitem [{\citenamefont {Ayari}\ \emph {et~al.}(2020)\citenamefont {Ayari},
  \citenamefont {Quick}, \citenamefont {Owschimikow}, \citenamefont
  {Christodoulou}, \citenamefont {Bertrand}, \citenamefont {Artemyev},
  \citenamefont {Moreels}, \citenamefont {Woggon}, \citenamefont {Jaziri},\
  and\ \citenamefont {Achtstein}}]{Ayari2020}%
  \BibitemOpen
  \bibfield  {author} {\bibinfo {author} {\bibfnamefont {S.}\ \bibnamefont
  {Ayari}}, \bibinfo {author} {\bibfnamefont {M.~T.}\ \bibnamefont
  {Quick}}, \bibinfo {author} {\bibfnamefont {N.}\ \bibnamefont
  {Owschimikow}}, \bibinfo {author} {\bibfnamefont {S.}\ \bibnamefont
  {Christodoulou}}, \bibinfo {author} {\bibfnamefont {G. H.~V.}\
  \bibnamefont {Bertrand}}, \bibinfo {author} {\bibfnamefont {M.}\
  \bibnamefont {Artemyev}}, \bibinfo {author} {\bibfnamefont {I.}\
  \bibnamefont {Moreels}}, \bibinfo {author} {\bibfnamefont {U.}\
  \bibnamefont {Woggon}}, \bibinfo {author} {\bibfnamefont {S.}\
  \bibnamefont {Jaziri}}, \ and\ \bibinfo {author} {\bibfnamefont
  {Alexander~W.}\ \bibnamefont {Achtstein}},\ }\bibfield  {title} {\enquote
  {\bibinfo {title} {Tuning trion binding energy and oscillator strength in a
  laterally finite 2d system: {CdSe} nanoplatelets as a model system for trion
  properties},}\ } {\bibfield  {journal}
  {\bibinfo  {journal} {Nanoscale}\ }\textbf {\bibinfo {volume} {12}},\
  \bibinfo {pages} {14448} (\bibinfo {year} {2020})}\BibitemShut
  {NoStop}%
\bibitem [{\citenamefont {Qiang}\ \emph {et~al.}(2020)\citenamefont {Qiang},
  \citenamefont {Golovatenko}, \citenamefont {Shornikova}, \citenamefont
  {Yakovlev}, \citenamefont {Rodina}, \citenamefont {Zhukov}, \citenamefont
  {Kalitukha}, \citenamefont {Sapega}, \citenamefont {Kaibyshev}, \citenamefont
  {Prosnikov}, \citenamefont {Christianen}, \citenamefont {Onushchenko},\ and\
  \citenamefont {Bayer}}]{qiang2020}%
  \BibitemOpen
  \bibfield  {author} {\bibinfo {author} {\bibfnamefont {G.}\ \bibnamefont
  {Qiang}}, \bibinfo {author} {\bibfnamefont {A.~A.}\ \bibnamefont
  {Golovatenko}}, \bibinfo {author} {\bibfnamefont {E.~V.}\ \bibnamefont
  {Shornikova}}, \bibinfo {author} {\bibfnamefont {D.~R.}\ \bibnamefont
  {Yakovlev}}, \bibinfo {author} {\bibfnamefont {A.~V.}\ \bibnamefont
  {Rodina}}, \bibinfo {author} {\bibfnamefont {E.~A.}\ \bibnamefont
  {Zhukov}}, \bibinfo {author} {\bibfnamefont {I.~V.}\ \bibnamefont
  {Kalitukha}}, \bibinfo {author} {\bibfnamefont {V.~F.}\ \bibnamefont
  {Sapega}}, \bibinfo {author} {\bibfnamefont {V.~K.}\ \bibnamefont
  {Kaibyshev}}, \bibinfo {author} {\bibfnamefont {M.~A.}\ \bibnamefont
  {Prosnikov}}, \bibinfo {author} {\bibfnamefont {P. C.~M.}\ \bibnamefont
  {Christianen}}, \bibinfo {author} {\bibfnamefont {A.~A.}\ \bibnamefont
  {Onushchenko}}, \ and\ \bibinfo {author} {\bibfnamefont {M.}\
  \bibnamefont {Bayer}},\ }\bibfield  {title} {\enquote {\bibinfo {title}
  {Polarized emission of cdse nanocrystals in magnetic field: the role of
  phonon-assisted recombination of the dark exciton},}\ }\href@noop {}
  {\bibfield  {journal} {\bibinfo  {journal} {arXiv:2011.02786}\ } (\bibinfo {year} {2020})}\BibitemShut {NoStop}%
\bibitem [{\citenamefont {Rodina}\ and\ \citenamefont
  {Efros}(2016)}]{Rodina2016}%
  \BibitemOpen
  \bibfield  {author} {\bibinfo {author} {\bibfnamefont {A.~V.}\ \bibnamefont
  {Rodina}}\ and\ \bibinfo {author} {\bibfnamefont {Al.~L.}\ \bibnamefont
  {Efros}},\ }\bibfield  {title} {\enquote {\bibinfo {title} {Radiative
  recombination from dark excitons in nanocrystals: Activation mechanisms and
  polarization properties},}\ }
  {\bibfield  {journal} {\bibinfo  {journal} {Phys. Rev. B}\ }\textbf {\bibinfo
  {volume} {93}},\ \bibinfo {pages} {155427} (\bibinfo {year}
  {2016})}\BibitemShut {NoStop}%
\bibitem [{\citenamefont {Cho}\ \emph {et~al.}(1975)\citenamefont {Cho},
  \citenamefont {Suga}, \citenamefont {Dreybrodt},\ and\ \citenamefont
  {Willmann}}]{PhysRevB.11.1512}%
  \BibitemOpen
  \bibfield  {author} {\bibinfo {author} {\bibfnamefont {K.}~\bibnamefont
  {Cho}}, \bibinfo {author} {\bibfnamefont {S.}~\bibnamefont {Suga}}, \bibinfo
  {author} {\bibfnamefont {W.}~\bibnamefont {Dreybrodt}}, \ and\ \bibinfo
  {author} {\bibfnamefont {F.}~\bibnamefont {Willmann}},\ }\bibfield  {title}
  {\enquote {\bibinfo {title} {Theory of degenerate $1s$ excitons in
  zinc-blende-type crystals in a magnetic field: Exchange interaction and cubic
  anisotropy},}\ } {\bibfield
  {journal} {\bibinfo  {journal} {Phys. Rev. B}\ }\textbf {\bibinfo {volume}
  {11}},\ \bibinfo {pages} {1512} (\bibinfo {year} {1975})}\BibitemShut
  {NoStop}%
\bibitem [{\citenamefont {Nirmal}\ \emph {et~al.}(1995)\citenamefont {Nirmal},
  \citenamefont {Norris}, \citenamefont {Kuno}, \citenamefont {Bawendi},
  \citenamefont {Efros},\ and\ \citenamefont {Rosen}}]{Nirmal1995}%
  \BibitemOpen
  \bibfield  {author} {\bibinfo {author} {\bibfnamefont {M.}~\bibnamefont
  {Nirmal}}, \bibinfo {author} {\bibfnamefont {D.J.}~\bibnamefont {Norris}},
  \bibinfo {author} {\bibfnamefont {M.}~\bibnamefont {Kuno}}, \bibinfo {author}
  {\bibfnamefont {M.G.}~\bibnamefont {Bawendi}}, \bibinfo {author} {\bibfnamefont
  {{Al.}~L.}\ \bibnamefont {Efros}}, \ and\ \bibinfo {author} {\bibfnamefont
  {M.}~\bibnamefont {Rosen}},\ }\bibfield  {title} {\enquote {\bibinfo {title}
  {{Observation of the ''dark exciton" in CdSe quantum dots}},}\ } {\bibfield  {journal} {\bibinfo
  {journal} {Phys. Rev. Lett.}\ }\textbf {\bibinfo {volume} {75}},\ \bibinfo
  {pages} {3728--3731} (\bibinfo {year} {1995})}\BibitemShut {NoStop}%
\bibitem [{\citenamefont {Johnston-Halperin}\ \emph {et~al.}(2001)\citenamefont
  {Johnston-Halperin}, \citenamefont {Awschalom}, \citenamefont {Crooker},
  \citenamefont {Efros}, \citenamefont {Rosen}, \citenamefont {Peng},\ and\
  \citenamefont {Alivisatos}}]{JohnstonHalperin2001}%
  \BibitemOpen
  \bibfield  {author} {\bibinfo {author} {\bibfnamefont {E.}~\bibnamefont
  {Johnston-Halperin}}, \bibinfo {author} {\bibfnamefont {D.~D.}\ \bibnamefont
  {Awschalom}}, \bibinfo {author} {\bibfnamefont {S.~A.}\ \bibnamefont
  {Crooker}}, \bibinfo {author} {\bibfnamefont {Al.~L.}\ \bibnamefont {Efros}},
  \bibinfo {author} {\bibfnamefont {M.}~\bibnamefont {Rosen}}, \bibinfo
  {author} {\bibfnamefont {X.}~\bibnamefont {Peng}}, \ and\ \bibinfo {author}
  {\bibfnamefont {A.~P.}\ \bibnamefont {Alivisatos}},\ }\bibfield  {title}
  {\enquote {\bibinfo {title} {{Spin spectroscopy of dark excitons in CdSe
  quantum dots to 60 T}},}\ }
  {\bibfield  {journal} {\bibinfo  {journal} {Phys. Rev. B}\ }\textbf {\bibinfo
  {volume} {63}},\ \bibinfo {pages} {205309} (\bibinfo {year}
  {2001})}\BibitemShut {NoStop}%
\bibitem [{\citenamefont {del {\'{A}}guila}\ \emph {et~al.}(2017)\citenamefont
  {del {\'{A}}guila}, \citenamefont {Pettinari}, \citenamefont {Groeneveld},
  \citenamefont {de~Mello~Doneg{\'{a}}}, \citenamefont {Vanmaekelbergh},
  \citenamefont {Maan},\ and\ \citenamefont
  {Christianen}}]{Granadosdelguila2017}%
  \BibitemOpen
  \bibfield  {author} {\bibinfo {author} {\bibfnamefont {A.~Granados}\
  \bibnamefont {del {\'{A}}guila}}, \bibinfo {author} {\bibfnamefont
  {G.}~\bibnamefont {Pettinari}}, \bibinfo {author} {\bibfnamefont
  {E.}~\bibnamefont {Groeneveld}}, \bibinfo {author} {\bibfnamefont
  {C.}~\bibnamefont {de~Mello~Doneg{\'{a}}}}, \bibinfo {author} {\bibfnamefont
  {D.}~\bibnamefont {Vanmaekelbergh}}, \bibinfo {author} {\bibfnamefont
  {J.~C.}\ \bibnamefont {Maan}}, \ and\ \bibinfo {author} {\bibfnamefont
  {P.~C.~M.}\ \bibnamefont {Christianen}},\ }\bibfield  {title} {\enquote
  {\bibinfo {title} {Optical spectroscopy of dark and bright excitons in {CdSe}
  nanocrystals in high magnetic fields},}\ } {\bibfield  {journal} {\bibinfo  {journal} {The
  J. Phys. Chem. C}\ }\textbf {\bibinfo {volume} {121}},\
  \bibinfo {pages} {23693} (\bibinfo {year} {2017})}\BibitemShut
  {NoStop}%
\bibitem [{\citenamefont {Adachi}(2005)}]{Adachi2005}%
  \BibitemOpen
  \bibfield  {author} {\bibinfo {author} {\bibfnamefont {Sadao}\ \bibnamefont
  {Adachi}},\ } {\emph {\bibinfo {title}
  {Properties of Group-{IV}, {III}-V and {II}-{VI} Semiconductors}}}\ (\bibinfo
   {publisher} {John Wiley {\&} Sons, Ltd},\ \bibinfo {year}
  {2005})\BibitemShut {NoStop}%
\bibitem [{\citenamefont {Benchamekh}\ \emph {et~al.}(2014)\citenamefont
  {Benchamekh}, \citenamefont {Gippius}, \citenamefont {Even}, \citenamefont
  {Nestoklon}, \citenamefont {Jancu}, \citenamefont {Ithurria}, \citenamefont
  {Dubertret}, \citenamefont {Efros},\ and\ \citenamefont
  {Voisin}}]{Benchamekh2014}%
  \BibitemOpen
  \bibfield  {author} {\bibinfo {author} {\bibfnamefont {R.}~\bibnamefont
  {Benchamekh}}, \bibinfo {author} {\bibfnamefont {N.~A.}\ \bibnamefont
  {Gippius}}, \bibinfo {author} {\bibfnamefont {J.}~\bibnamefont {Even}},
  \bibinfo {author} {\bibfnamefont {M.~O.}\ \bibnamefont {Nestoklon}}, \bibinfo
  {author} {\bibfnamefont {J.-M.}\ \bibnamefont {Jancu}}, \bibinfo {author}
  {\bibfnamefont {S.}~\bibnamefont {Ithurria}}, \bibinfo {author}
  {\bibfnamefont {B.}~\bibnamefont {Dubertret}}, \bibinfo {author}
  {\bibfnamefont {{Al.}~L.}\ \bibnamefont {Efros}}, \ and\ \bibinfo {author}
  {\bibfnamefont {P.}~\bibnamefont {Voisin}},\ }\bibfield  {title} {\enquote
  {\bibinfo {title} {Tight-binding calculations of image-charge effects in
  colloidal nanoscale platelets of CdSe},}\ } {\bibfield  {journal} {\bibinfo  {journal} {Phys.
  Rev. B}\ }\textbf {\bibinfo {volume} {89}},\ \bibinfo {pages} {035307}
  (\bibinfo {year} {2014})}\BibitemShut {NoStop}%
\bibitem [{\citenamefont {Wang}\ \emph {et~al.}(2015)\citenamefont {Wang},
  \citenamefont {Bouet}, \citenamefont {Glazov}, \citenamefont {Amand},
  \citenamefont {Ivchenko}, \citenamefont {Palleau}, \citenamefont {Marie},\
  and\ \citenamefont {Urbaszek}}]{Wang_2015}%
  \BibitemOpen
  \bibfield  {author} {\bibinfo {author} {\bibfnamefont {G}~\bibnamefont
  {Wang}}, \bibinfo {author} {\bibfnamefont {L}~\bibnamefont {Bouet}}, \bibinfo
  {author} {\bibfnamefont {M~M}\ \bibnamefont {Glazov}}, \bibinfo {author}
  {\bibfnamefont {T}~\bibnamefont {Amand}}, \bibinfo {author} {\bibfnamefont
  {E~L}\ \bibnamefont {Ivchenko}}, \bibinfo {author} {\bibfnamefont
  {E}~\bibnamefont {Palleau}}, \bibinfo {author} {\bibfnamefont
  {X}~\bibnamefont {Marie}}, \ and\ \bibinfo {author} {\bibfnamefont
  {B}~\bibnamefont {Urbaszek}},\ }\bibfield  {title} {\enquote {\bibinfo
  {title} {Magneto-optics in transition metal diselenide monolayers},}\ } {\bibfield  {journal} {\bibinfo
  {journal} {2D Materials}\ }\textbf {\bibinfo {volume} {2}},\ \bibinfo {pages}
  {034002} (\bibinfo {year} {2015})}\BibitemShut {NoStop}%
\end{thebibliography}

%

\end{document}